\documentclass[twocolumn]{aastex63}

\shorttitle{MAPS III: Chemical Substructures}
\graphicspath{{./}{figures/}}

\newcommand{\clean}{CLEAN}

\begin{document}

\title{Molecules with ALMA at Planet-forming Scales (MAPS) III: Characteristics of Radial Chemical Substructures}

\correspondingauthor{Charles J. Law}
\email{charles.law@cfa.harvard.edu}

\author[0000-0003-1413-1776]{Charles J. Law}
\affiliation{Center for Astrophysics \textbar\, Harvard \& Smithsonian, 60 Garden St., Cambridge, MA 02138, USA}

\author[0000-0002-8932-1219]{Ryan A. Loomis}
\affiliation{National Radio Astronomy Observatory, 520 Edgemont Rd., Charlottesville, VA 22903, USA}

\author[0000-0003-1534-5186]{Richard Teague}
\affiliation{Center for Astrophysics \textbar\, Harvard \& Smithsonian, 60 Garden St., Cambridge, MA 02138, USA}

\author[0000-0001-8798-1347]{Karin I. \"Oberg}
\affiliation{Center for Astrophysics \textbar\, Harvard \& Smithsonian, 60 Garden St., Cambridge, MA 02138, USA}

\author[0000-0002-1483-8811]{Ian Czekala}
\altaffiliation{NASA Hubble Fellowship Program Sagan Fellow}
\affiliation{Department of Astronomy and Astrophysics, 525 Davey Laboratory, The Pennsylvania State University, University Park, PA 16802, USA}
\affiliation{Center for Exoplanets and Habitable Worlds, 525 Davey Laboratory, The Pennsylvania State University, University Park, PA 16802, USA}
\affiliation{Center for Astrostatistics, 525 Davey Laboratory, The Pennsylvania State University, University Park, PA 16802, USA}
\affiliation{Institute for Computational \& Data Sciences, The Pennsylvania State University, University Park, PA 16802, USA}
\affiliation{Department of Astronomy, 501 Campbell Hall, University of California, Berkeley, CA 94720-3411, USA}

\author[0000-0003-2253-2270]{Sean M. Andrews}
\affiliation{Center for Astrophysics \textbar\, Harvard \& Smithsonian, 60 Garden St., Cambridge, MA 02138, USA}

\author[0000-0001-6947-6072]{Jane Huang}
\altaffiliation{NASA Hubble Fellowship Program Sagan Fellow}
\affiliation{Center for Astrophysics \textbar\, Harvard \& Smithsonian, 60 Garden St., Cambridge, MA 02138, USA}
\affiliation{Department of Astronomy, University of Michigan, 323 West Hall, 1085 South University Avenue, Ann Arbor, MI 48109, USA}

\author[0000-0003-3283-6884]{Yuri Aikawa}
\affiliation{Department of Astronomy, Graduate School of Science, The University of Tokyo, Tokyo 113-0033, Japan}

\author[0000-0002-2692-7862]{Felipe Alarc\'on}
\affiliation{Department of Astronomy, University of Michigan, 323 West Hall, 1085 South University Avenue, Ann Arbor, MI 48109, USA}

\author[0000-0001-7258-770X]{Jaehan Bae}
\altaffiliation{NASA Hubble Fellowship Program Sagan Fellow}
\affil{Earth and Planets Laboratory, Carnegie Institution for Science, 5241 Broad Branch Road NW, Washington, DC 20015, USA}
\affiliation{Department of Astronomy, University of Florida, Gainesville, FL 32611, USA}

\author[0000-0003-4179-6394]{Edwin A.\ Bergin}
\affiliation{Department of Astronomy, University of Michigan, 323 West Hall, 1085 South University Avenue, Ann Arbor, MI 48109, USA}

\author[0000-0002-8716-0482]{Jennifer B. Bergner} 
\altaffiliation{NASA Hubble Fellowship Program Sagan Fellow}
\affiliation{University of Chicago Department of the Geophysical Sciences, Chicago, IL 60637, USA}

\author[0000-0002-8692-8744]{Yann Boehler}
\affiliation{Univ. Grenoble Alpes, CNRS, IPAG, F-38000 Grenoble, France}

\author[0000-0003-2014-2121]{Alice S. Booth}
\affiliation{Leiden Observatory, Leiden University, 2300 RA Leiden, the Netherlands}
\affiliation{School of Physics and Astronomy, University of Leeds, Leeds, UK, LS2 9JT}

\author[0000-0003-4001-3589]{Arthur D. Bosman}
\affiliation{Department of Astronomy, University of Michigan, 323 West Hall, 1085 South University Avenue, Ann Arbor, MI 48109, USA}

\author[0000-0002-0150-0125]{Jenny K. Calahan} 
\affiliation{Department of Astronomy, University of Michigan, 323 West Hall, 1085 South University Avenue, Ann Arbor, MI 48109, USA}

\author[0000-0002-2700-9676]{Gianni Cataldi}
\affiliation{Department of Astronomy, Graduate School of Science, The University of Tokyo, Tokyo 113-0033, Japan}
\affiliation{National Astronomical Observatory of Japan, 2-21-1 Osawa, Mitaka, Tokyo 181-8588, Japan}

\author[0000-0003-2076-8001]{L. Ilsedore Cleeves}
\affiliation{Department of Astronomy, University of Virginia, Charlottesville, VA 22904, USA}

\author[0000-0002-2026-8157]{Kenji Furuya}
\affiliation{National Astronomical Observatory of Japan, 2-21-1 Osawa, Mitaka, Tokyo 181-8588, Japan}

\author[0000-0003-4784-3040]{Viviana V. Guzm\'{a}n}
\affiliation{Instituto de Astrof\'isica, Pontificia Universidad Cat\'olica de Chile, Av. Vicu\~na Mackenna 4860, 7820436 Macul, Santiago, Chile}

\author[0000-0003-1008-1142]{John~D.~Ilee}
\affil{School of Physics and Astronomy, University of Leeds, Leeds, UK, LS2 9JT}

\author[0000-0003-1837-3772]{Romane Le Gal}
\affiliation{Center for Astrophysics \textbar\, Harvard \& Smithsonian, 60 Garden St., Cambridge, MA 02138, USA}
\affiliation{Univ. Grenoble Alpes, CNRS, IPAG, F-38000 Grenoble, France}
\affiliation{IRAP, Universit\'{e} de Toulouse, CNRS, CNES, UT3, 31400 Toulouse, France}
\affiliation{IRAM, 300 rue de la piscine, F-38406 Saint-Martin d'H\`{e}res, France}

\author[0000-0002-7616-666X]{Yao Liu}
\affiliation{Purple Mountain Observatory \& Key Laboratory for Radio Astronomy, Chinese Academy of Sciences, Nanjing 210023, China}

\author[0000-0002-7607-719X]{Feng Long}
\affiliation{Center for Astrophysics \textbar\, Harvard \& Smithsonian, 60 Garden St., Cambridge, MA 02138, USA}

\author[0000-0002-1637-7393]{Fran\c cois M\'enard}
\affiliation{Univ. Grenoble Alpes, CNRS, IPAG, F-38000 Grenoble, France}

\author[0000-0002-7058-7682]{Hideko Nomura}
\affiliation{National Astronomical Observatory of Japan, 2-21-1 Osawa, Mitaka, Tokyo 181-8588, Japan}

\author[0000-0001-8642-1786]{Chunhua Qi}
\affiliation{Center for Astrophysics \textbar\, Harvard \& Smithsonian, 60 Garden St., Cambridge, MA 02138, USA}

\author[0000-0002-6429-9457]{Kamber R. Schwarz}
\altaffiliation{NASA Hubble Fellowship Program Sagan Fellow}
\affiliation{Lunar and Planetary Laboratory, University of Arizona, 1629 E. University Blvd, Tucson, AZ 85721, USA}

\author[0000-0002-5991-8073]{Anibal Sierra} \affiliation{Departamento de Astronom\'ia, Universidad de Chile, Camino El Observatorio 1515, Las Condes, Santiago, Chile}

\author[0000-0002-6034-2892]{Takashi Tsukagoshi} \affiliation{National Astronomical Observatory of Japan, 2-21-1 Osawa, Mitaka, Tokyo 181-8588, Japan}

\author[0000-0003-4099-6941]{Yoshihide Yamato}
\affiliation{Department of Astronomy, Graduate School of Science, The University of Tokyo, Tokyo 113-0033, Japan}

\author[0000-0002-2555-9869]{Merel L. R. van 't Hoff}
\affiliation{Department of Astronomy, University of Michigan, 323 West Hall, 1085 South University Avenue, Ann Arbor, MI 48109, USA}

\author[0000-0001-6078-786X]{Catherine Walsh}
\affiliation{School of Physics and Astronomy, University of Leeds, Leeds, UK, LS2 9JT}

\author[0000-0003-1526-7587]{David J. Wilner}
\affiliation{Center for Astrophysics \textbar\, Harvard \& Smithsonian, 60 Garden St., Cambridge, MA 02138, USA}

\author[0000-0002-0661-7517]{Ke Zhang}
\altaffiliation{NASA Hubble Fellow}
\affiliation{Department of Astronomy, University of Michigan, 323 West Hall, 1085 South University Avenue, Ann Arbor, MI 48109, USA}
\affiliation{Department of Astronomy, University of Wisconsin-Madison, 475 N Charter St, Madison, WI 53706}

%% AASTeX 6.3 has the new \collaboration and \nocollaboration commands to provide the collaboration status of a group of authors. These commands can be used either before or after the list of corresponding authors. The argument for \collaboration is the collaboration identifier. Authors are encouraged to surround collaboration identifiers with ()s. The \nocollaboration command takes no argument and exists to indicate that the nearby authors are not part of surrounding collaborations.

% detailed view of the environment where planets form
\begin{abstract}
The Molecules with ALMA at Planet-forming Scales (MAPS) Large Program provides a detailed, high resolution (${\sim}$10--20~au) view of molecular line emission in five protoplanetary disks at spatial scales relevant for planet formation. Here, we present a systematic analysis of chemical substructures in 18~molecular lines toward the MAPS sources: IM~Lup, GM~Aur, AS~209, HD~163296, and MWC~480. We identify more than 200 chemical substructures, which are found at nearly all radii where line emission is detected. A wide diversity of radial morphologies --- including rings, gaps, and plateaus --- is observed both within each disk and across the MAPS sample. This diversity in line emission profiles is also present in the innermost 50~au. Overall, this suggests that planets form in varied chemical environments both across disks and at different radii within the same disk. Interior to 150~au, the majority of chemical substructures across the MAPS disks are spatially coincident with substructures in the millimeter continuum, indicative of physical and chemical links between the disk midplane and warm, elevated molecular emission layers. Some chemical substructures in the inner disk and most chemical substructures exterior to 150~au cannot be directly linked to dust substructure, however, which indicates that there are also other causes of chemical substructures, such as snowlines, gradients in UV~photon fluxes, ionization, and radially-varying elemental ratios. This implies that chemical substructures could be developed into powerful probes of different disk characteristics, in addition to influencing the environments within which planets assemble. This paper is part of the MAPS special issue of the Astrophysical Journal Supplement.
\end{abstract}
%% See the online documentation for the full list of available subject keywords and the rules for their use.

\section{Introduction} \label{sec:intro}

Protoplanetary disks provide the constituent materials necessary for forming planets. The colliding and coalescing of dust grains leads to the formation of pebbles, which grow into planetesimals and ultimately planets \citep[e.g.,][]{Mordasini08}, while the spatial distribution of ice and gas sets the volatile compositions of incipient planets \citep{Oberg11}. The diversity of known exoplanetary systems \citep[e.g.,][]{Batalha13} may originate, at least in part, due to differences in the gas and dust distribution observed across protoplanetary disks \citep[e.g.,][]{Mordasini12}. Disk observations can thus provide crucial constraints on the formation locations of planets \citep{Zhu14, Zhang18} and the processes by which initial gas and dust distributions evolve into planetary systems \citep{Birnstiel15, vanderMarel15, Perez15, Andrews20_ann_rev}. Moreover, the organic compositions of planets are linked to the chemistry of their parental disks \citep[][]{Oberg16, Cridland16, Cridland17}, which makes a detailed understanding of the chemical environment in which young planets form of particular interest to origins of life studies.

Disk chemistry is regulated by a combination of inherited material and \textit{in situ} processes that depend on density, temperature, and radiation fields. While models of chemical structures often assume smoothly-decreasing surface densities and temperatures \citep{Hughes08, Andrews09}, disks are now known to be highly-structured in their dust \citep{ALMA15, Andrews16, Long18} and gas \citep{Isella_ea_2016, Teague_ea_2017, Huang18TWHya}. The DSHARP program \citep{Huang18, Andrews18}, and subsequent observations \citep[e.g.,][]{Facchini20, Cieza21}, showed that dust substructures at 1-to-10 au scales in the form of rings, gaps, and spirals are ubiquitous in protoplanetary disks. Disks also possess complex gas distributions and exhibit gradients in C/N/O ratios and organic molecules \citep[e.g.,][]{Bergin16, Cleeves16, Iseall16PhRvL, vanderMarel16, Huang17, Bergner18, Huang18TWHya, Kastner18, Bergner19, Garufi20, Pegues20, Facchini21PDS, Booth21_CH3OH}. However, the majority of molecular line observations have been limited to coarser angular resolutions (${\sim}$0\farcs5--1\farcs0), which trace physical scales of 50--150~au at typical distances (${\sim}$100--150~pc) of nearby disks \citep[e.g.,][]{Dartois03, Pietu07, Chapillon12, Mathews13, Gregorio13, Flaherty17, Salinas17, LeGal19}. The relatively small number of studies at high spatial resolutions (${<}$0\farcs3) have focused on CO \citep{Fedele17, pinte18, Isella18, Favre19, Rosotti20, Wolfer21}, and perhaps one or two additional molecules such as C$_2$H \citep{Bergin16, Miotello19}, CN \citep{vanTerwisga19, Teague_Loomis_2020}, HCO$^+$ \citep{Long_Zach18, Tsukagoshi19, Huang20}, H$_2$CO \citep{Podio19}, and CS \citep{Rosotti21, Nomura21}. Thus, the detailed structure of the gas component of disks remains largely unexplored, especially toward the inner, planet-forming regions (${<}100$~au). 

Hence, the relationship between chemical and dust structure at small scales is unclear. As the distribution of dust strongly impacts the chemistry \citep[][]{Cleeves16, Facchini17, vanderMarel18}, it is expected that the presence of dust substructures will also alter local chemical environments. This is because the total surface area of dust present throughout the disk is linked to many physical and chemical processes, such as the disk thermal structure, thermal coupling between the gas and solid phases, and balance between freeze-out and desorption. Additionally, grain growth and vertical settling affect the penetration depth of UV photons, which in turn alters the chemistry and gas temperatures \citep[e.g.,][]{Fogel11, Akimkin13, Cleeves16}.

A systematic analysis of a wide set of molecular lines is required in order to assess the relationship between chemical and dust substructures in disks and establish a clearer understanding of the chemical environments in which planets form. To this end, we quantitatively characterize the properties of chemical substructures observed as part of the Molecules with ALMA at Planet-forming Scales (MAPS) Large Program \citep{oberg20}. In Section \ref{sec:generation_moment_profiles}, we discuss the generation of moment maps and radial profiles. In Section \ref{sec:disk_features_section}, we describe how we measured the locations, widths, and relative contrasts of the observed chemical substructures. We present aggregate properties of substructures in Section \ref{sec:aggregate_properties} and examine spatial trends in their locations and discuss possible origins in Section \ref{sec:origins_of_chemical_substr}. We summarize our findings in Section \ref{sec:conclusions} and provide a listing of all available data products in Section \ref{sec:VADPs_listing}.

\section{Generation of Moment Maps and Radial Profiles} \label{sec:generation_moment_profiles}

\subsection{Observations} \label{sec:observations}

\begin{figure*}
\centering
\includegraphics[width=\linewidth]{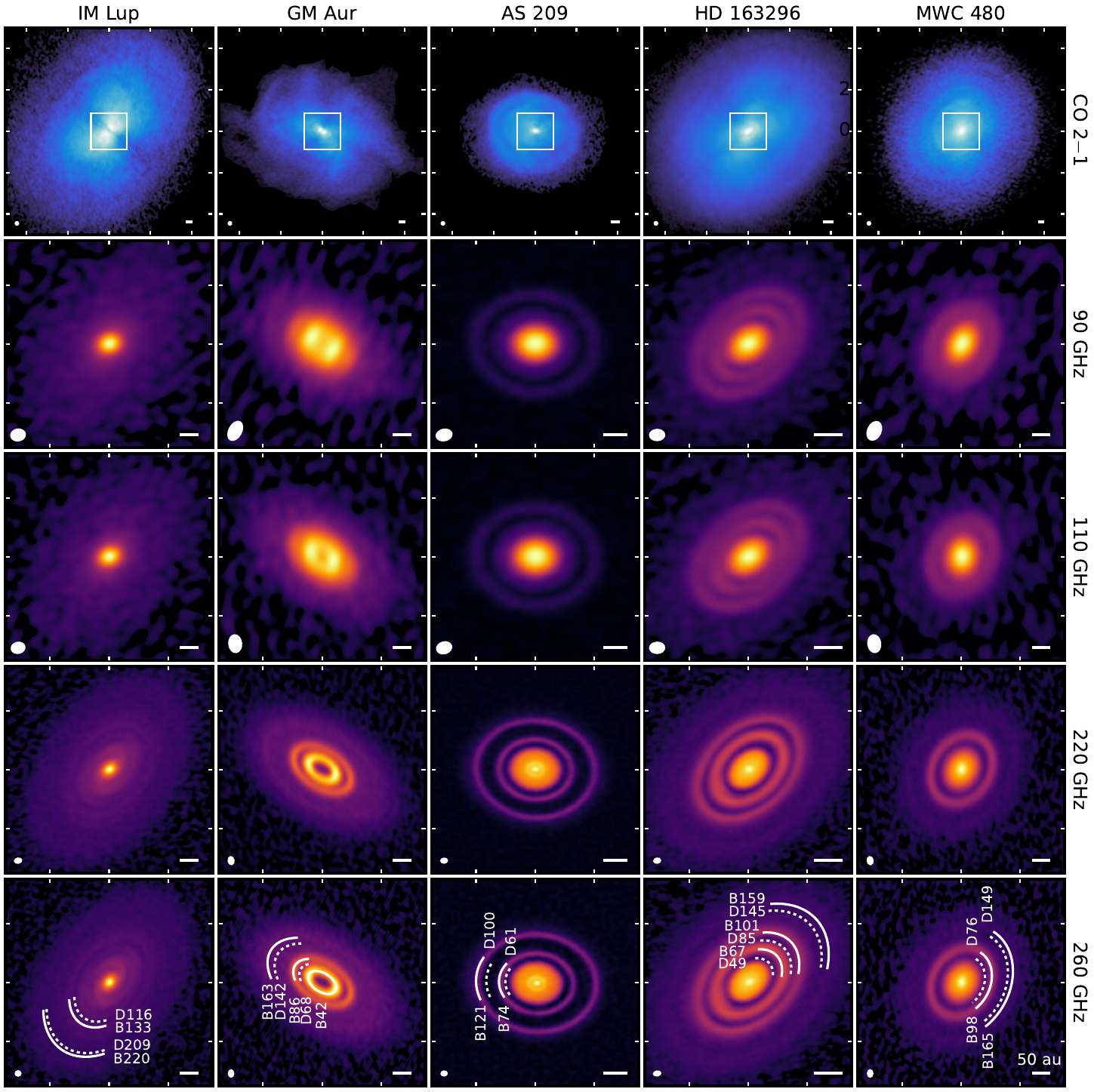}
\caption{Zeroth moment maps of CO 2$-$1 and continuum images for the MAPS sample, ordered from left to right by increasing stellar mass (see Table 1 in \citet{oberg20}). Axes are angular offsets from the disk center and the white rectangle overlaid on the CO 2$-$1 zeroth moment maps defines the field of view of the continuum images. Each tick mark is 2$^{\prime \prime}$ and 1$^{\prime \prime}$ for the CO 2$-$1 and continuum images, respectively. Color stretches were individually optimized and applied to each panel to increase the visibility of substructures. Care should thus be taken when comparing between panels, and instead, we recommend using the corresponding radial profiles in Figure \ref{fig:CO_wcont_radial_profiles} for this purpose. Continuum substructures, as described in Section \ref{sec:continuum_substructures}, are labeled on the 260~GHz continuum images following the nomenclature of \citet{Huang18}. Rings and gaps are shown as solid and dotted arcs, respectively, with azimuthal extents chosen for maximal visual clarity. The synthesized beam and a scale bar indicating 50~au is shown in the lower left and right corner, respectively, of each panel.}
\label{fig:CO_with_continuum_Moment0}
\end{figure*}

The MAPS Large Program (2018.1.01055.L) targeted the protoplanetary disks around IM~Lup, GM~Aur, AS~209, HD~163296, and MWC~480 in four spectral setups in ALMA Bands 6 and 3. Figure \ref{fig:CO_with_continuum_Moment0} shows an overview of each disk in CO $2-1$ and continuum emission. The analysis presented here is based on the fiducial images, as described in \citet{oberg20}, which have 0\farcs15 and 0\farcs30 circularized beams for lines in Bands 6 and 3, respectively. For those transitions covered in Band 6 that were either marginally detected or lacked sufficient signal-to-noise ratio (SNR), we instead used the corresponding tapered (0\farcs30) images \citep[see Section 6.2,][]{czekala20}. The correction for a significantly non-Gaussian dirty beam, i.e., the ``JvM correction'' first described in \citet{Jorsater95}, is salient to the following discussion and is explained in detail in \citet{czekala20}, together with the full imaging procedure. Briefly, the application of the ``JvM-correction'' correctly scales the residuals in the image cube to be in units consistent with the CLEAN model. This ensures that the starting point for the moment map generation, the CLEANed image, is in the correct units of Jy\,\{$\mathrm{CLEAN\,beam}^{-1}$\}. \citet{oberg20} provides details about the observational setup and calibration, as well as basic information about each image, including the JvM-corrected RMS noise level.

We focus this work on 18 lines, listed in Table \ref{tab:ProfSelection}, that are sufficiently bright and spatially extended to allow for an analysis of radial substructures. We analyzed only the brightest component of those transitions with multiple hyperfine components, namely C$_2$H N=3--2, J=$\frac{7}{2}$--$\frac{5}{2}$, F=4--3; C$_2$H N=1--0, J=$\frac{3}{2}$--$\frac{1}{2}$, F=2--1; c-C$_3$H$_2$ (J$_{\rm{K}_{\rm{a}}, {\rm{K}_{\rm{c}}}}$)=7$_{07}$--6$_{16}$/7$_{17}$--6$_{06}$, HCN J=3--2, F=3--2; and HCN J=1--0, F=2--1. Subsequently, we refer to these lines as C$_2$H 3--2, 1--0; c-C$_3$H$_2$ 7--6; and HCN 3--2, 1--0. Due to difficulties in separating the closely-spaced F=$\frac{3}{2}$--$\frac{1}{2}$ and F=$\frac{5}{2}$--$\frac{3}{2}$ hyperfine lines of the CN N=1--0, J=$\frac{3}{2}$--$\frac{1}{2}$ transition, we instead combined them to increase the SNR and improve radial substructure identification. From now on, we refer to these combined lines as CN 1$-$0. Additional details about the CN lines are in \citet{bergner20}. We also combined the blended CH$_3$CN J=12$-$11, K=0 and K=1 lines \citep[see][]{ilee20}, which we simply designate as CH$_3$CN 12$-$11 for the remainder of this work. For simplicity, we likewise label the H$_2$CO (J$_{\rm{K}_{\rm{a}}, {\rm{K}_{\rm{c}}}}$)=3$_{03}-$2$_{02}$ line as H$_2$CO 3$-$2. A comprehensive set of observed transitions is presented in \citet{oberg20}, and detailed analyses of weaker and less spatially extended lines not discussed here can be found in \citet{aikawa20, cataldi20, ilee20, legal20, zhang20}.

In the following subsections, we describe the creation of a set of publicly available Value-Added Data Products (VADPs), namely moment maps and radial intensity profiles. Although line image cubes in principle contain maximal information, the creation of such products is necessary to reduce the overall dimensionality and more intuitively visualize and interpret the data.

\subsection{Moment maps} \label{sec:moment_maps_generation}
A map of the velocity-integrated intensity, or ``zeroth moment map,'' is often useful as a summary representation of an image cube. In this subsection, we describe the process by which we generate moment maps and describe the non-uniform noise distribution that frequently occurs in protoplanetary disk applications.

In its simplest form, a moment map is generated by collapsing an image cube along the velocity dimension to produce a two-dimensional representation of the velocity-integrated flux. Whereas image cubes have units of $\mathrm{Jy\,beam}^{-1}$, moment maps have units of $\mathrm{Jy\,beam}^{-1}\,\mathrm{km\,s}^{-1}$. For sources with complex position-position-velocity morphologies, it is common to first apply a mask to the image cube, so as to prevent regions known to be free of source emission from contributing noise to the moment map.

We adopted a Keplerian mask identical to the one used during the CLEANing process (for more details, see \citet{czekala20}). We did not use a flux threshold for pixel inclusion, i.e., sigma clipping, to ensure accurate flux recovery. We used the Python package \texttt{bettermoments} \citep{Teague18_bettermoments} to generate zeroth moment maps from the non-primary-beam corrected image cubes. 

\begin{figure}[]
\centering
\includegraphics[width=0.75\linewidth]{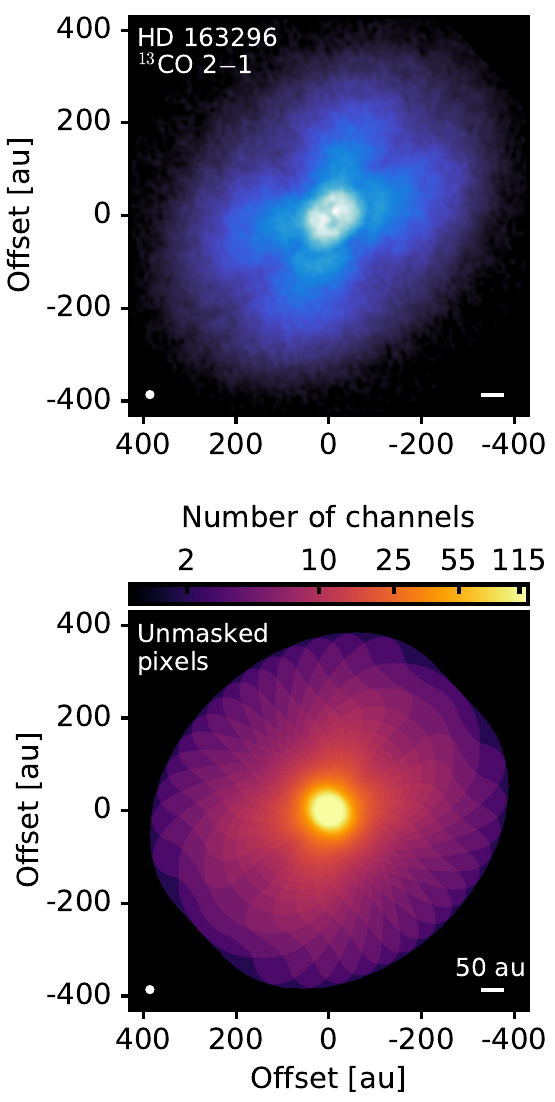}
\caption{Zeroth moment map (top) of $^{13}$CO 2--1 in HD~163296 and a map of unmasked pixels (bottom) used in its generation. The colorscale shows the number of unmasked channels that were summed to create the zeroth moment map. A log10 color stretch has been applied to highlight those disk regions at larger radii with comparatively fewer summed channels. Spatial discontinuities from the Keplerian masking process are evident.}
\label{fig:MomentMapsMask}
\end{figure}

While the use of a mask can substantially improve the visual appearance of a moment map, as shown in \citet{Teague19RNAAS}, it can also introduce strong spatial variance in the noise distribution. As an example, Figure \ref{fig:MomentMapsMask} shows the number of unmasked channels that were summed to create the zeroth moment map for $^{13}$CO 2--1 in HD~163296. Assuming that each channel in the image cube is independent and has the same noise distribution, the noise in the moment map grows $\propto \sqrt{N}$, where $N$ is the number of channels summed. On the other hand, the signal in the moment map will not necessarily grow with $\propto N$, because the sky brightness is not uniform across the image cube.

The discontinuous noise distribution created by the Keplerian masks occasionally imprinted arc-like artifacts in the central few arcseconds of zeroth moment maps created from weak and moderately-bright line image cubes. Such artifacts are the result of channelization and have no effect on the flux properties of the final moment maps, so long as the uncertainties are correctly accounted for. All subsequent quantitative analysis, including the generation of radial intensity profiles, was done using these unclipped and Keplerian masked zeroth moment maps.

\begin{figure*}[]
\centering
\includegraphics[width=\linewidth]{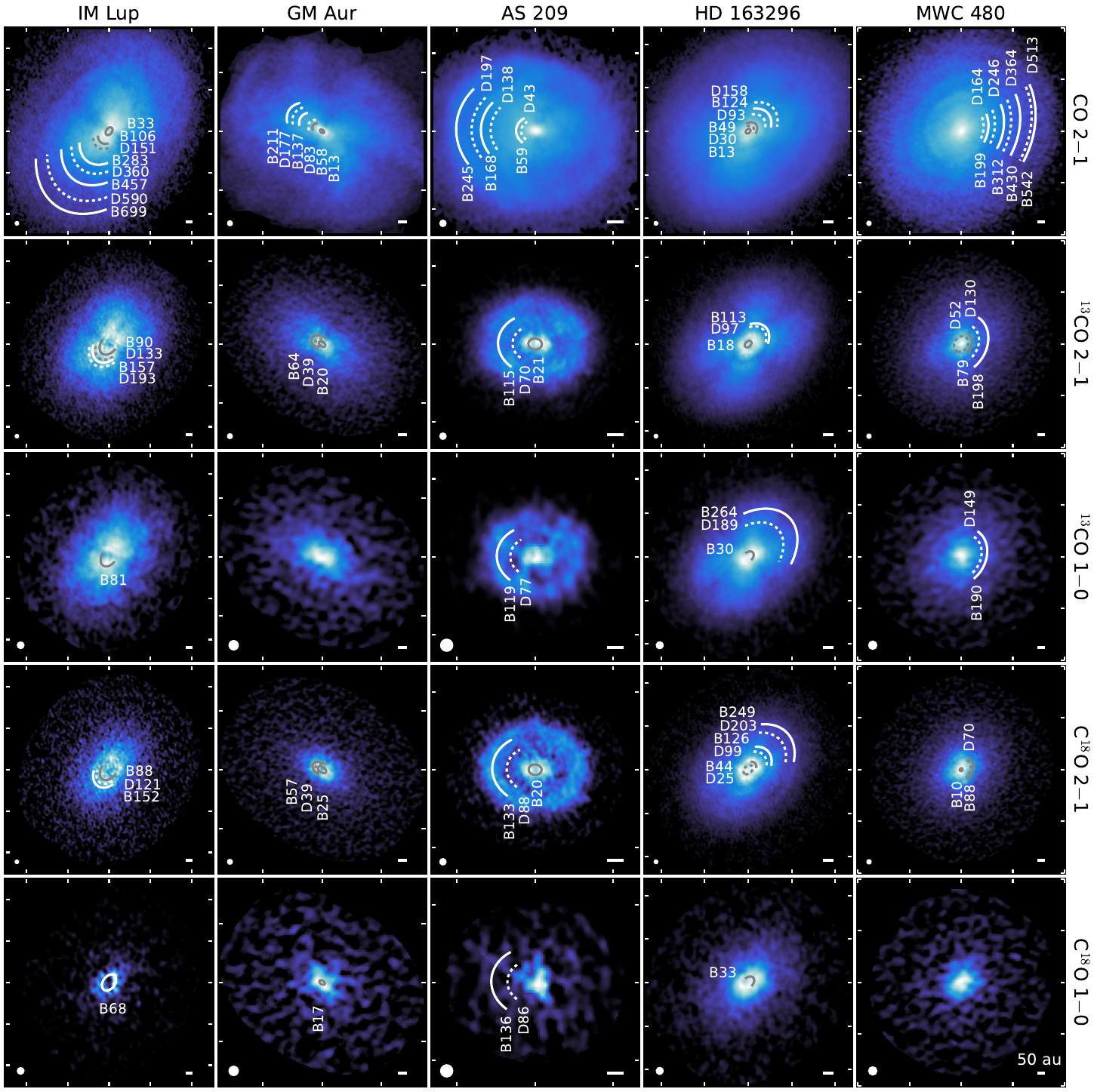}
\caption{Zeroth moment maps of CO, $^{13}$CO, and C$^{18}$O lines for the MAPS sample, ordered from left to right by increasing stellar mass (see Table 1 in \citet{oberg20}). Axes are angular offsets from the disk center with each white tick mark represents a spacing of 2$^{\prime \prime}$. Color stretches were individually optimized and applied to each panel to increase the visibility of substructures. Care should thus be taken when comparing between panels, and instead, we recommend using the corresponding radial profiles in Figure \ref{fig:CO_Radial_Profiles} for this purpose. Chemical substructures from Table \ref{tab:SubstrProp} in the form of rings and gaps are marked by solid and dotted arcs, respectively, with azimuthal extents and colors chosen for maximal visual clarity. Several inner low-contrast CO 2--1 substructures in IM~Lup (D50, B68, D80) and HD~163296 (D71, B81) are omitted for visual clarity. The synthesized beam and a scale bar indicating 50~au is shown in the lower left and right corner, respectively, of each panel.}
\label{fig:CO_Moment0}
\end{figure*}

\begin{figure*}[]
\centering
\includegraphics[width=\linewidth]{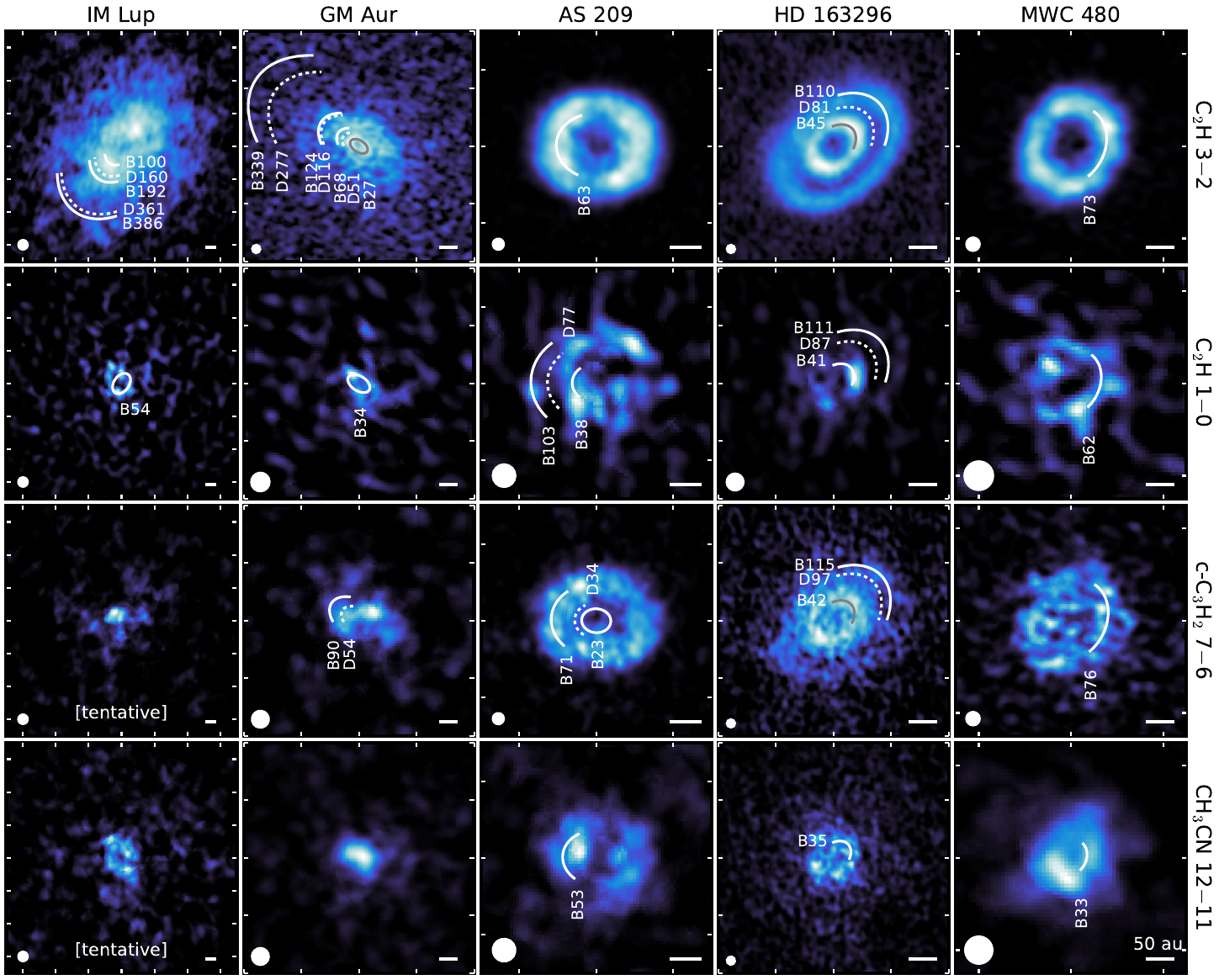}
\caption{Zeroth moment maps of C$_2$H, c-C$_3$H$_2$, CH$_3$CN lines for the MAPS sample, ordered from left to right by increasing stellar mass (see Table 1 in \citet{oberg20}). Axes are angular offsets from the disk center with each white tick mark represents a spacing of 1$^{\prime \prime}$. Color stretches were individually optimized and applied to each panel to increase the visibility of substructures. Care should thus be taken when comparing between panels, and instead, we recommend using the corresponding radial profiles in Figure \ref{fig:C2H_Radial_Profiles} for this purpose. Chemical substructures from Table \ref{tab:SubstrProp} in the form of rings and gaps are marked by solid and dotted arcs, respectively, with azimuthal extents and colors chosen for maximal visual clarity. The synthesized beam and a scale bar indicating 50~au is shown in the lower left and right corner, respectively, of each panel.}
\label{fig:C2H_Moment0}
\end{figure*}

\begin{figure*}[]
\centering
\includegraphics[width=\linewidth]{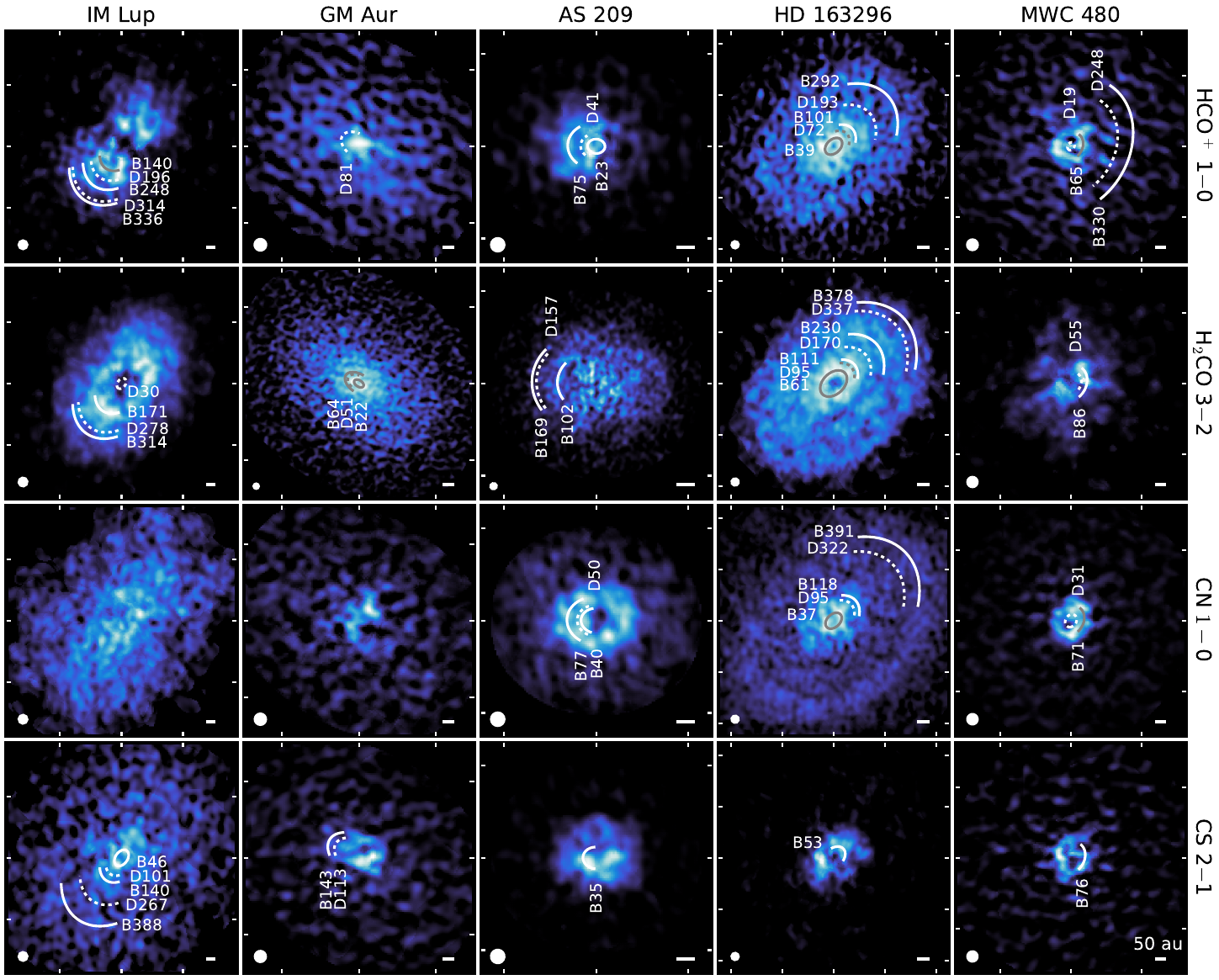}
\caption{Zeroth moment maps of HCO$^+$, H$_2$CO, CN, and CS lines for the MAPS sample, ordered from left to right by increasing stellar mass (see Table 1 in \citet{oberg20}). Axes are angular offsets from the disk center with each white tick mark represents a spacing of 2$^{\prime \prime}$. Color stretches were individually optimized and applied to each panel to increase the visibility of substructures. Care should thus be taken when comparing between panels, and instead, we recommend using the corresponding radial profiles in Figure \ref{fig:H2CO_Radial_Profiles} for this purpose. Chemical substructures from Table \ref{tab:SubstrProp} in the form of rings and gaps are marked by solid and dotted arcs, respectively, with azimuthal extents and colors chosen for maximal visual clarity. The synthesized beam and a scale bar indicating 50~au is shown in the lower left and right corner, respectively, of each panel.}
\label{fig:H2CO_Moment0}
\end{figure*}

\begin{figure*}
\centering
\includegraphics[width=\linewidth]{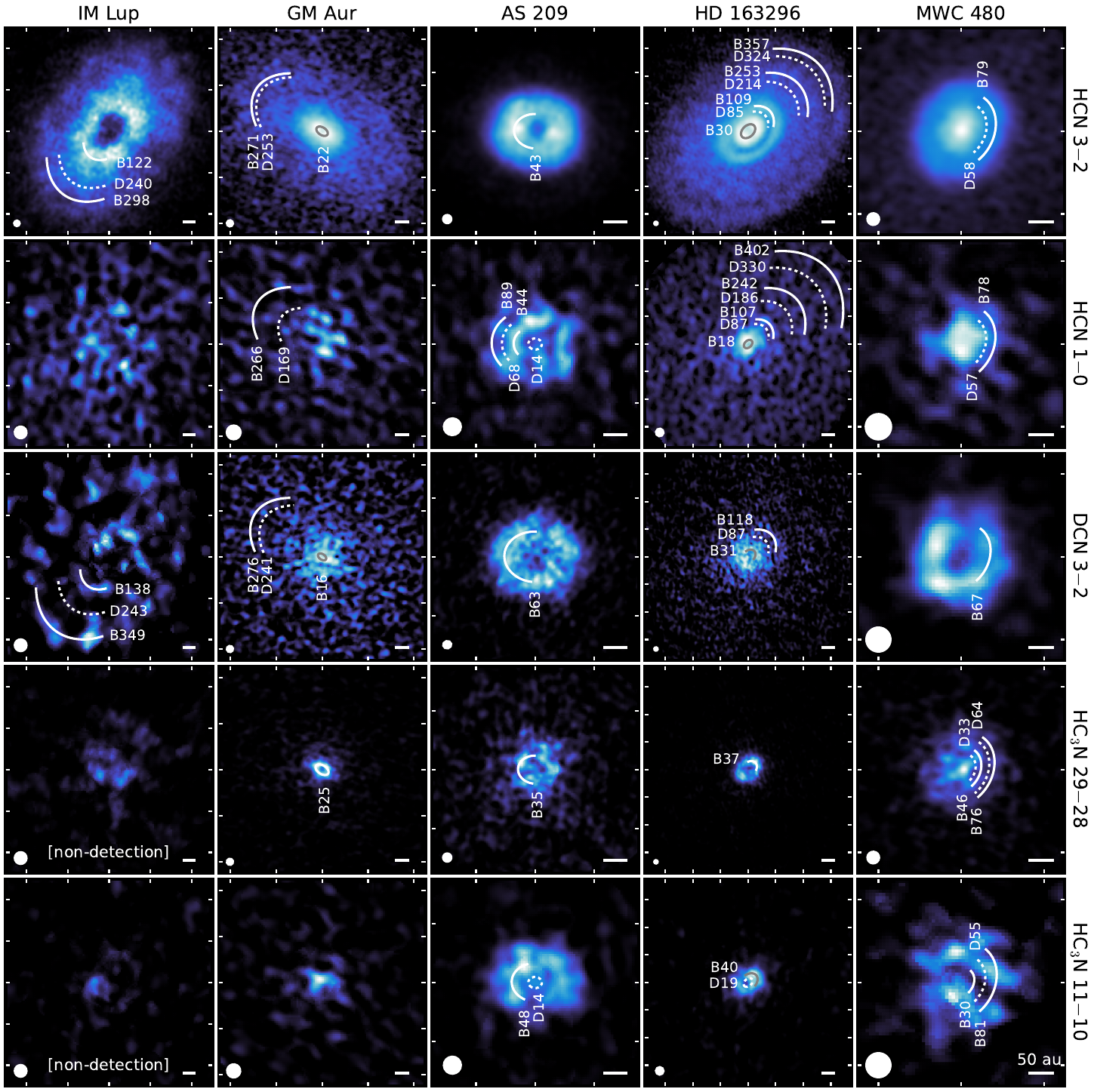}
\caption{Zeroth moment maps of HCN, DCN, and HC$_3$N lines for the MAPS sample, ordered from left to right by increasing stellar mass (see Table 1 in \citet{oberg20}). Axes are angular offsets from the disk center with each white tick mark represents a spacing of 1$^{\prime \prime}$. Color stretches were individually optimized and applied to each panel to increase the visibility of substructures. Care should thus be taken when comparing between panels, and instead, we recommend using the corresponding radial profiles in Figure \ref{fig:HCN_Radial_Profiles} for this purpose. Chemical substructures from Table \ref{tab:SubstrProp} in the form of rings and gaps are marked by solid and dotted arcs, respectively, with azimuthal extents and colors chosen for maximal visual clarity. The synthesized beam and a scale bar indicating 50~au is shown in the lower left and right corner, respectively, of each panel.}
\label{fig:HCN_Moment0}
\end{figure*}

While quantitatively correct, unclipped zeroth moment maps may sometimes be visually misleading due to similarities between arc-like artifacts and real substructures. To address this, we also generated a set of ``hybrid" zeroth moment maps using an approach similar to the auto-masking routine employed within CASA \citep{Kepley_ea_2020}. This combines Keplerian \clean\ masks with smoothed intensity-based masks, which is described in more detail in Appendix~\ref{sec:app:moment_map_generation}. These hybrid maps mitigate or remove the majority of these artifacts and thus better visualize radial structures compared to the zeroth moment maps generated directly from the Keplerian masks. Figures \ref{fig:CO_Moment0}, \ref{fig:C2H_Moment0}, \ref{fig:H2CO_Moment0}, and \ref{fig:HCN_Moment0} show these hybrid zeroth moment maps on a line-by-line basis, while source-specific galleries are found in Appendix \ref{sec:app:disk_specific_zeroth_moment_map}. As the sigma clipping used to generate these hybrid zeroth moment maps artificially reduces integrated intensities, we emphasize these maps are only presentational in nature and are never used for quantitative analysis.

In addition to zeroth moment maps, we also produced maps of the spectral line maximum intensity, or ``peak intensity map," and of the intensity-weighted average velocity, or ``rotation map". Peak intensity maps were generated using the ``quadratic" method of \texttt{bettermoments}, which fits a quadratic curve to the channel with the peak intensity and two adjacent channels \citep{Teague18_bettermoments}. This approach better recovers the true line peak when the line profile is only sparsely sampled, unlike traditional eighth moment maps, which are limited by the spectral resolution of the data. A full set of peak intensity maps are shown in Appendix \ref{sec:app:peak_intensity_maps}, which although not directly analyzed in this work, are provided for completeness. Rotation maps of the line center were also created using the ``quadratic" method of \texttt{bettermoments}, which produces a less biased map for highly-flared disks than first moment maps. These maps are not shown here but see \citet{teague20} for a detailed presentation and discussion of CO rotation maps.

This set of maps -- zeroth moment, rotation, peak intensity -- are provided as VADPs and are available to the community through our dedicated website hosted by ALMA (\url{https://almascience.nrao.edu/alma-data/lp/maps}). Moment maps were generated for all lines covered in MAPS \citep[see Tables 2 and 3,][]{oberg20}, not just those considered here, and for all available angular resolutions \citep[see Table 5,][]{oberg20}. A corresponding set of maps, derived as described above, for the non-continuum-subtracted images are also provided. As all maps are generated from \texttt{bettermoments}, they also include a corresponding map of statistical uncertainty for each measured quantity, as described in \citet{Teague18_bettermoments}. Scripts to generate the hybrid zeroth moment maps used for presentational purposes will also be made publicly available.

\subsection{Radial intensity profiles} \label{sec:radial_profiles_generation}

A radial line intensity profile provides a helpful one-dimensional representation of emission in protoplanetary disks as a function of radius. In this subsection, we describe the process by which we generate radial profiles, including details of the deprojection process and extraction methods, and how we select those profiles which best reveal observed chemical substructures.

We generated radial profiles using the \texttt{radial\_profile} function in the Python package \texttt{GoFish} \citep{Teague19JOSS} to deproject the zeroth moment maps. Radial bin sizes are calculated as 1/4 of the full-width at half-maximum (FWHM) of the synthesized beam, which corresponds to about 1.5-2 pixels. The uncertainty of the measured intensity in each radial bin is estimated as the standard error on the mean in the annulus or arc over which the emission was averaged. An advantage of this empirical error estimate is that it includes uncertainties related to the entire imaging and moment map generation process. However, uncertainties are artificially larger in regions with large intrinsic azimuthal variation. One such example is molecular emission from regions highly elevated above the disk midplane, which manifests as a large `X' morphology in many of the panels showing CO and $^{13}$CO in Figure~\ref{fig:CO_Moment0}. This `X' shape arises from spatially separated isovelocity contours in the inter-axis regions \citep[e.g., see Figure 4,][]{Keppler19}, which allows emission from both sides of the disk to reach the observer.

In addition to extracting radial intensity profiles from the zeroth moment maps, we tested the method used in \citet{Teague_Loomis_2020}. There, the authors first corrected for the velocity structure of the disk, before azimuthally averaging the spectra and then integrating the line profiles \citep[e.g.,][]{Yen16} in \texttt{GoFish}. There were negligible differences between these methods, and we opted to use the radial profiles of the zeroth moment maps for simplicity.

For each line, we generated an azimuthally-averaged profile and a set of profiles extracted along the major axis of each disk, where emission was averaged within varying azimuthal wedges, namely $\pm$15$^{\circ}$, $\pm$30$^{\circ}$, and $\pm$45$^{\circ}$. We manually selected wedge sizes that maximized the relative contrasts of individual substructures, while still maintaining high fidelity. Narrower wedges taken along the disk major axis often resulted in features with sharper contrasts and were used whenever the SNR allowed. This is the result of a lower effective spatial resolution along the minor axis of an inclined disk, which if included, can smear radial features. However, if decreasing the wedge size did not lead to the emergence of any new features or the sharpening of existing substructures, we used progressively larger wedge sizes, up to a complete azimuthal average, to improve the SNR and overall smoothness of profiles. The selections for each radial profile are summarized in Table \ref{tab:ProfSelection}.

\begin{figure}
\centering
\includegraphics[width=\linewidth]{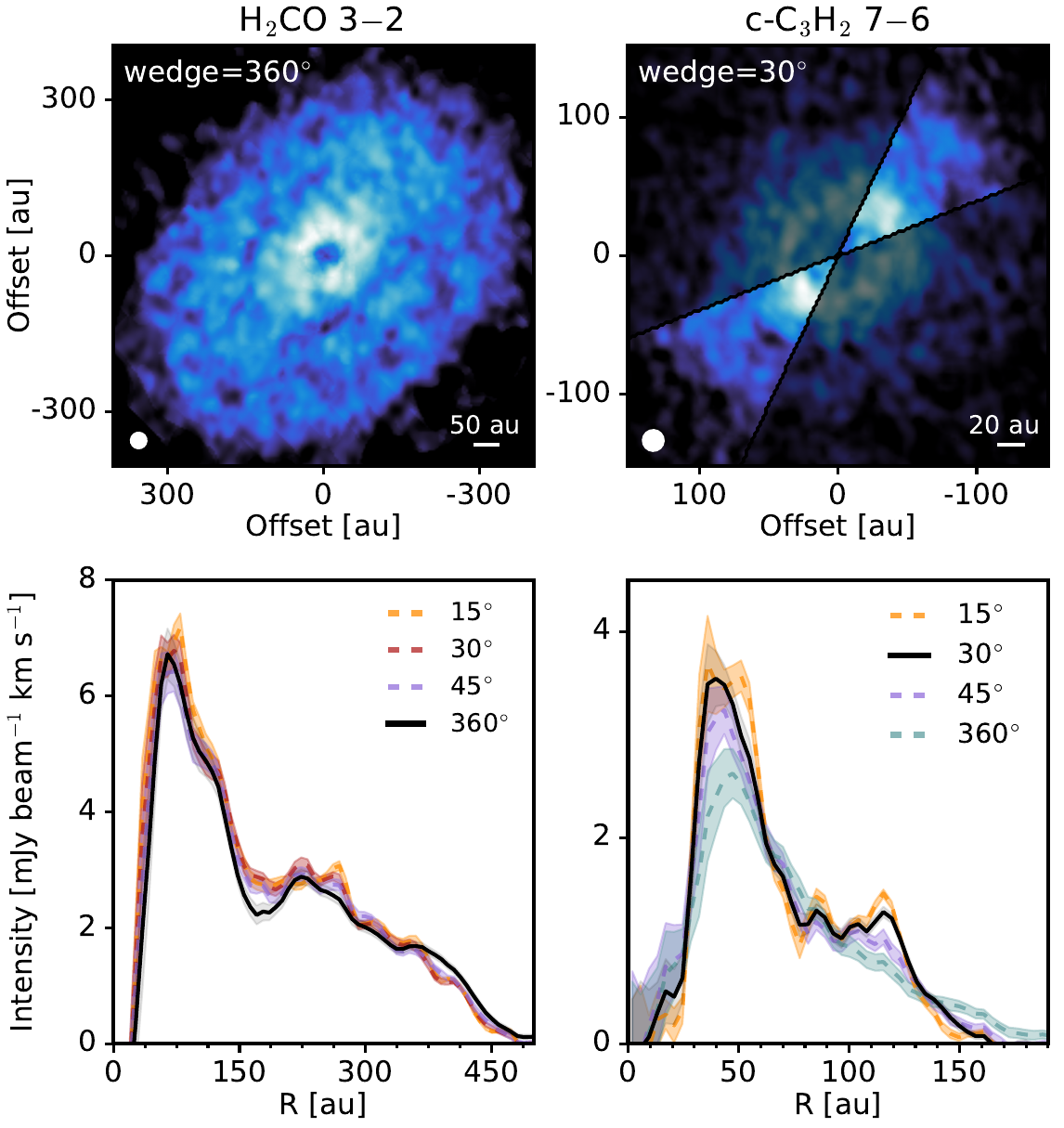}
\caption{Effects of wedge sizes on radial intensity profiles. The top row shows zeroth moment maps of H$_2$CO 3--2 (left column) and c-C$_3$H$_2$ 7--6 (right column) in HD~163296. The synthesized beam and a scale bar indicating either 20 or 50 au is shown in the lower left and right corner, respectively, of each panel. Axes are labeled as offsets in au from the disk center. The bottom row shows the radial intensity profiles. Radial profiles, as indicated by the legend color, are either azimuthally-averaged or extracted from a wedge size of $\pm$15$^{\circ}$, $\pm$30$^{\circ}$, $\pm$45$^{\circ}$ along the disk major axis. The selected profile, as in Table \ref{tab:ProfSelection}, is indicated as a solid black line. Shaded regions show the 1$\sigma$ scatter at each radial bin (i.e., arc or annulus) divided by the ratio of the square root of bin circumference and FWHM of the synthesized beam.}
\label{fig:wedges_methods}
\end{figure}

Figure \ref{fig:wedges_methods} shows an example of this process. For medium-to-strong lines with well-defined substructures, such as the inner emission ring at ${\sim}$50~au in c-C$_3$H$_2$ 7--6 in HD~163296, narrow wedges result in higher contrast features and more accurate determinations of radial locations. Narrow wedges also often amplify substructures not evident when using wider azimuthal wedges. This is the case for the outer ring at ${\sim}$110~au in c-C$_3$H$_2$ 7--6, which is not present in the azimuthally averaged radial profile, but is clearly seen when using a $\pm$30$^{\circ}$ wedge. For lines with smoother, more extended radial morphologies, like H$_2$CO 3--2 in HD~163296, an azimuthally averaged profile is sometimes most effective in identifying features, e.g., the dip at ${\sim}$170~au.

In cases where line emission is originating from a layer substantially higher than the disk midplane, we must take this emitting surface into account to accurately deproject the observations into annuli of constant radius. We deprojected radial profiles using the derived surfaces from \citet{law20a}, as indicated in Table \ref{tab:ProfSelection}, for those lines with meaningful constraints on their emission surfaces, namely CO 2--1, $^{13}$CO 2--1, HCN 3--2, and C$_2$H 3--2 (see Appendix \ref{sec:app:surface_and_wedges_assumptions}). Otherwise, for simplicity, we assumed that the line emission is arising from the midplane, i.e., $z/r=0.0$. Further testing confirmed that the radial intensity profiles of those lines lacking explicit emission surface determinations are consistent for any reasonable choice of assumed surfaces (e.g., $z/r = 0.0, 0.1, 0.2$).

The AS~209 disk suffers from cloud absorption at $v_{\rm{LSR}}~\lesssim~5$~km~s$^{-1}$ \citep{Oberg11_DISCS}. This results in reduced CO 2--1 flux toward the west half of its disk \citep{Huang16_AS209, Guzman18_dsharp}, which is also clearly seen in the MAPS data (Figure \ref{fig:CO_Moment0}). In addition to CO 2--1, significant absorption is present in HCO$^+$ 1--0, while more modest east-to-west flux asymmetries are noted in C$_2$H 1--0 and HCN 1$-$0. For these lines, we adopted an asymmetric $\pm55^{\circ}$ wedge, as in \citet{Teague18_AS209}, that was applied to the uncontaminated eastern half. Otherwise, no obvious azimuthal asymmetries were identified and all other lines were assumed to be azimuthally symmetric, but see \citet{legal20} for a exploration of potential asymmetries in CS.

Figures \ref{fig:CO_Radial_Profiles}, \ref{fig:C2H_Radial_Profiles}, \ref{fig:H2CO_Radial_Profiles}, and \ref{fig:HCN_Radial_Profiles} show the set of radial intensity profiles selected here to highlight radial chemical substructures. These radial profiles, along with those generated from all combinations of wedge sizes, are provided as publicly-available VADPs. Radial profiles for all lines covered in MAPS, including those not analyzed here \citep[see Tables 2 and 3,][]{oberg20}, and for all imaged angular resolutions \citep[see Table 5,][]{oberg20}, are also available. Additionally, as for the moment maps, a corresponding set of radial profiles for the non-continuum-subtracted images are included. See Section \ref{sec:VADPs_listing} for more details and a full listing of available VADPs.

%clearpage

\startlongtable
\centerwidetable
\begin{deluxetable*}{l|cccccccccccccccccc}
\tablecaption{Summary of Radial Profiles \label{tab:ProfSelection}}
\tablewidth{0pt}
\tablehead{
\colhead{}\vline & \multicolumn2c{IM~Lup} & \multicolumn2c{GM~Aur} &  \multicolumn2c{AS~209} & \multicolumn2c{HD~163296} & \multicolumn2c{MWC~480} \\
\colhead{Line$^a$} \vline & \colhead{Type} & \colhead{Surface}  & \colhead{Type} & \colhead{Surface} & \colhead{Type} & \colhead{Surface} & \colhead{Type} & \colhead{Surface} & \colhead{Type}  & \colhead{Surface}}
\startdata
CO 2$-$1	&30$^{\circ}$/$0\farcs15$	& Y	&15$^{\circ}$/$0\farcs15$	& Y	&55$^{\circ,\,d}$/$0\farcs15$	& Y	&15$^{\circ}$/$0\farcs15$	& Y	&15$^{\circ}$/$0\farcs15$	& Y	&\\
$^{13}$CO 2$-$1	&360$^{\circ}$/$0\farcs15$	& Y	&30$^{\circ}$/$0\farcs15$	& Y	&30$^{\circ}$/$0\farcs15$	&N	&15$^{\circ}$/$0\farcs15$	& Y	&360$^{\circ}$/$0\farcs15$	&N	&\\
$^{13}$CO 1$-$0	&360$^{\circ}$/$0\farcs30$	&N	&45$^{\circ}$/$0\farcs30$	&N	&30$^{\circ}$/$0\farcs30$	&N	&30$^{\circ}$/$0\farcs30$	&N	&45$^{\circ}$/$0\farcs30$	&N	&\\
C$^{18}$O 2$-$1	&45$^{\circ}$/$0\farcs15$	&N	&30$^{\circ}$/$0\farcs15$	&N	&30$^{\circ}$/$0\farcs15$	&N	&45$^{\circ}$/$0\farcs15$	&N	&360$^{\circ}$/$0\farcs15$	&N	&\\
C$^{18}$O 1$-$0	&45$^{\circ}$/$0\farcs30$	&N	&45$^{\circ}$/$0\farcs30$	&N	&45$^{\circ}$/$0\farcs30$	&N	&30$^{\circ}$/$0\farcs30$	&N	&360$^{\circ}$/$0\farcs30$	&N	&\\
C$_2$H 3$-$2	&45$^{\circ}$/$0\farcs30$	&N	&45$^{\circ}$/$0\farcs15$	&N	&30$^{\circ}$/$0\farcs15$	&N	&15$^{\circ}$/$0\farcs15$	&Y\,$^e$	&30$^{\circ}$/$0\farcs15$	&N	&\\
C$_2$H 1$-$0	&360$^{\circ}$/$0\farcs30$	&N	&360$^{\circ}$/$0\farcs30$	&N	&55$^{\circ,\,d}$/$0\farcs30$	&N	&30$^{\circ}$/$0\farcs30$	&N	&360$^{\circ}$/$0\farcs30$	&N	&\\
c-C$_3$H$_2$ 7$-$6	&360$^{\circ}$/$0\farcs30$	&N	&30$^{\circ}$/$0\farcs30$	&N	&360$^{\circ}$/$0\farcs15$	&N	&30$^{\circ}$/$0\farcs15$	&N	&30$^{\circ}$/$0\farcs30$	&N	&\\
H$_2$CO 3$-$2	&45$^{\circ}$/$0\farcs30$	&N	&30$^{\circ}$/$0\farcs15$	&N	&30$^{\circ}$/$0\farcs15$	&N	&360$^{\circ}$/$0\farcs30$	&N	&30$^{\circ}$/$0\farcs30$	&N	&\\
HCO$^+$ 1$-$0	&30$^{\circ}$/$0\farcs30$	&N	&30$^{\circ}$/$0\farcs30$	&N	&55$^{\circ,\,d}$/$0\farcs30$	&N	&360$^{\circ}$/$0\farcs30$	&N	&30$^{\circ}$/$0\farcs30$	&N	&\\
CS 2$-$1	&30$^{\circ}$/$0\farcs30$	&N	&360$^{\circ}$/$0\farcs30$	&N	&45$^{\circ}$/$0\farcs30$	&N	&360$^{\circ}$/$0\farcs30$	&N	&360$^{\circ}$/$0\farcs30$	&N	&\\
CN 1$-$0\,$^b$	&360$^{\circ}$/$0\farcs30$	&N	&360$^{\circ}$/$0\farcs30$	&N	&30$^{\circ}$/$0\farcs30$	&N	&30$^{\circ}$/$0\farcs30$	&N	&30$^{\circ}$/$0\farcs30$	&N	&\\
HCN 3$-$2	&360$^{\circ}$/$0\farcs15$	&N	&30$^{\circ}$/$0\farcs15$	&N	&30$^{\circ}$/$0\farcs15$	&N	&30$^{\circ}$/$0\farcs15$	&Y\,$^e$	&30$^{\circ}$/$0\farcs15$	&N	&\\
HCN 1$-$0	&360$^{\circ}$/$0\farcs30$	&N	&45$^{\circ}$/$0\farcs30$	&N	&55$^{\circ,\,d}$/$0\farcs30$	&N	&30$^{\circ}$/$0\farcs30$	&N	&30$^{\circ}$/$0\farcs30$	&N	&\\
DCN 3$-$2	&360$^{\circ}$/$0\farcs30$	&N	&360$^{\circ}$/$0\farcs15$	&N	&360$^{\circ}$/$0\farcs15$	&N	&30$^{\circ}$/$0\farcs30$	&N	&360$^{\circ}$/$0\farcs30$	&N	&\\
HC$_3$N 29$-$28	&360$^{\circ}$/$0\farcs30$	&N	&360$^{\circ}$/$0\farcs15$	&N	&360$^{\circ}$/$0\farcs15$	&N	&30$^{\circ}$/$0\farcs15$	&N	&360$^{\circ}$/$0\farcs15$	&N	&\\
HC$_3$N 11$-$10	&360$^{\circ}$/$0\farcs30$	&N	&360$^{\circ}$/$0\farcs30$	&N	&45$^{\circ}$/$0\farcs30$	&N	&30$^{\circ}$/$0\farcs30$	&N	&45$^{\circ}$/$0\farcs30$	&N	&\\
CH$_3$CN 12$-$11\,$^c$	&360$^{\circ}$/$0\farcs30$	&N	&360$^{\circ}$/$0\farcs30$	&N	&360$^{\circ}$/$0\farcs30$	&N	&30$^{\circ}$/$0\farcs15$	&N	&360$^{\circ}$/$0\farcs30$	&N	&\\
\enddata
\tablecomments{Type indicates the wedge size of the radial profile used in this analysis and FWHM of the synthesized beam of the image. Profiles were extracted along the disk major axis in an azimuthal wedge twice (i.e., $\pm$) that of the listed value, except for those listed as 360$^{\circ}$, which denote an azimuthally-averaged profile. Surface choices (Y/N) are taken from \citet{law20a}.}
\tablenotetext{a}{Only the brightest component of those transitions with multiple hyperfine components are considered, unless otherwise noted. For further details of selected lines, see Section \ref{sec:observations}.}
\tablenotetext{b}{The closely-spaced F=$\frac{3}{2}$--$\frac{1}{2}$ and F=$\frac{5}{2}$--$\frac{3}{2}$ hyperfine lines of the CN N=1--0, J=$\frac{3}{2}$--$\frac{1}{2}$ transition have been combined for increased SNR, see \citet{bergner20}.}
\tablenotetext{c}{The blended K=0 and K=1 lines of the CH$_3$CN J=12$-$11 transition have been combined for increased SNR, see \citet{ilee20}.}
\tablenotetext{d}{An asymmetric 55$^{\circ}$ wedge, as in \citet{Teague18_AS209}, was used to avoid cloud absorption present in AS~209.}
\tablenotetext{e}{A constant $z/r = 0.1$ emitting surface was assumed.}
\end{deluxetable*}

\section{Characterization of Disk Features} \label{sec:disk_features_section}

\subsection{Radial locations of substructures}
\label{sec:radial_locations}

We adopt a chemical substructure nomenclature analogous to that established for annular dust substructure \citep[][]{Huang18, Huang20}. Each substructure is labeled with its radial location rounded to the nearest whole number in astronomical units and is preceded by either ``B" (for ``bright") or ``D" (for ``dark") depending on if the emission represents a local maximum or minimum, respectively. These features are also frequently referred to as ``rings" or ``gaps," respectively \citep[e.g.,][]{Oberg15, Bergin16}. In a few cases, e.g., single isolated rings, the term substructure is a misnomer but is a useful convention for the purposes of a homogeneous comparison. Below, we describe the procedure used to identify, characterize, and label these substructures.

For each intensity profile exhibiting radial substructure, we model the profile as a sum of one or more Gaussian profiles using the Levenberg-Marquardt minimization implementation in \texttt{LMFIT} \citep{LMFIT}. Before fitting, the number of component Gaussian profiles was fixed via visual inspection. The fitted centers of each Gaussian are taken to be the radial location of each feature and are reported in Table \ref{tab:SubstrProp}. The majority of lines are well-suited to this approach due to the high contrasts and well-separated nature of their substructural features. Even in cases when components overlap, Gaussian decomposition captures the underlying features. On occasion, it was necessary to manually restrict the fitting range to better reproduce the observed profiles. This was most often necessary in cases where plateau-like emission was located on one side of an emission ring, resulting in incorrectly skewed fits. In cases such as this, accurate determinations of the radial location of line peaks or gaps were prioritized and attempts were not made to fully reproduce highly skewed or asymmetric features. Substructures displaying notable deviations from Gaussian shapes are discussed in more detail in Section \ref{sec:tentative_substructures}.

\begin{figure}[!ht]
\centering
\includegraphics[width=\linewidth]{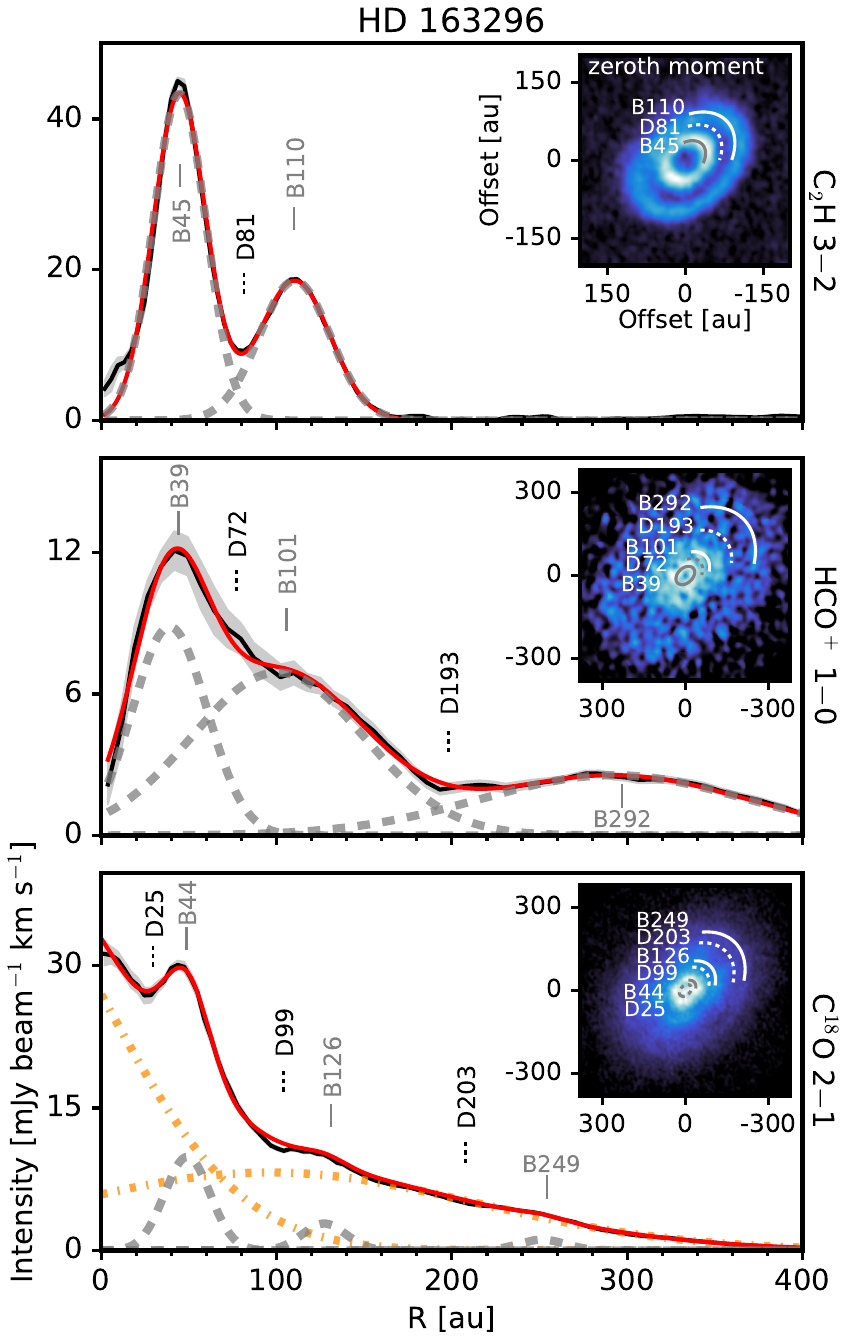}
\caption{Example Gaussian decomposition for C$_2$H 3--2 (top), HCO$^+$ 1--0 (middle), and C$^{18}$O 2--1 (bottom) in HD~163296. The red solid line indicates the composite fit, while individual Gaussian profiles are shown as dashed gray lines. The orange dash-dotted lines in the bottom panel are Gaussians used to remove the broad and smoothly decreasing background in C$^{18}$O 2--1. The intensities of the low-amplitude Gaussians at ${\sim}$130 and 250~au have each been multiplied by a factor of 3 for visual clarity. The insets show zeroth moment maps with axes labeled as offsets in au from the disk center. Chemical substructures are marked by solid and dotted arcs, indicating bright and dark features, respectively.}
\label{fig:Fitting_Example}
\end{figure}

Unlike the majority of other species, the CO lines are not composed of well-separated, distinct features, but instead of numerous low-contrast features on top of a broad power-law-like background. As a result, it was often necessary to first fit and remove this broad component to accurately characterize the substructural features. This was done by fitting either an exponential power-law component or one or more broad Gaussians. Figure \ref{fig:Fitting_Example} shows an example of this Gaussian decomposition process. 

While Gaussian profiles provide a natural characterization of emission rings, they do not as readily describe the radial locations of gaps. Only in some cases, i.e., CO lines, where we could fit and remove the underlying smooth profile, was it possible to directly fit gaps with (inverted) Gaussian profiles. However, for the majority of gaps, we instead report the local minimum of each emission gap as its radial location. The positional uncertainty of each minima is estimated as the width of one radial bin (from 4~au to 13~au, depending on transition frequency and source distance).

For certain lines, e.g., HCO$^+$ 1--0, H$_2$CO 3--2 and CN 1--0, as shown in Figure \ref{fig:H2CO_Radial_Profiles}, there are regions of the radial profiles which represent bona fide annular substructures, i.e., deviations from a smooth profile, but are not in the form of distinct emission rings or gaps. Such features are often referred to as either emission ``plateau'' or ``shoulders'' \citep[e.g.,][]{Huang18TWHya, Huang18, Huang20}. For consistency, we define those deviations that have relatively narrow radial extents as shoulders, e.g., B192, C$_2$H 3$-$2 in IM~Lup (Figure \ref{fig:C2H_Radial_Profiles}); B46 and B76 in HC$_3$N 29--28 in MWC~480 (Figure \ref{fig:HCN_Radial_Profiles}), while those that display nearly constant excess emission out to large radii as plateau, e.g., HCN 1--0 in IM~Lup (Figure \ref{fig:HCN_Radial_Profiles}); HCO$^+$ 1--0 in GM~Aur (Figure \ref{fig:H2CO_Radial_Profiles}). Emission plateaus, which lack a single well-defined radial position, are not explicitly listed in Table \ref{tab:SubstrProp}, but a few prominent examples are instead noted in Table \ref{tab:tentative_radial_substructures}. In contrast, emission shoulders are more well-defined and, when possible, were characterized using Gaussian profiles; otherwise, their radial positions were catalogued visually.

Following \citet{Huang18}, the inner and outer edges of an emission shoulder were denoted with the prefixes ``D" and ``B," respectively, followed by the radial location in au rounded to the nearest integer. Even in cases when the outer edge of a ``B" substructure was well-fit with a Gaussian profile, the inner edge still needed to be visually identified. As this method is more subjective than either Gaussian fitting or local extrema identification, approximate locations are listed in Table \ref{tab:SubstrProp} without formal error estimates. However, the uncertainties should be less than a synthesized beam. 

\begin{figure*}[]
\centering
\includegraphics[width=\linewidth]{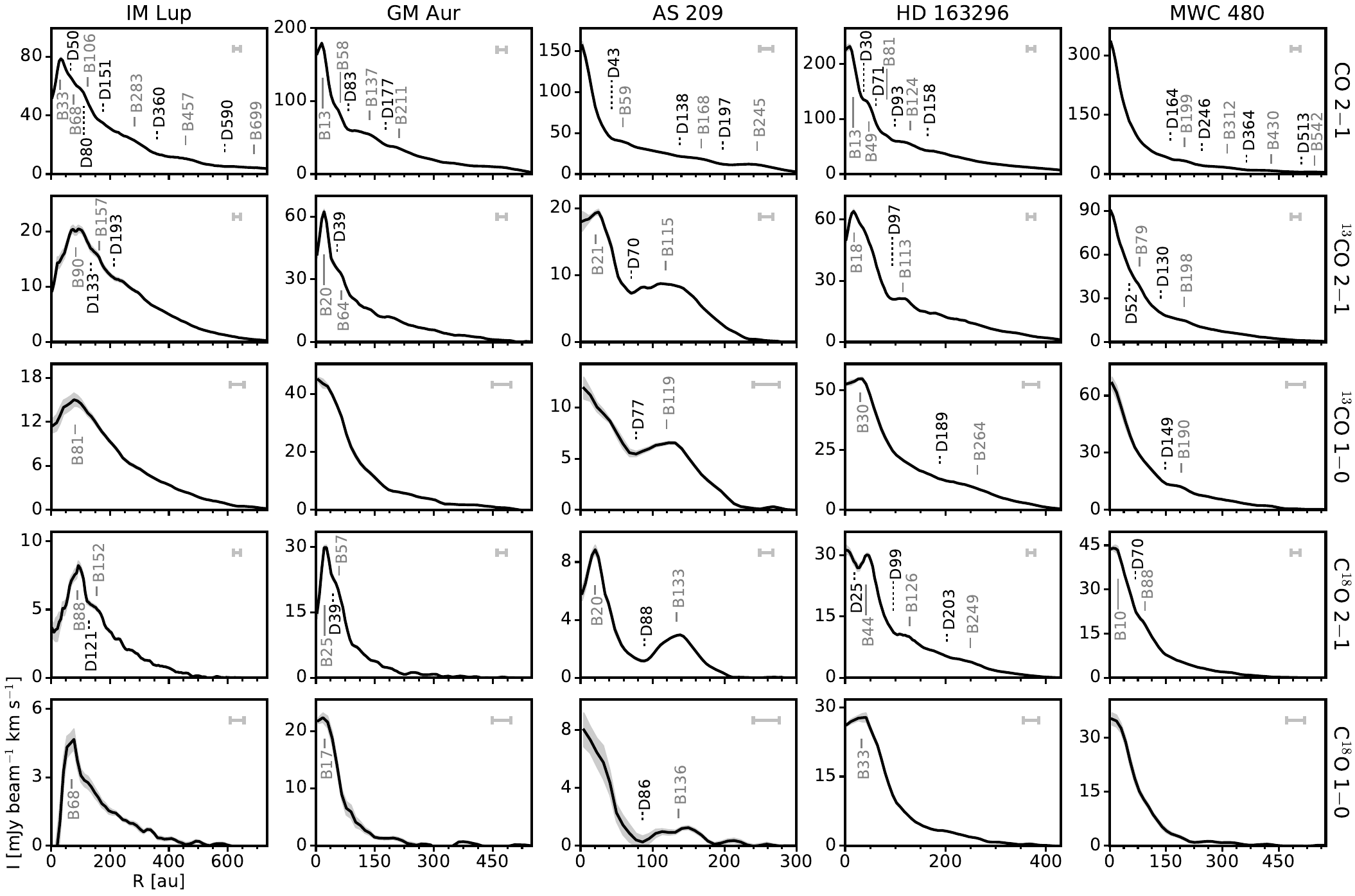}
\caption{Deprojected radial intensity profiles of CO lines, as indicated in Table \ref{tab:ProfSelection}, for the MAPS sample, ordered from left to right by increasing stellar mass. Gray shaded regions show the 1$\sigma$ scatter at each radial bin (i.e., arc or annulus) divided by the ratio of the square root of bin circumference and FWHM of the synthesized beam. Solid gray lines mark emission rings and dotted black lines mark gaps, as listed in Table \ref{tab:SubstrProp}. The FWHM of the synthesized beam is shown by a horizontal bar in the upper right corner of each panel.}
\label{fig:CO_Radial_Profiles}
\end{figure*}

\begin{figure*}[]
\centering
\includegraphics[width=\linewidth]{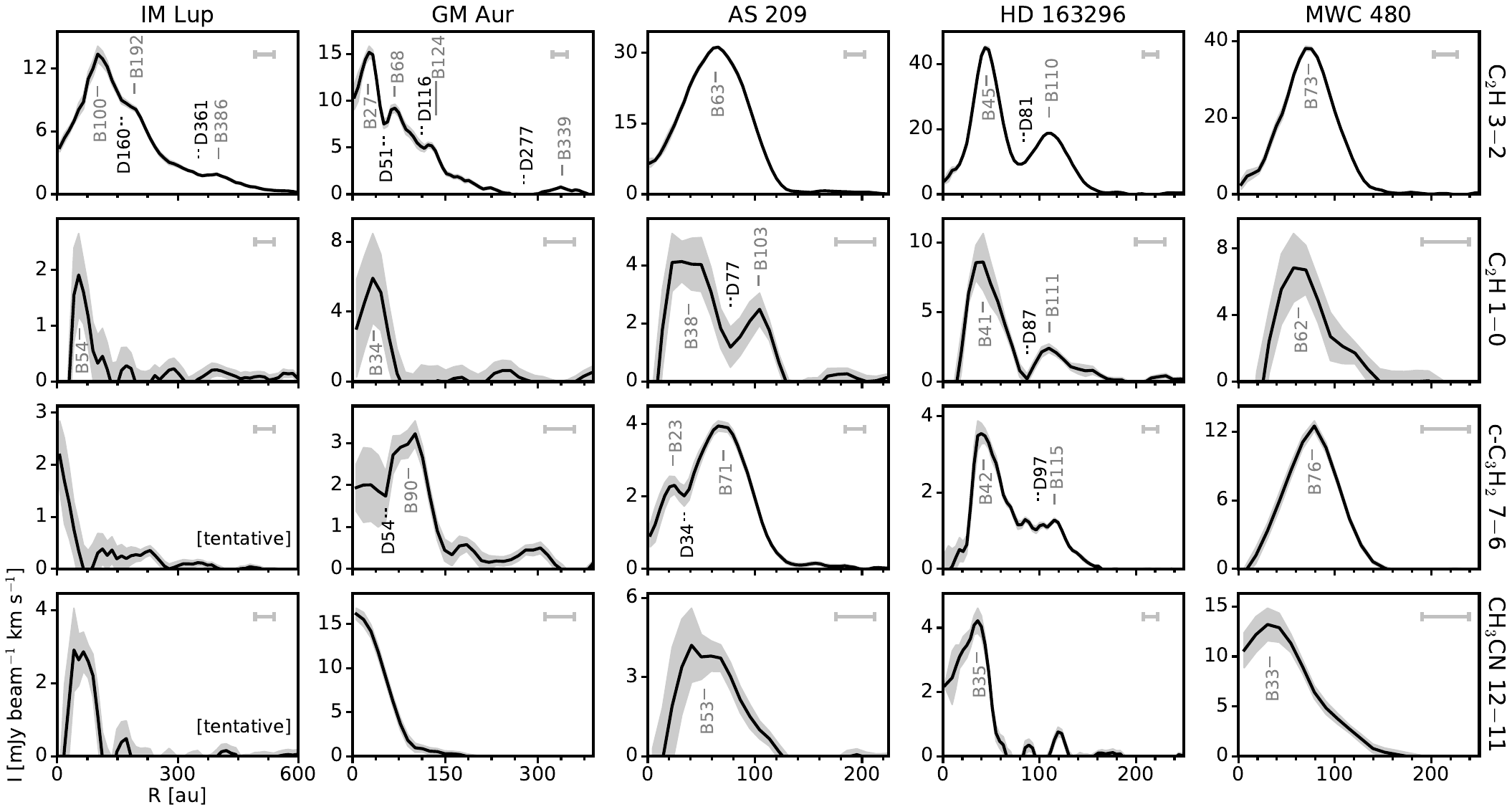}
\caption{Deprojected radial intensity profiles of C$_2$H, c-C$_3$H$_2$, and CH$_3$CN lines, as indicated in Table \ref{tab:ProfSelection}. An apparent emission ring at 60~au is present in CH$_3$CN 12--11 in IM~Lup, but as this line is only tentatively detected \citep{ilee20}, we do not label this substructure and omit it from further analysis. Otherwise, as in Figure \ref{fig:CO_Radial_Profiles}.}
\label{fig:C2H_Radial_Profiles}
\end{figure*}

\begin{figure*}[]
\centering
\includegraphics[width=\linewidth]{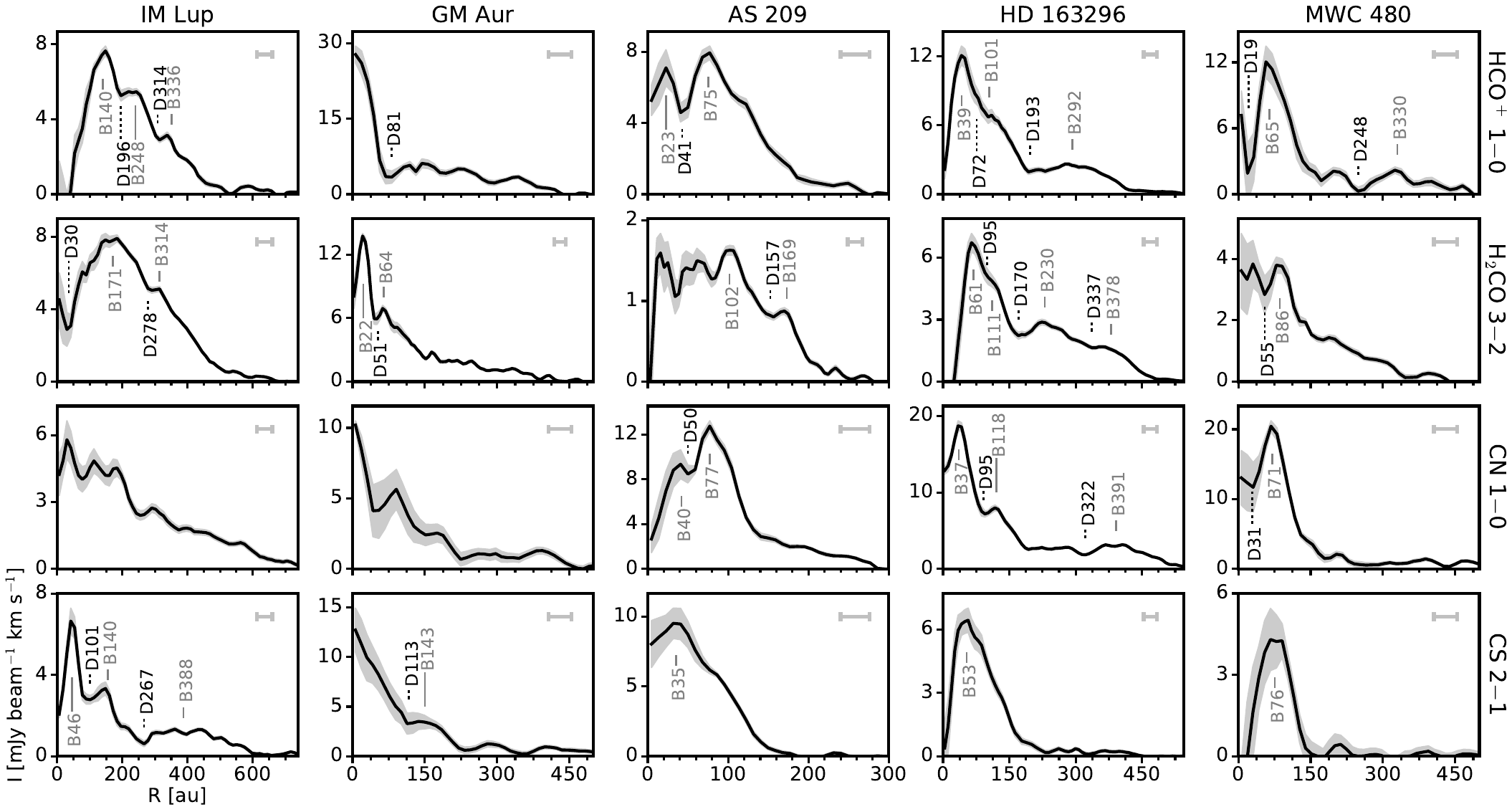}
\caption{Deprojected radial intensity profiles of HCO$^+$, H$_2$CO, CN, and CS lines, as indicated in Table \ref{tab:ProfSelection}. Otherwise, as in Figure \ref{fig:CO_Radial_Profiles}.}
\label{fig:H2CO_Radial_Profiles}
\end{figure*}

\begin{figure*}[]
\centering
\includegraphics[width=\linewidth]{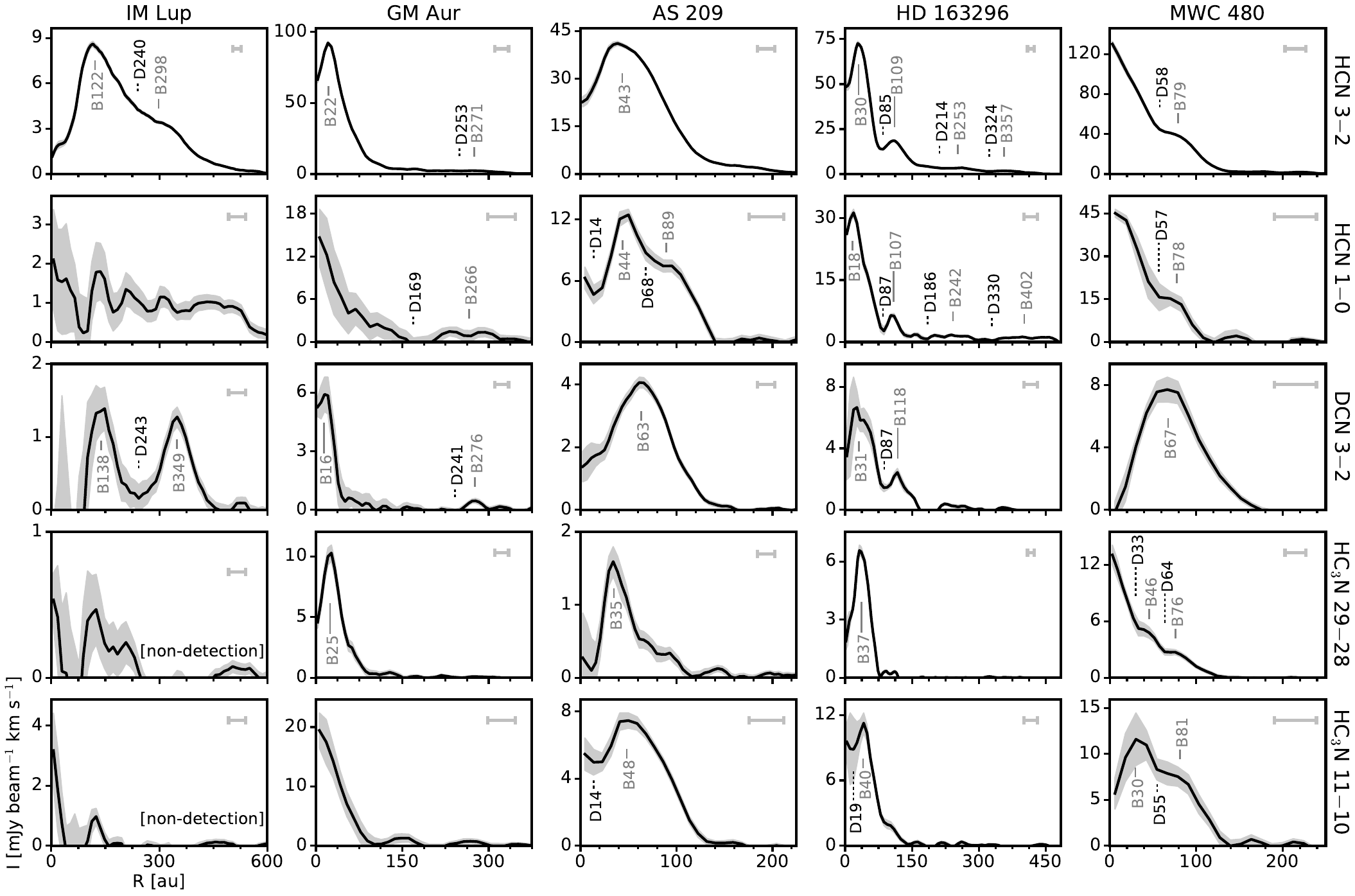}
\caption{Deprojected radial intensity profiles of HCN, DCN, and HC$_3$N lines, as indicated in Table \ref{tab:ProfSelection}. Otherwise, as in Figure \ref{fig:CO_Radial_Profiles}.}
\label{fig:HCN_Radial_Profiles}
\end{figure*}

Measurements that are derived from Gaussian fittings are indicated as ``G" in Table \ref{tab:SubstrProp}, while those based on the identification of local extrema in the radial profiles are labeled ``R". Visual identifications are denoted as ``V." Figures \ref{fig:CO_Radial_Profiles}, \ref{fig:C2H_Radial_Profiles}, \ref{fig:H2CO_Radial_Profiles}, and \ref{fig:HCN_Radial_Profiles} show the labeled radial intensity profiles. Figure \ref{fig:CO_log_scale} in Appendix \ref{sec:app:log_CO_profiles} provides logarithmically-scaled radial intensity profiles for all CO lines to more clearly show low-contrast substructures, especially those at large radii.

\subsection{Widths and depths of substructures} \label{sec:widths_and_contrast}

Defining the widths and depths of substructures is less straightforward than identifying their radial locations. For emission rings, which were modeled as Gaussian profiles, the FWHMs were taken to be the ring widths. For all gaps, we instead followed the empirical procedure outlined in \citet{Huang18}. In brief, substructure widths were defined by the radial locations where the intensity is equal to the mean intensity of a consecutive ring-gap pair. For rings with emission profiles that can be modeled as isolated Gaussians, this definition reduces to approximately the FWHM, and as such, makes it a comparable metric for gap widths. A detailed description of this procedure, including treatment of various special cases, is found in \citet{Huang18}. 

The relative contrast of an adjacent gap-ring pair is defined by their intensity ratio. Specifically, gap depth is given as $I_d / I_b$, where the gap intensity $I_d$ is the intensity value at the radial position of the gap and $I_b$ is the intensity value at the radial position of the ring directly outside the gap. In a few cases, there was no suitable ring outside of the gap (D14 in HCN 1--0, HC$_3$N 11--10; D34 in c-C$_3$H$_2$ 7--6; and D50 in CN 1--0 for AS~209) and we used the ring interior for calculating the depth. Substructure widths and depths are listed in Table \ref{tab:SubstrProp}. We subsequently refer to gap depths according to their decrease in fractional intensity with deeper gaps having lower intensity ratios, e.g., an intensity ratio of $I_d / I_b$=0.2 indicates a gap depth of 80\%.

Beam effects are not explicitly accounted for in these definitions and likely result in underestimating gap widths and overestimating ring widths compared to their true values. Similarly, gap depths may also be underestimated, as beam convolution reduces peak intensities and fills in gaps. While for clearly resolved features, the effects of beam smearing should be small, it becomes significant for features which have widths comparable to the beam size. Nonetheless, the adopted conventions are still useful for comparing substructures across the MAPS sample.

\subsection{Additional and tentative substructures} \label{sec:tentative_substructures}

\begin{figure}[!ht]
\centering
\includegraphics[width=0.75\linewidth]{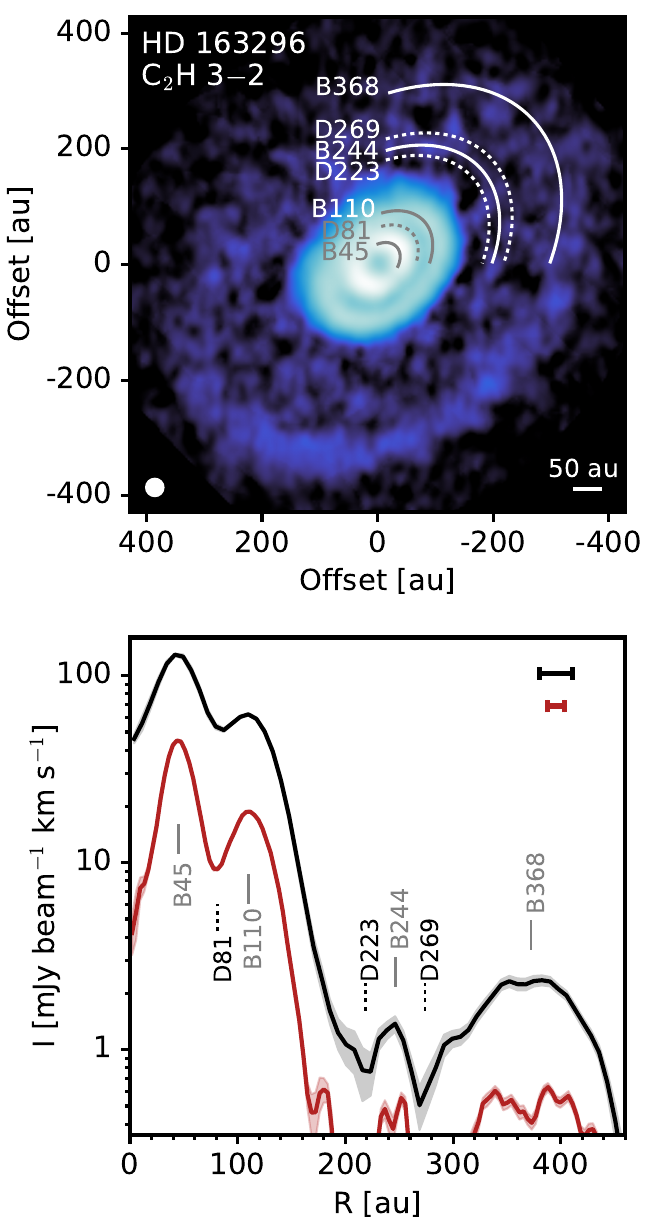}
\caption{Zeroth moment map (top) and radial intensity profile (bottom) of C$_2$H 3$-$2 in HD~163296. An arcsinh color stretch has been applied to the zeroth moment map. The red and black curves show the 0\farcs15 and 0\farcs30 resolution profiles, respectively.}
\label{fig:HD16_C2H_outerRing}
\end{figure}

To ensure that we did not miss the presence of extended but low SNR features in the radial profiles listed in Table \ref{tab:ProfSelection}, we inspected all tapered (0\farcs30) profiles\footnote{A tentative outer ring at ${\sim}$400~au is seen in CS 2--1 in GM~Aur (see \citet{legal20}), but we do not include it in this analysis.} and zeroth moment maps. The process revealed two additional C$_2$H 3$-$2 emission rings (B244, B368) in HD~163296, as shown in Figure \ref{fig:HD16_C2H_outerRing}. This outermost ring was previously detected by \citet{Bergner19}, while the narrower ring at 244~au is newly detected in the MAPS observations. As indicated in Table \ref{tab:SubstrProp}, we used the 0\farcs30 tapered resolution radial profile to fit both of these rings, as they are too low SNR in the 0\farcs15 resolution image for a robust characterization.

A few lines exhibit suggestive emission shoulders, which were not distinctive enough to be considered as independent substructural features, but may still indicate the presence of additional, marginally resolved emission rings. A full listing of such emission shoulders is provided in Table \ref{tab:tentative_radial_substructures}. We also catalogued lines that exhibit prominent emission plateaus, e.g., H$_2$CO 3--2, HCO$^+$ 1--0 in GM~Aur (Figure \ref{fig:H2CO_Radial_Profiles}); HCN 1--0 in IM~Lup (Figure \ref{fig:HCN_Radial_Profiles}). All MAPS disks exhibit diffuse, radially-extended CN 1--0 emission, which is discussed further in \citet{bergner20}.

The radial intensity profiles of a few emission rings deviate from Gaussian or otherwise symmetric profiles. Asymmetric profiles such as these may be the result of two unresolved rings or reflect true ring asymmetries. While we did not attempt a detailed characterization of such asymmetries, some notable instances are listed in Table \ref{tab:tentative_radial_substructures}. In particular, HCN 3$-$2, HC$_3$N 11--10 in AS~209 (Figure \ref{fig:HCN_Radial_Profiles}) both display asymmetric tails toward larger radii in their emission rings. More modest asymmetries are seen in several other lines, such as CS 2--1 in HD~163296 and DCN 3--2 in MWC~480.

\subsection{Annular continuum substructures} \label{sec:continuum_substructures}

All disks have existing high angular resolution observations of their millimeter continua \citep{Long18, Huang18} and sometimes in several ALMA bands \citep[e.g.,][]{Huang20}. However, due to the sensitivity of the MAPS observations, we detected new substructures in the outer continuum disks of IM~Lup (D209, B220) and MWC~480 (D149, B165). In IM~Lup, this outer ring had been tentatively seen in the lower resolution (0\farcs3) observations of \citet{Cleeves16}. In MWC~480, this additional dust ring, although not seen in previous imaging, had been inferred from visibility model fitting \citep{Long18, Liu19}.

\begin{figure*}[]
\centering
\includegraphics[width=\linewidth]{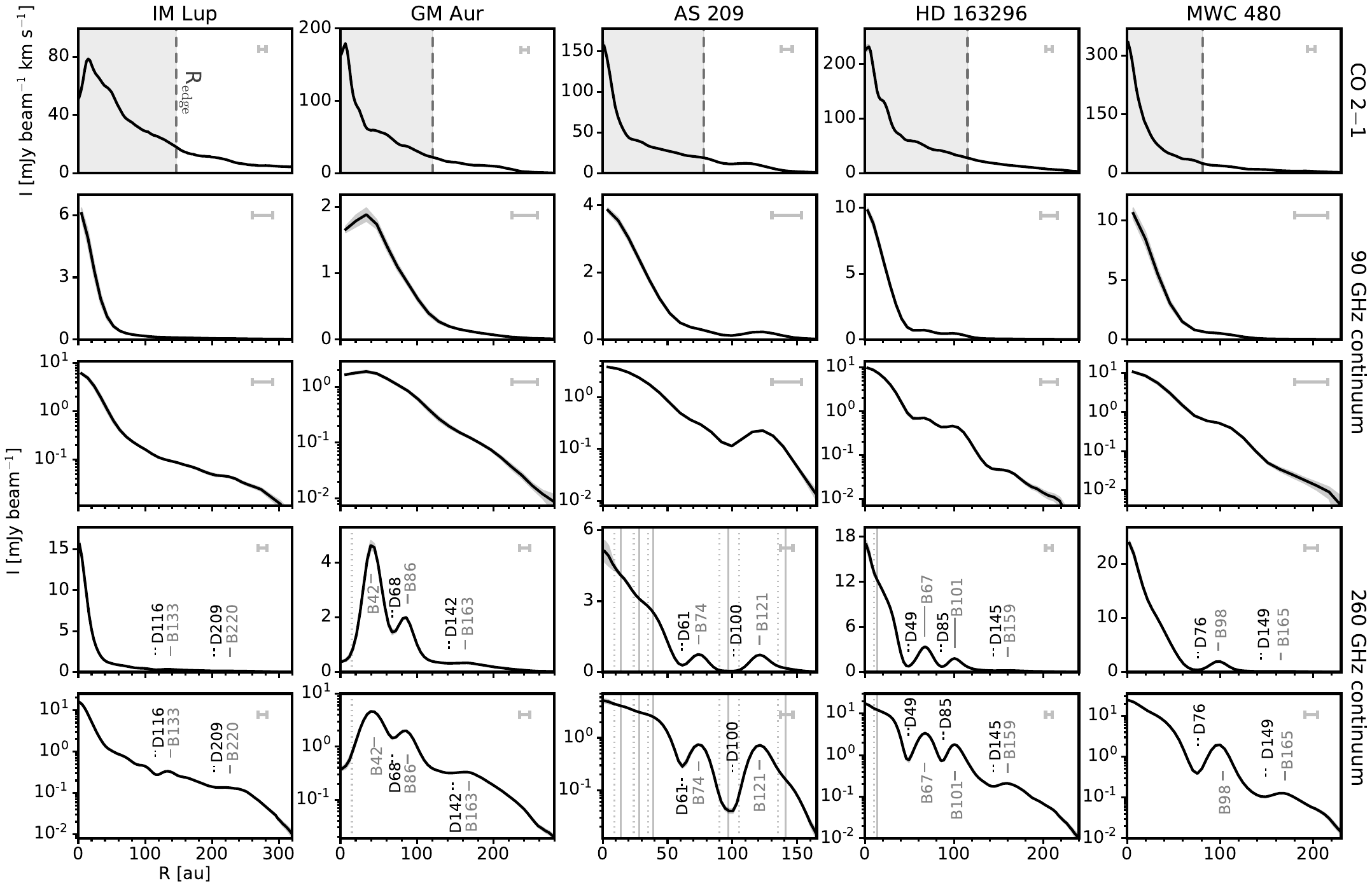}
\caption{Deprojected radial intensity profiles of the 90 GHz and 260 GHz continua compared to CO 2--1, ordered from left to right by increasing stellar mass. The outer edge of the millimeter continuum (R$_{\rm{edge}}$) is shown as a gray shaded region in the CO 2--1 profiles. Both linear scale and log scale continuum radial intensity profiles are shown. Gray shaded regions show the 1$\sigma$ scatter at each radial bin (i.e., arc or annulus) divided by the ratio of the square root of bin circumference and FWHM of the synthesized beam. Solid gray lines mark emission rings and dotted black lines mark gaps, as listed in Table \ref{tab:ContinuumSubstrProp}. Vertical solid and dotted lines indicate millimeter continuum rings and gaps, respectively, that are unresolved in our observations. The FWHM (i.e., 0$\farcs$15 for CO 2--1 and minor axis for continuum images) of the synthesized beam is shown by a horizontal bar in the upper right corner of each panel.}
\label{fig:CO_wcont_radial_profiles}
\end{figure*}

The characteristics of these new continuum substructures are reported in Table \ref{tab:ContinuumSubstrProp}. We generated continuum radial profiles and identified annular features, as in Sections \ref{sec:radial_profiles_generation} and \ref{sec:radial_locations}--\ref{sec:widths_and_contrast}, respectively. All annular substructures were characterized using azimuthally-averaged profiles to increase SNR in the outer radii and with the 260~GHz continuum, which possesses the highest angular resolution (${\sim}$0\farcs1). For all disks, the other three continuum frequency settings were inspected, but none revealed any additional substructures not present in the 260~GHz continuum, as illustrated in Figure \ref{fig:CO_wcont_radial_profiles}.

Although GM~Aur has extensive continuum observations \citep{Huang20}, the individual annular substructures lack reported widths and depths. As a result, we refit all substructures self-consistently and found that all radial locations were within 2--5~au of those reported in \citet{Huang20}. Similarly, previous millimeter continuum observations exist for MWC~480 \citep{Long18, Liu19}, but the MAPS observations have a higher spatial resolution and improved RMS, as detailed in \citet{sierra20}. We refit the MWC~480 continuum substructures and found differences in derived radial locations of no more than 3~au.

To self-consistently compare continuum and chemical substructures, we refit those continuum substructures which have been previously catalogued in DSHARP at higher spatial resolutions. All refitted values, and their corresponding DSHARP names, are marked in Table \ref{tab:ContinuumSubstrProp}. In Table \ref{tab:ContinuumSubstrProp}, we also list those substructures identified by \citet{Huang18, Huang20} at small radii or with very narrow widths that we did not detect due to our lower spatial resolution. Overall, we find close agreement between the radial positions reported in DSHARP and those derived from our refitting process with a maximum difference of any individual substructure of no more than 4~au. This consistency indicates that the derived substructure characteristics do not depend strongly on the details of the fitting process. We also emphasize that refitted values are solely for the purposes of self-consistent comparison and that those previously derived from higher angular resolution observations ultimately represent more accurate radial locations and widths. An overview of continuum radial intensity profiles for the MAPS sources with labeled continuum substructures is shown in Figure \ref{fig:CO_wcont_radial_profiles}. For a more detailed analysis of the continuum, see \citet{sierra20}.

\subsection{Gas disk radii} \label{sec:gas_disk_radii}

\begin{figure*}[!ht]
\centering
\includegraphics[width=\linewidth]{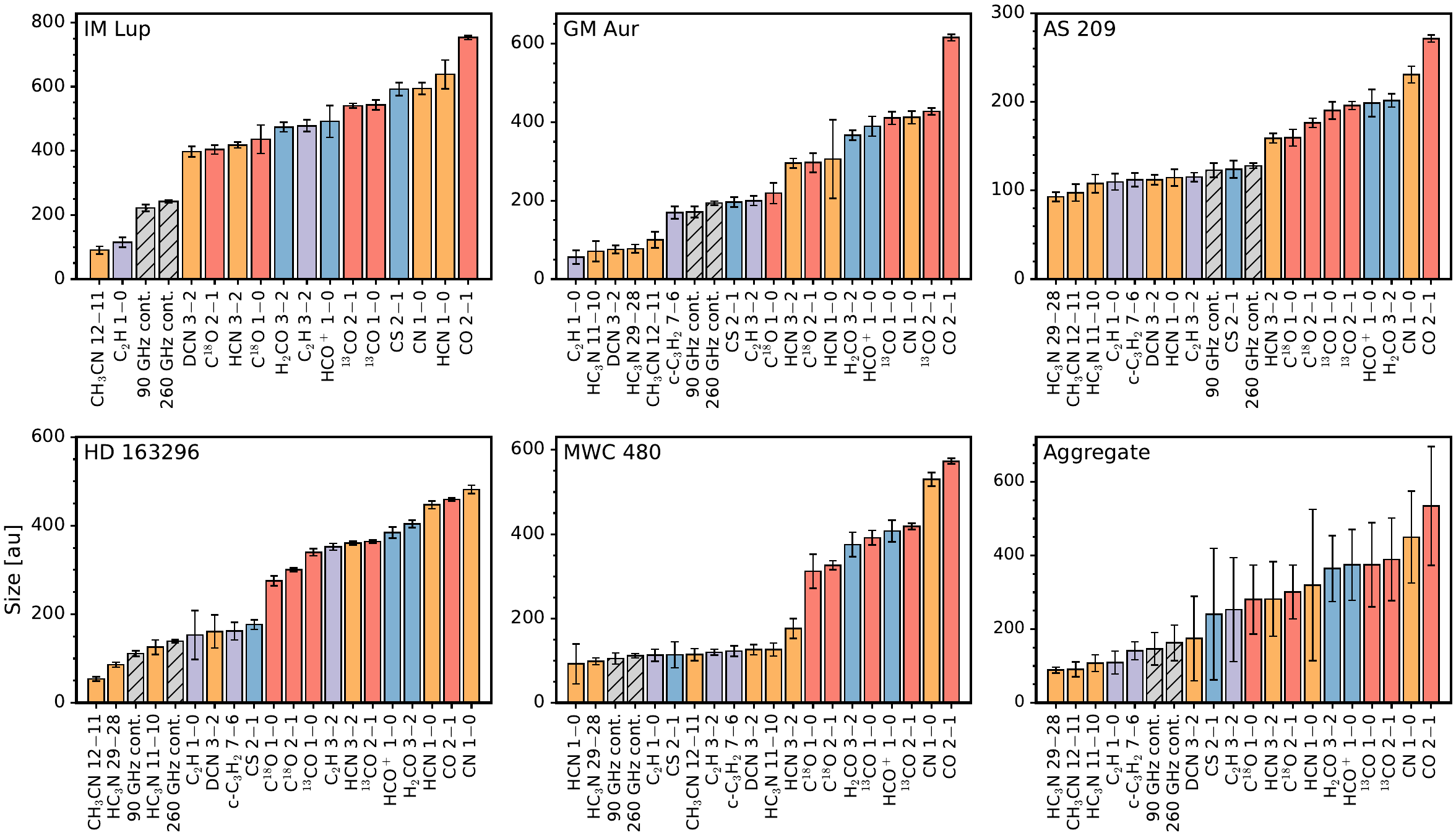}
\caption{Gas disk size for all lines organized by increasing sizes within each disk. Sizes are color-coded by species, as described in Section \ref{sec:gas_disk_radii}. The size of the 90~GHz and 260~GHz continuum disks are shown as gray, hatched bars for comparison. The aggregate panel shows mean sizes of each line across the MAPS disks with error bars showing the standard deviation. Disk sizes are defined as the radius containing 90\% of total flux.}
\label{fig:disk_sizes}
\end{figure*}

\begin{figure*}[!ht]
\centering
\includegraphics[width=\linewidth]{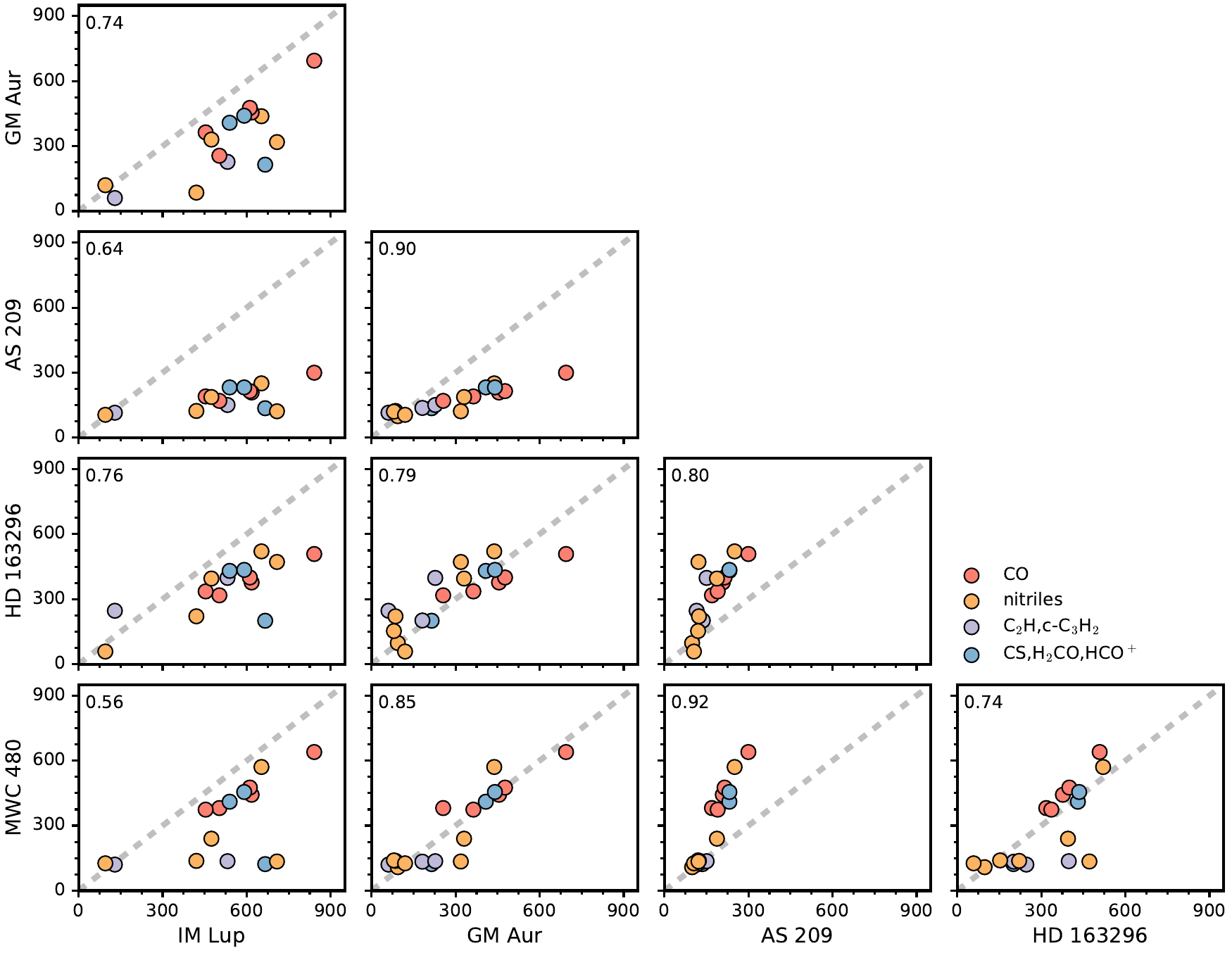}
\caption{Gas disk size of chemical species in au for each disk plotted against one another. Spearman correlation coefficients are displayed in the upper left corners of each scatter plot. A one-to-one size ratio is shown as a gray dashed line. Sizes are color-coded by species according to the legend. In general, the distribution of sizes in different species is quite consistent among disks.}
\label{fig:Spearman_disk_size}
\end{figure*}

\textit{Note: The calculation of gas disk sizes in the published article initially omitted the necessary factor of $2\pi r$ when computing the cumulative sums, i.e., the integral of a radial function has an areal element of $2\pi r dr$. Please see the associated Erratum for a full update of this Section; while the text is unchanged, we have corrected the values shown in Table \ref{tab:disksizes} and Figures  \ref{fig:disk_sizes}-\ref{fig:Spearman_disk_size}.}

To explore the relative size of line emission in the MAPS disks, we computed the radius of the gas disk for each line. We used a method similar to that of \citet{Ansdell18}. We first measured a total line flux, taken as the asymptotic value of the azimuthally-averaged radial intensity profiles. We then define R$_{\rm{gas}}$ as the radius which encloses 90\% of this total flux. The measured values are listed in Table \ref{tab:disksizes} and shown in Figure \ref{fig:disk_sizes}. The estimated uncertainty includes the uncertainty in radial location equal to one bin, as for the radial features, and the uncertainty in the line fluxes.

\begin{deluxetable*}{lcccccccc}
\tablecaption{Gas Disk Sizes \label{tab:disksizes}}
\tablewidth{0pt}
\tablehead{
\colhead{Line} & \multicolumn5c{Disk Size (au)}\\ \cline{2-7}
\colhead{} & \colhead{IM Lup} & \colhead{GM Aur} & \colhead{AS 209} & \colhead{HD 163296} & \colhead{MWC 480}}
\startdata
CO 2$-$1		&753~$\pm$~6		&616~$\pm$~9		&272~$\pm$~4		&459~$\pm$~4		&573~$\pm$~7		&\\
$^{13}$CO 2$-$1		&540~$\pm$~7		&427~$\pm$~9		&196~$\pm$~5		&364~$\pm$~4		&419~$\pm$~7		&\\
$^{13}$CO 1$-$0		&543~$\pm$~16		&410~$\pm$~16		&190~$\pm$~10		&340~$\pm$~8		&392~$\pm$~17		&\\
C$^{18}$O 2$-$1		&404~$\pm$~14		&297~$\pm$~24		&176~$\pm$~5		&301~$\pm$~4		&326~$\pm$~11		&\\
C$^{18}$O 1$-$0		&436~$\pm$~45		&219~$\pm$~26		&160~$\pm$~9		&275~$\pm$~11		&312~$\pm$~40		&\\
C$_2$H 3$-$2		&478~$\pm$~18		&200~$\pm$~12		&115~$\pm$~5		&352~$\pm$~8		&120~$\pm$~7		&\\
C$_2$H 1$-$0		&114~$\pm$~15		&56~$\pm$~17		&110~$\pm$~9		&153~$\pm$~55		&113~$\pm$~14		&\\
c-C$_3$H$_2$ 7$-$6		&$\ldots$		&169~$\pm$~16		&112~$\pm$~8		&162~$\pm$~20		&122~$\pm$~13		&\\
H$_2$CO 3$-$2		&474~$\pm$~15		&366~$\pm$~13		&202~$\pm$~7		&404~$\pm$~9		&375~$\pm$~29		&\\
HCO$^+$ 1$-$0		&492~$\pm$~50		&389~$\pm$~25		&199~$\pm$~15		&385~$\pm$~13		&408~$\pm$~26		&\\
CS 2$-$1		&592~$\pm$~21		&197~$\pm$~13		&124~$\pm$~10		&177~$\pm$~11		&114~$\pm$~31		&\\
HCN 3$-$2		&418~$\pm$~9		&295~$\pm$~13		&159~$\pm$~5		&361~$\pm$~4		&176~$\pm$~24		&\\
HCN 1$-$0		&638~$\pm$~45		&306~$\pm$~100		&115~$\pm$~9		&447~$\pm$~9		&92~$\pm$~47		&\\
DCN 3$-$2		&397~$\pm$~17		&76~$\pm$~10		&112~$\pm$~6		&161~$\pm$~37		&126~$\pm$~12		&\\
HC$_3$N 29$-$28		&$\ldots$		&77~$\pm$~11		&93~$\pm$~5		&86~$\pm$~6		&98~$\pm$~8		&\\
HC$_3$N 11$-$10		&$\ldots$		&71~$\pm$~26		&108~$\pm$~10		&126~$\pm$~16		&126~$\pm$~15		&\\
CN 1$-$0		&594~$\pm$~18		&412~$\pm$~16		&231~$\pm$~9		&482~$\pm$~9		&530~$\pm$~16		&\\
CH$_3$CN 12$-$11		&90~$\pm$~12		&100~$\pm$~21		&98~$\pm$~10		&54~$\pm$~5		&114~$\pm$~14		&\\
90 GHz continuum		&222~$\pm$~10		&171~$\pm$~14		&123~$\pm$~8		&111~$\pm$~6		&105~$\pm$~13		&\\
260 GHz continuum		&242~$\pm$~4		&194~$\pm$~5		&128~$\pm$~3		&139~$\pm$~3		&111~$\pm$~5		&\\
\enddata
\tablecomments{Disk size was computed as the radius which encloses 90\% of the total disk flux (see Section \ref{sec:gas_disk_radii}). Note that this is often smaller than the total radial extent of an emission line due to the presence of diffuse, low flux emission at large radii.}

\end{deluxetable*}

Gas disk sizes span a wide range for individual transitions within a disk and across the MAPS sample. IM~Lup has the largest disk with the most extended lines having R$_{\rm{gas}} \gtrsim 450$~au, while the AS~209 disk is the most compact, having a maximum R$_{\rm{gas}}$ of only ${\sim}$200~au. In general, the CO lines and CN 1--0, HCO$^+$ 1--0, and H$_2$CO 3--2 are the largest, while the nitriles are the smallest, with HC$_3$N 29--28 and CH$_3$CN 12--11 typically presenting the most compact radii in all disks. The hydrocarbons C$_2$H and c-C$_3$H$_2$ are also found in the bottom quartile of sizes. No clear trends are found when comparing Band 3 and Band 6 transitions from the same molecule, and in general, R$_{\rm{gas}}$ values are not systematically larger in Band 3 or Band 6, as would have been expected if we were resolution or sensitivity limited, respectively. Only in a few instances of especially weak lines, i.e., C$_2$H 1--0 in IM~Lup and GM~Aur, did we find substantially smaller sizes in Band 3 versus the Band 6 line of the same molecule, which for these particular cases, may suggest artificially smaller Band 3 lines due to insufficient sensitivity.

Lines are color-coded in Figure \ref{fig:disk_sizes}, and in subsequent analysis, according to the following groupings: CO isotopologues (red), nitriles (orange), hydrocarbons C$_2$H, c-C$_3$H$_2$ (purple), and CS, H$_2$CO, HCO$^+$ (blue). The first three groupings are, in part, motivated by chemical similarity, as each nitrile has a $-$C$\equiv$N functional group and hydrocarbons are exclusively made up of hydrogen and carbon atoms. These categories are also in qualitative agreement with radial emission morphologies visually identified when comparing radial profiles. The grouping of CS, H$_2$CO, and HCO$^+$ is, however, one of convenience, as these molecules are not chemically similar to other species in our sample, nor with one another.

Figure \ref{fig:disk_sizes} suggests that O-poor organic chemistry (e.g., the hydrocarbons C$_2$H, c-C$_3$H$_2$ and nitriles HC$_3$N, CH$_3$CN) is, on average, quite compact, while CO and its inorganic and organic derivatives are extended. This should result in a large scaled C/O gradient across the disk and implies that the inner 100~au of the disk, which is most relevant for planet formation, is more C-rich than perhaps disk-averaged line emission would suggest \citep[for further discussion, see][]{alarcon20, bosman20_C_over_O}. In further support of this interpretation, the formation of complex nitriles, such as HC$_3$N and CH$_3$CN, has been shown to be efficient at elevated C/O ratios \citep{Legal19_co_ratio}. This effect is most pronounced in AS~209, MWC~480, and to a lesser degree, GM~Aur, which have nearly bimodal size distributions between the extended CO and related species versus the compact complex nitriles. In fact, for all MAPS disks, the complex nitriles are no larger than ${\sim}$120~au in size --- and are often comparable to that of the continuum extent --- despite the wide variations in CO disk sizes. This suggests an association with the millimeter continuum, where these molecules may be more easily destroyed at radii beyond the pebble disk due to, e.g., less shielding from radiation or increased gas-phase O-chemistry. 

To explore the relationship between the continuum and gas disk sizes, we included the size\footnote{Multiple definitions for continuum disk size exist \citep[e.g.,][]{Tripathi17, Long18, Huang18} and the 90\% flux definition is chosen here for consistent comparison with the molecular lines. The location of the outermost edge of the continuum emission can often be over twice as large due to the presence of diffuse, low flux emission at large radii, cf. R$_{\rm{edge}}$ in Section \ref{sec:outer_edge_of_mm_continuum_disk} and Table \ref{tab:ContinuumSubstrProp}.} of the 90~GHz and 260~GHz continuum, measured as for the molecular lines, in Figure \ref{fig:disk_sizes}. The 260~GHz continuum disk is typically ${\sim}$20--35\% larger than that of the 90~GHz. The two exceptions to this trend are MWC~480, where they are nearly equal, and IM~Lup, where the continuum disk at 260~GHz is 55\% larger than at 90~GHz. The continua are smaller than nearly every line in the MAPS disks, except for GM~Aur, where they are larger than about one-third of the lines considered here. We also calculated molecular line-to-dust size ratios, which spanned a wide range of 0.4 to 6 across individual lines in the MAPS disks. The ratios associated with the complex nitriles are typically ${<}$1.5, reflecting their compact spatial distributions that do not extend much beyond the continuum disk. In contrast, the CO lines are much more extended with ratios between 2--6. In general, we find large disk-to-disk variations in line size (${>}$100~au) with the exception of the complex nitriles and c-C$_3$H$_2$, which have size variations of only ${\sim}$40~au among the MAPS disks.

To assess how similar or different disks are in their gas sizes, we compared the rank ordering of lines within each disk against one another. To do so, we computed Spearman correlation coefficients for each pair of disks, as shown in Figure \ref{fig:Spearman_disk_size}. All pairs of disks are positively correlated, indicating that while the substructure patterns vary dramatically across disks, the relative radial size distribution of lines is similar between the different sources. In fact, this similarity is nearly at a one-to-one ratio among the GM~Aur, HD~163296, and MWC~480 disks.

\section{Properties of radial substructures} \label{sec:aggregate_properties}

\subsection{Distribution of radial substructure locations} \label{sec:radial_distribution}

Figures \ref{fig:CO_Moment0}--\ref{fig:HCN_Moment0} and \ref{fig:CO_Radial_Profiles}--\ref{fig:HCN_Radial_Profiles} show that substructures are observed at almost all radii where line emission is detected from ${\lesssim}$10~au to over 500~au, although the majority occur within 200~au. There is a wide range in the number of features seen across the different sources. The HD~163296 disk possesses the most chemical substructures with multiple emission rings and gaps in numerous lines, while IM~Lup and GM~Aur have smoother radial intensity profiles with relatively fewer well-defined substructures. AS~209 and MWC~480 show a single bright ring in the majority of lines and the occasional presence of emission shoulders and lower contrast substructures. Variations in the number of features observed within a single disk across different lines are also common, with HD~163296 demonstrating the most variability. For instance in HD~163296, HCN 3--2 has a set of four well-defined emission rings, but HC$_3$N 29--28 only has a single isolated ring. In contrast, AS~209 is the most consistent in its relative number of substructures across lines.

\begin{figure*}[!ht]
\centering
\includegraphics[width=\linewidth]{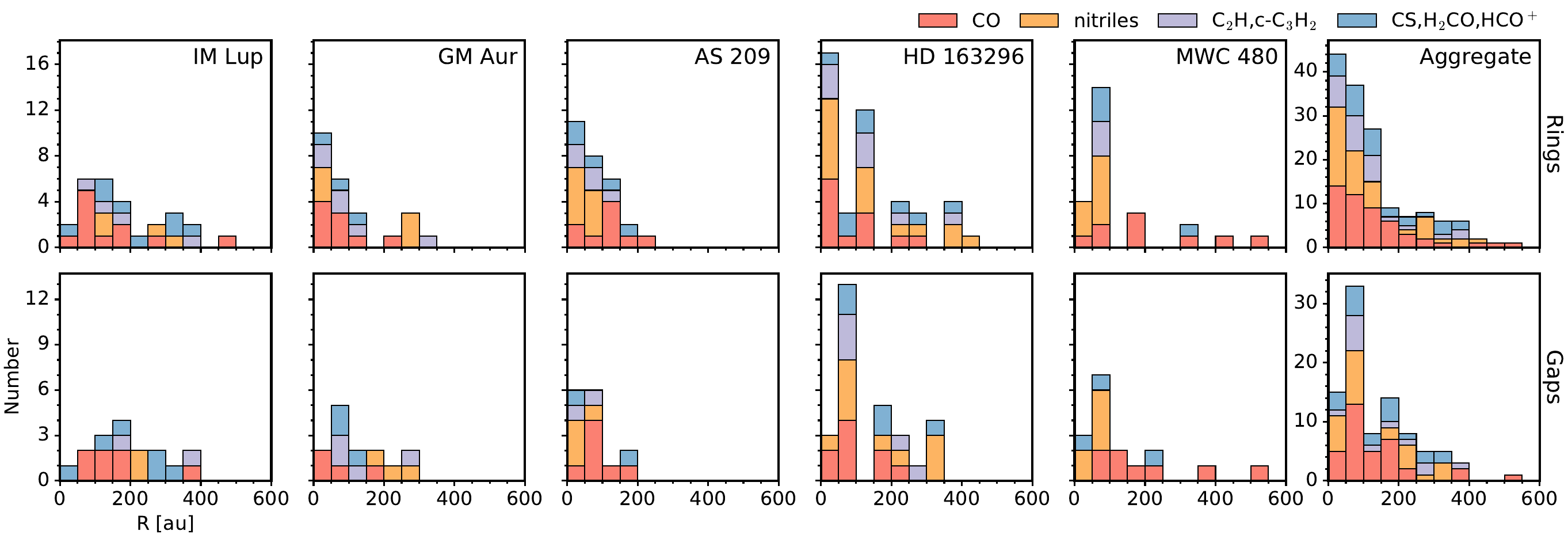}
\caption{Histogram of radial locations of emission rings (top row) and gaps (bottom row) for each disk and in aggregate (columns). Substructures are color-coded by species according to the legend.}
\label{fig:r0_histogram}
\end{figure*}

Figure \ref{fig:r0_histogram} shows histograms of the positions of radial substructures in the MAPS sample. The number of rings and gaps in each disk generally decreases, although not monotonically, as a function of radius. As before, substructures are color-coded according to species. Each type of species displays radial substructure and when compared in aggregate, the relative number of rings contributed by each group is approximately constant in radius. In particular, in the inner 150~au, rings arising from each group occur in equal proportion, while the distribution of gaps has a modest deficit in substructures from hydrocarbons.

\subsection{Distribution of substructure widths and depths} \label{sec:distr_widths_contrasts}

The measured widths and depths of substructures span a relatively wide range. Substructures have widths from ${<}10$~au to over 200~au, but the majority of features are less than 100~au wide. The deepest gaps have depths as low as 90\%, but most gaps are considerably shallower with depths of ${\sim}$10--30\%. The majority of extremely low contrast gaps, those on the order of a few percent, have widths that are comparable or smaller than the synthesized beam. Thus, their apparent shallowness may be a consequence of limitations in angular resolution. No trends with radius or species are identified in the gap depths of any MAPS source. Instead, there is significant variation in substructure depth at all radii and also among similar species.

\begin{figure*}[!ht]
\centering
\includegraphics[width=\linewidth]{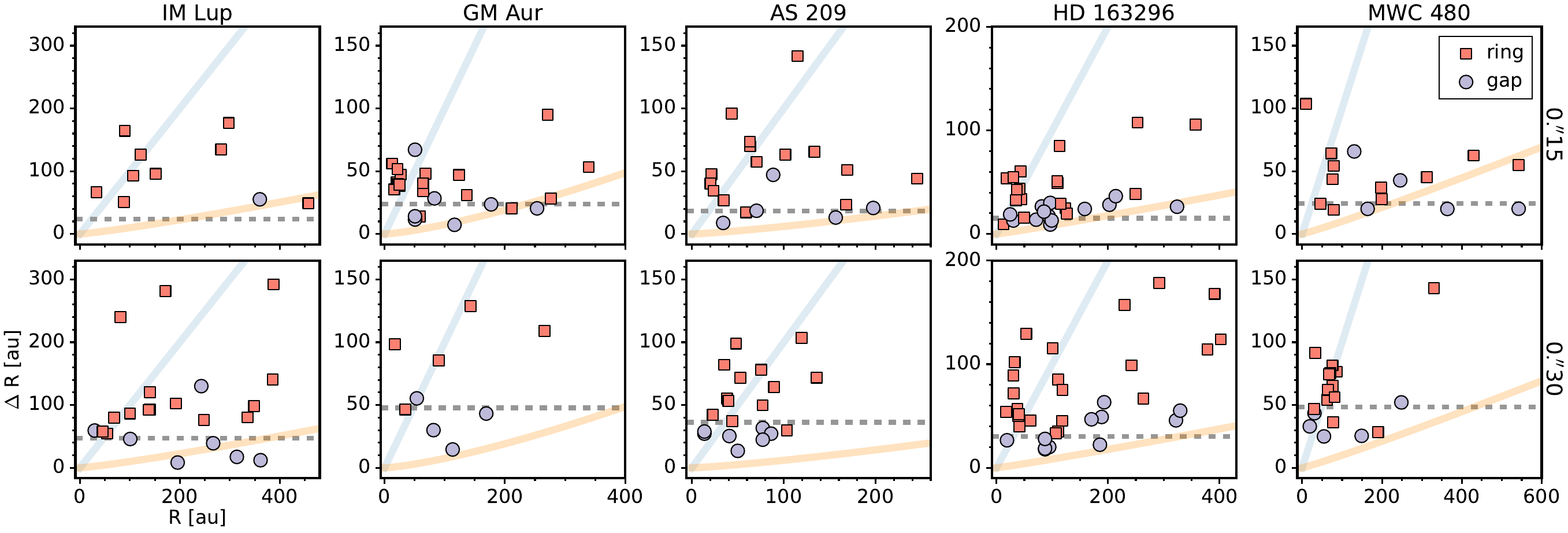}
\caption{Substructure width as a function of radial location for lines at 0$\farcs$15 (top row) and 0$\farcs$3 (bottom row) resolution. Red squares denote rings and purple circles mark gaps. The gray dashed horizontal lines correspond to $\theta_{\rm{b}} \times d$, which divides resolved features above the line from those that are unresolved or marginally resolved below or at the line. A constant $\Delta R / R = 1$ is shown as a solid blue line. The pressure scale height for each disk from \citet{zhang20} is shown as a solid orange line.}
\label{fig:delta_r_plot}
\end{figure*}

Figure \ref{fig:delta_r_plot} shows substructure widths relative to their radial locations. Most features are spatially-resolved with measured widths that are larger than the FWHM of the beam, as shown by the gray dashed lines in Figure \ref{fig:delta_r_plot}. However, the smallest measured widths are almost entirely located within the inner 100~au and should be treated as upper limits since they are often not clearly resolved. Gaps observed at 0$\farcs$3 resolution, particularly those in IM~Lup, are still only marginally-resolved even at radii larger than 100~au. We find no systematic differences in feature widths measured using either the 0$\farcs$15 or 0$\farcs$3 resolution images.

Substantial variation in substructure widths are observed within individual disks. IM~Lup and HD~163296 have the largest range of ${\sim}$200~au between their widest and narrowest features, while GM~Aur and MWC~480 have a spread of no more than ${\sim}100$~au. In each disk, substructure widths generally increase with radius. This is unsurprising, as the physical size of disk structures grows with distance from the central star due to increases in local scale height \citep{Chiang97}. However, the ratio between width and radial position of all substructures decreases toward larger radius. Provided that some of these lines trace the gas distribution, substructure widths provide constraints as to their origins, e.g., the width of gaps opened by planets of a given mass scale with radius \citep{Kanagawa16}.

We also compare substructure widths with disk pressure scale heights \citep{zhang20}, shown as solid orange lines in Figure \ref{fig:delta_r_plot}. Nearly all rings are substantially wider than pressure scale heights, while gaps are often no greater than ${\sim}$2 scale heights, and in some cases, are comparable to or smaller than the scale height. These relatively narrow gap widths are considerably smaller than what is expected from planet-disk interactions \citep{Kanagawa16, Yun19} and may instead indicate that some molecular gaps are due to local density/temperature changes or steep chemical gradients across phase transition regions (e.g., snowlines).

\subsection{Source-specific description of substructures} \label{sec:source_specific_substructure}

In addition to considering the aggregate properties of substructures, we briefly summarize the distribution of substructures, including salient trends or notable features, for each MAPS disk below: 

\subsubsection{IM~Lup} \label{sec:IM_Lup_Summary}
IM~Lup is the only MAPS source that shows spiral structures in its millimeter continuum \citep{Huang18}, but no corresponding spirals are seen in molecular line emission. Among the MAPS disks, IM~Lup possesses the largest radial extent (${\sim}$700~au) in CO 2--1, which has been explained by the presence of a photoevaporative wind \citep{Haworth17}. IM~Lup has a unique line emission distribution with a central depression, broad ring-like structure, and plateau of diffuse emission extending out to large radii (${\sim}$600~au) in nearly all lines. This morphology is best illustrated by C$_2$H 3$-$2 (Figures \ref{fig:C2H_Moment0} and \ref{fig:C2H_Radial_Profiles}) and HCN 3$-$2 (Figures \ref{fig:HCN_Moment0} and \ref{fig:HCN_Radial_Profiles}) and is also seen, to a lesser degree, in HCO$^+$ 1--0 and H$_2$CO 3--2 (Figures \ref{fig:H2CO_Moment0} and \ref{fig:H2CO_Radial_Profiles}). Plateau-like emission, although at low SNR, is also observed in HCN 1--0 and CN 1--0. DCN 3--2 has a broad double-ringed emission morphology, similar to previous observations of double rings in DCO$^+$ 3--2 \citep{Oberg15, Huang17} and N$_2$D$^+$ 3--2 \citep{cataldi20}. IM~Lup has the lowest SNR emission for each line in our sample and is the only disk with non-detected transitions in the set of lines considered here (see \citealt{ilee20} for more details). As IM~Lup is the youngest \citep[0.2--1.3~Myr;][]{Alcala17} MAPS source, the origin of some of these unique features may be a consequence of its youth.

\subsubsection{GM~Aur} \label{sec:GM_Aur_Summary}
The GM~Aur disk is classified as a transitional disk due to its central dust and gas cavity \citep[][]{Calvet05, Dutrey08, Hughes09}. Subsequent observations detected inner continuum emission and resolved this dust cavity into an annular gap at 15~au with a corresponding but more compact gas cavity \citep{Huang20}. The MAPS observations confirm this, as GM~Aur has a central dip and bright, compact inner ring at ${\sim}$15--30~au in all lines at 0\farcs15 resolution. This is best illustrated by the inner HCN 3--2 and HC$_3$N 29--28 (Figures \ref{fig:HCN_Moment0} and \ref{fig:HCN_Radial_Profiles}) and H$_2$CO 3--2 (Figures \ref{fig:H2CO_Moment0} and \ref{fig:H2CO_Radial_Profiles}) emission rings. In contrast, the 0$\farcs$3 resolution profiles, e.g., HCN 1--0 and HC$_3$N 11--10, show a smoothly rising profile in the inner disk. This is likely a resolution effect, and the presence of an inner ring in these lines could be confirmed with higher angular resolution observations. For a more detailed discussion of the inner regions of GM~Aur, see Section \ref{sec:inner_disk}. Beyond this inner compact ring, some lines, e.g., HCN 3--2, decrease smoothly with radius, while others such as C$_2$H 3--2 (Figures \ref{fig:C2H_Moment0} and \ref{fig:C2H_Radial_Profiles}) show the presence of two additional, narrow rings (B68, B124). An outer ring at ${\sim}$300~au is seen in HCN 3--2, 1--0; C$_2$H 3--2; and DCN 3--2, while diffuse emission out to ${\sim}$450~au is present in HCO$^+$ 1--0, H$_2$CO 3--2, and CN 1--0 (Figures \ref{fig:H2CO_Moment0} and \ref{fig:H2CO_Radial_Profiles}). GM~Aur also exhibits dramatic spiral arms in CO 2--1 (Figure \ref{fig:CO_Moment0}; see \citet{huang20_spiral_arms} for more details), which are not seen in any other lines in either GM~Aur or across the MAPS sample.

\subsubsection{AS~209} \label{sec:AS_209_Summary}
AS~209 has the most compact MAPS disk, with many lines not extending beyond ${\sim}$200~au and an outer CO 2--1 radius of no more than ${\sim}$300~au. We identify three gaps in CO 2--1, but do not detect the gap at 74~au reported in the higher spatial resolution (0\farcs08) observations of \citet{Guzman18_dsharp}. Notably, it is the only MAPS disk with high contrast substructures in its CO isotopologues, namely outer $^{13}$CO and C$^{18}$O emission rings at ${\sim}$120--130~au (Figures \ref{fig:CO_Moment0} and \ref{fig:CO_Radial_Profiles}), as seen in previous observations \citep{Huang16_AS209, Favre19}. These well-defined rings suggest that these lines are less optically thick relative to CO isotopologues at the same radii in other MAPS disks. Thus, the rings and gaps in AS~209 are most likely a result of local variations of CO abundance and gas column density \citep{alarcon20, zhang20}. The majority of non-CO lines in AS~209 take the form of a central depression and a broad single ring at ${\sim}$50--80~au, but many lines also exhibit low contrast emission shoulders, e.g., c-C$_3$H$_2$ 7--6 (Figure \ref{fig:C2H_Radial_Profiles}); HCO$^+$ 1--0, CN 1--0 (Figure \ref{fig:H2CO_Radial_Profiles}); and HCN 1--0 (Figure \ref{fig:HCN_Radial_Profiles}). This hints at the presence of additional but unresolved narrow rings. If this single ring is in fact two (or more) narrow rings, this would more closely mirror the continuum structure, which is in the form of a tightly nested set of narrow concentric rings \citep{Guzman18_dsharp}.

\subsubsection{HD~163296} \label{sec:HD_163296_Summary}
HD~163296 has the largest number of unique substructures among the MAPS disks and shows a well-defined, multi-ringed emission morphology in the majority of lines. Indications of rings seen by \citet[][]{Bergner19} are now confirmed by the MAPS observations, which show four well-defined rings in HCN 3--2 (Figures \ref{fig:HCN_Moment0} and \ref{fig:HCN_Radial_Profiles}) and C$_2$H 3--2 (Figure \ref{fig:HD16_C2H_outerRing}). Most lines show one (CH$_3$CN 12--11, HC$_3$N 29--28, 11--10, CS 2--1), two (c-C$_3$H$_2$ 7--6, DCN 3--2), or three (CN 1--0) rings, which are approximately radially coincident. The outermost emission rings in C$_2$H 3--2, HCN 3--2, 1--0, CN 1--0 occur at large radii ${\sim}$400~au and are the most radially-extended non-CO substructures seen in the MAPS disks. A ring-like feature (B44) is also present in C$^{18}$O 2--1 (Figures \ref{fig:CO_Moment0} and \ref{fig:CO_Radial_Profiles}), while HCO$^+$ 1--0 and H$_2$CO 3--2 (Figures \ref{fig:H2CO_Moment0} and \ref{fig:H2CO_Radial_Profiles}) show blended ring-like structures, some of which had been seen previously in \citet{Huang17, Carney17, Guzman18}.

\subsubsection{MWC~480} \label{sec:MWC_480_Summary}
MWC~480 shows the greatest morphological variations between hydrocarbons and nitriles. C$_2$H and c-C$_3$H$_2$ (Figures \ref{fig:C2H_Moment0} and \ref{fig:C2H_Radial_Profiles}) are in the form of a single ring, while HCN and HC$_3$N (Figures \ref{fig:HCN_Moment0} and \ref{fig:HCN_Radial_Profiles}) have centrally peaked profiles with shallow gaps. This latter distribution is unique among non-CO lines across the MAPS disks and is best described as superimposed emission plateaus of different intensities, where the gaps mark the transition regions. DCN 3--2 does not follow this trend and is instead in the form of a single emission ring, similar to C$_2$H and c-C$_3$H$_2$ rather than the other nitriles. A single ring is also evident in HCO$^+$ 1--0, H$_2$CO 3--2, CN 1--0, and CS 2--1 (Figures \ref{fig:H2CO_Moment0} and \ref{fig:H2CO_Radial_Profiles}) and is radially coincident with the hydrocarbon ring. Despite its compact radial extent in most lines, MWC~480 has extended (${\sim}600$~au) and structured CO 2--1 emission in the form of four concentric bright-dark features. Further discussion of the origins and nature of these CO 2--1 substructures is found in \citet{teague20}.

\section{Origins of chemical substructure} \label{sec:origins_of_chemical_substr}

The presence of rings, gaps, and other substructures in molecular line emission may be the result of various chemical effects, including variations in C/O ratios, freeze-out onto grains in the disk midplane, thermal desorption in dust substructures, and UV-driven production or selective photodissociation in the disk atmospheres \citep[e.g.,][]{Teague_ea_2017, Cazzoletti18, Miotello19}. Substructures can also result from local deviations in disk physical structure, either in density or temperature, e.g., from planet-disk interactions or alterations in dust properties \citep[][]{Bae17, Guzman18_dsharp, Huang20}. Non-LTE and excitation effects have a direct effect on observed emission intensities \citep{Pavlyuchenkov07}. Many of these topics are the subject of other MAPS papers \citep{aikawa20, alarcon20, bergner20, bosman20_C_over_O, calahan20b, cataldi20, guzman20, schwarz20, teague20, zhang20}, while here we instead aim to empirically explore and connect spatial trends in chemical substructures with their potential chemical and physical origins. We also briefly comment on some notable trends and describe those of particular interest in more detail below.

\subsection{Spatial links between chemical substructures} \label{sec:chemical_trends}

We want to assess the relative similarity of molecular line emission profiles within each disk and across the entire MAPS sample. In this context, similarity means consistent radial morphologies, namely the shapes and locations of gaps, rings, and emission shoulders. To quantify this kind of similarity, we compared pairs of radial profiles and calculated their radially-integrated absolute differences. The profiles were first normalized such that the peak brightness of the first profile (corresponding to the line with the brighter absolute peak intensity) was set to unity, and then the second radial profile was scaled to minimize the difference between the two profiles. We found that this approach is effective at identifying profiles that look similar by eye, i.e., it does not overly penalize profiles with different relative fluxes but otherwise similar morphologies, such as HCN 3--2 and DCN 3--2 or HCN 3--2 and C$_2$H 3--2 in HD~163296. 

\begin{figure*}
\centering
\includegraphics[width=0.47\linewidth]{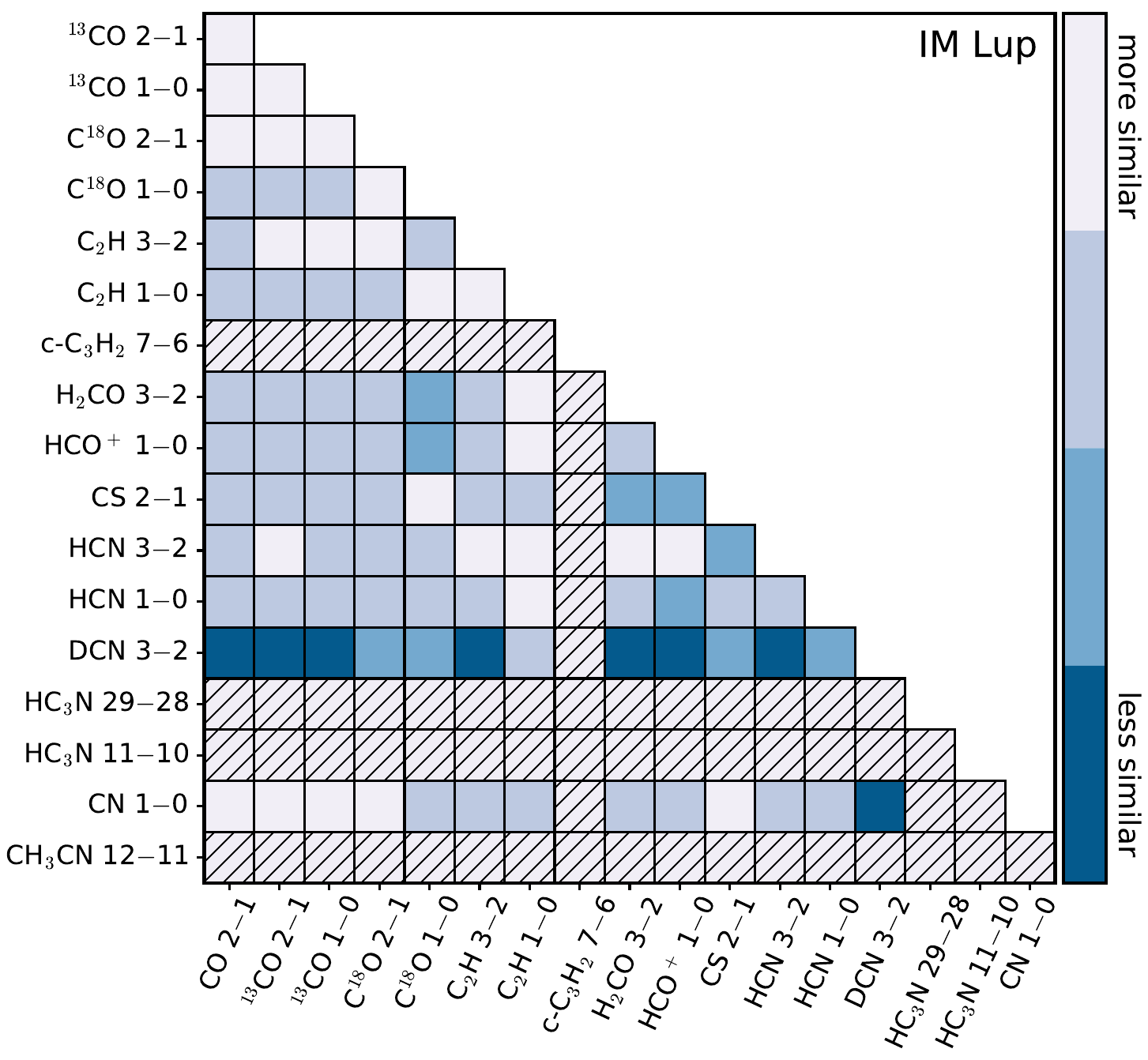}
\includegraphics[width=0.47\linewidth]{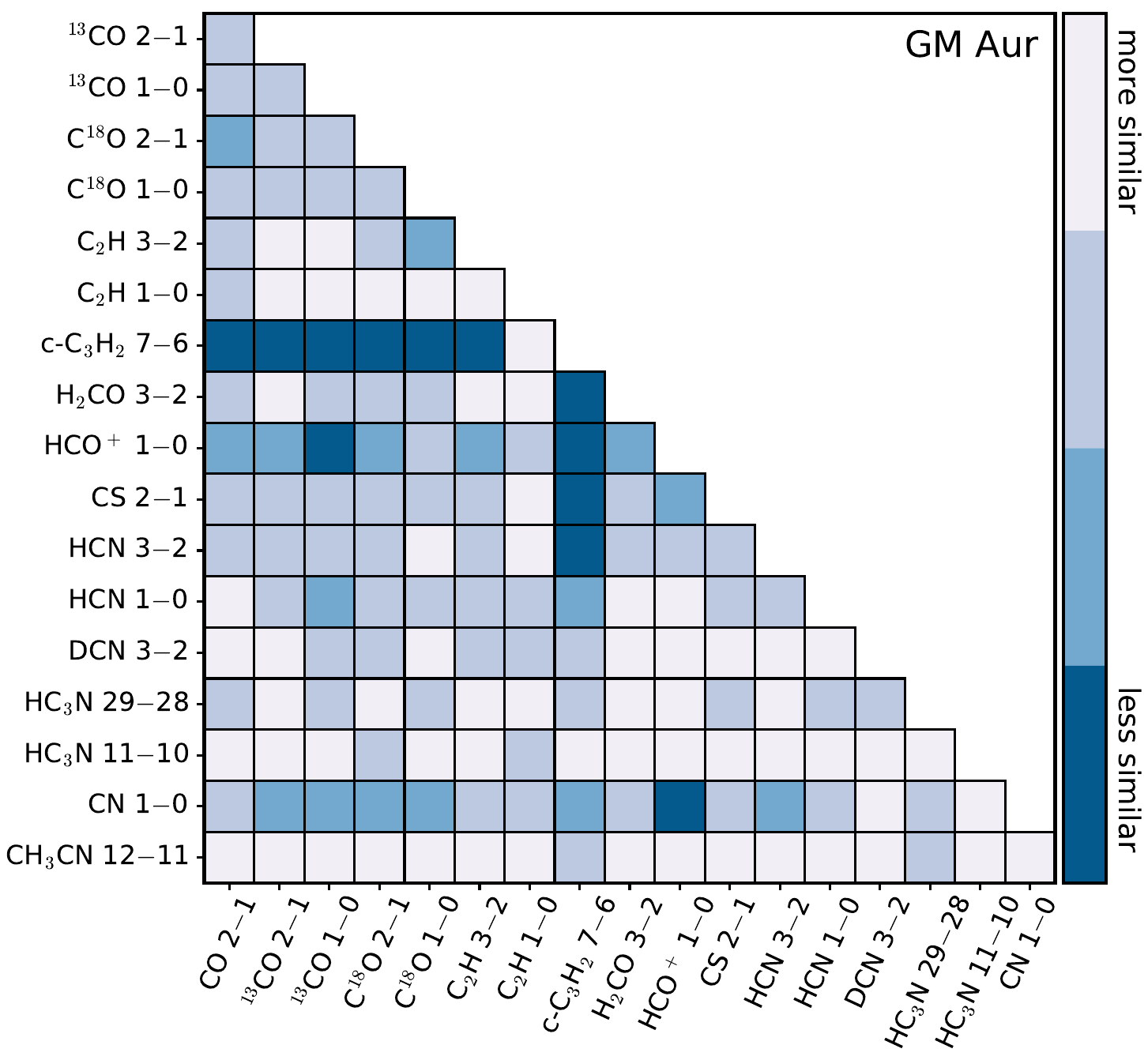}\\
\includegraphics[width=0.47\linewidth]{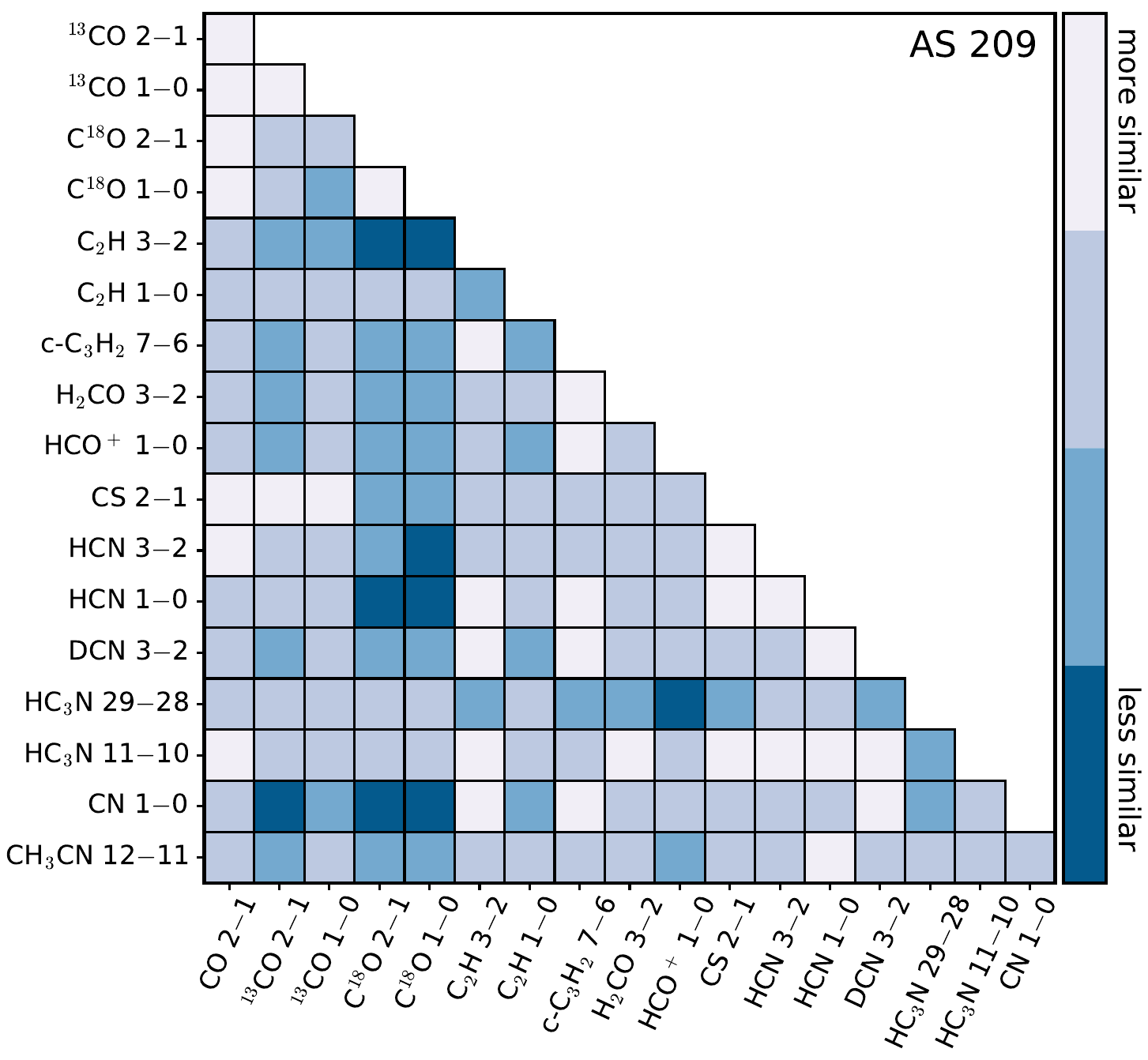}
\includegraphics[width=0.47\linewidth]{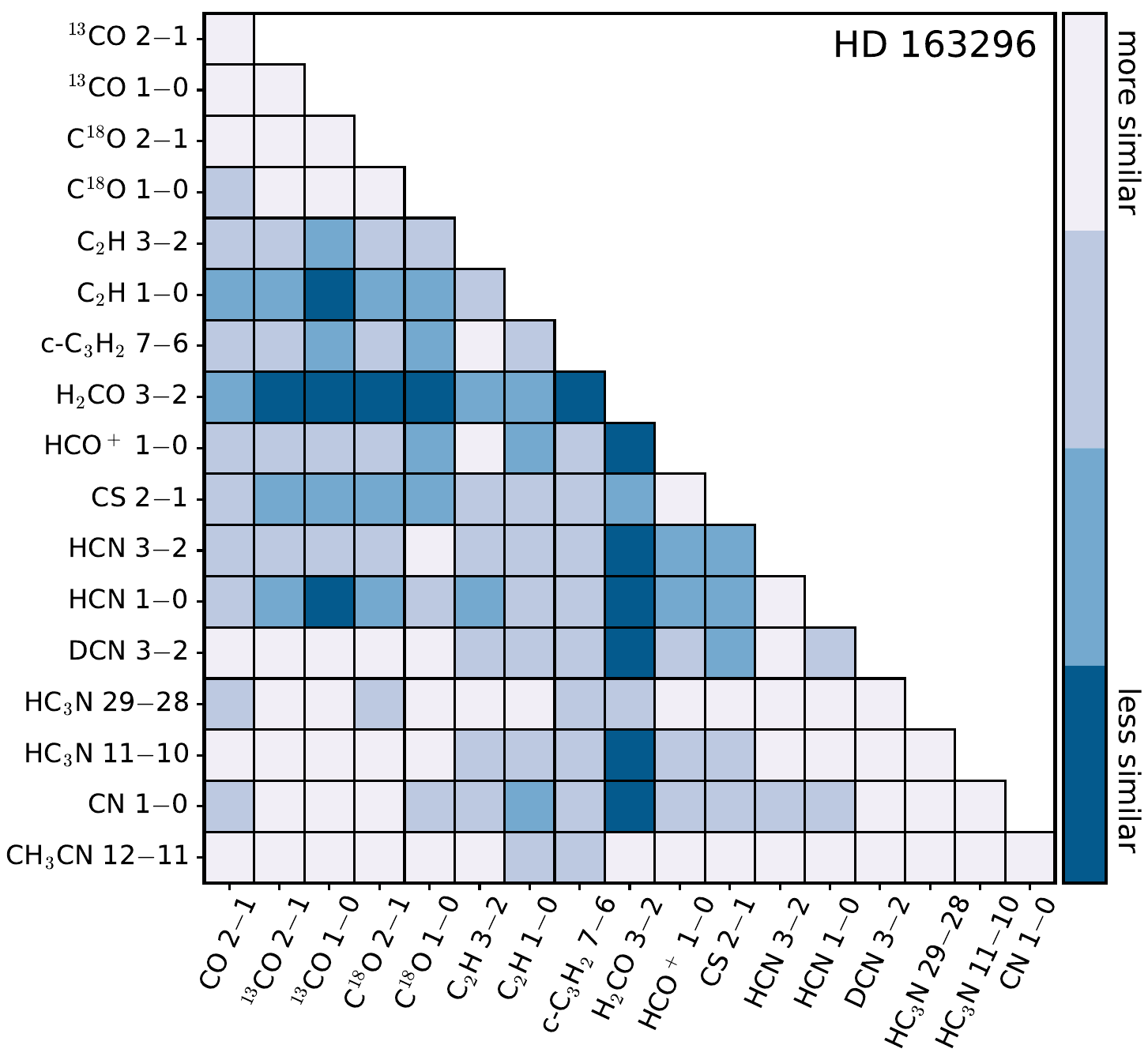}\\
\includegraphics[width=0.47\linewidth]{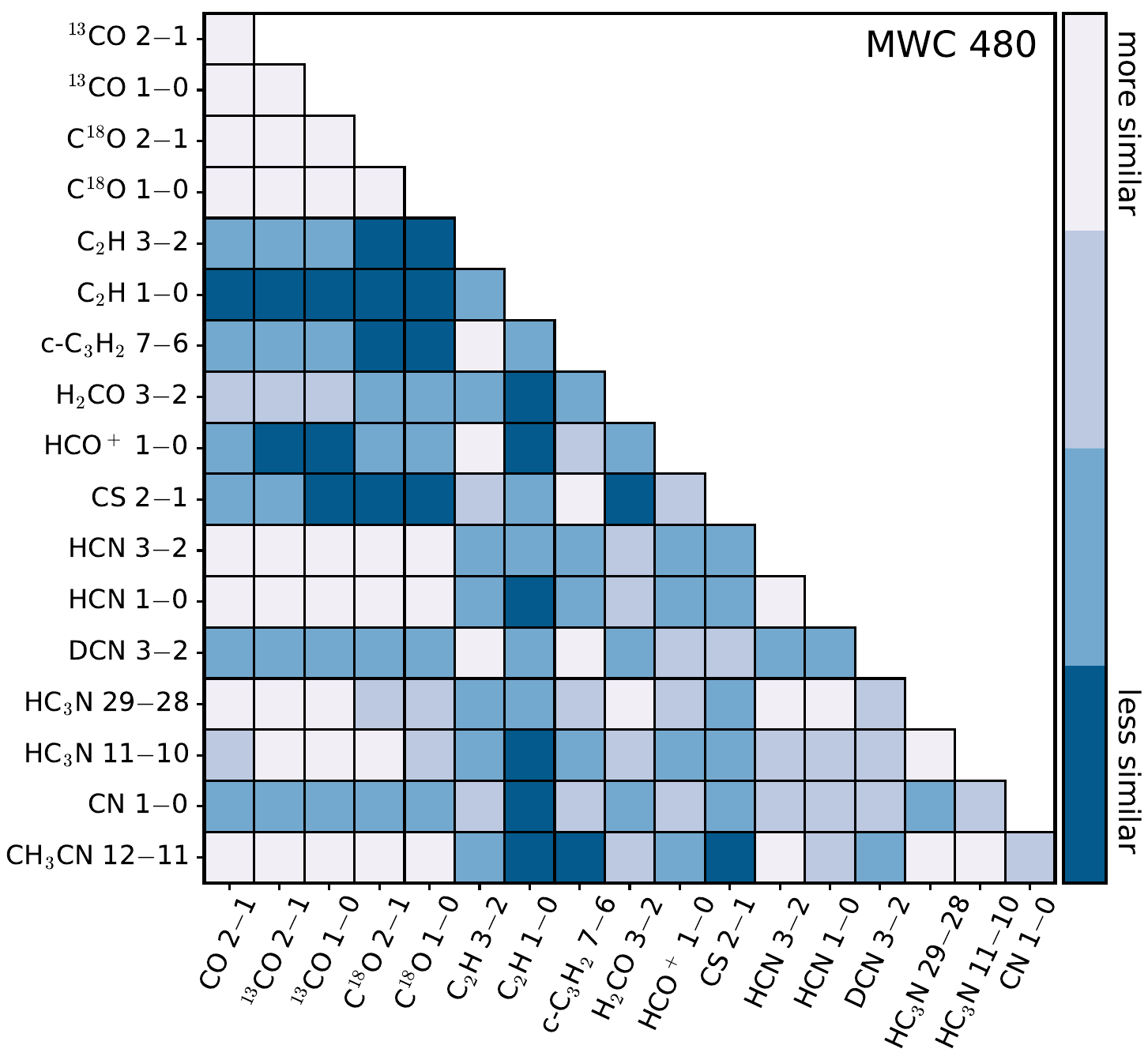} 
\includegraphics[width=0.47\linewidth]{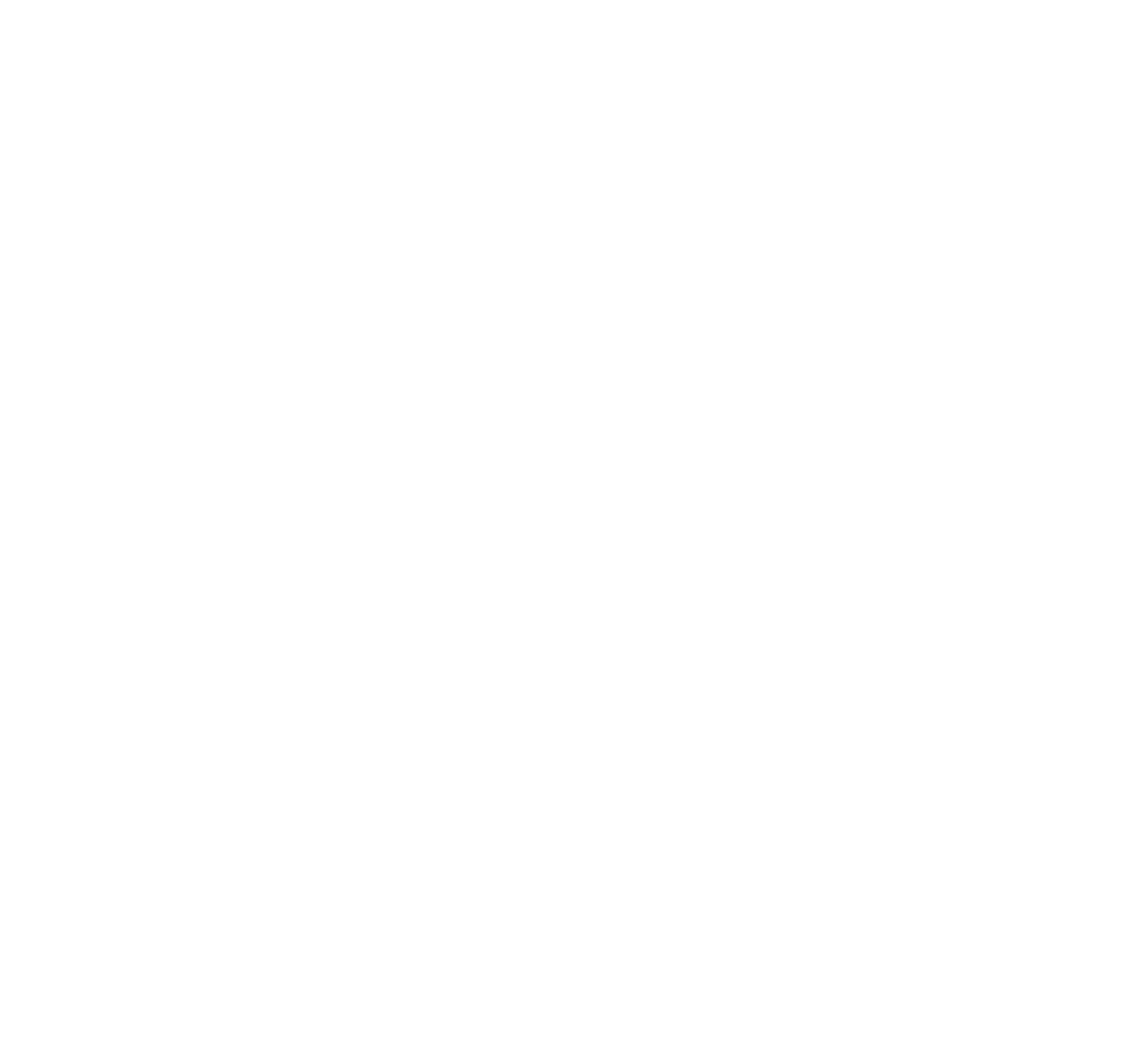}
\caption{Integrated differences between all pairs of lines within each MAPS disk. Darker colors indicate lines with less similar radial morphologies, while lighter colors show those that are more similar. Hatched squares designate tentative and non-detected lines, or those lacking sufficient SNR for a robust comparison.}
\label{fig:intra_disk_heatmap}
\end{figure*}

Figure \ref{fig:intra_disk_heatmap} shows the results for all pairs of lines in each disk. Pairs of lines with profiles that are more dissimilar in their emission morphologies are shown in darker colors, while those that are more similar are shown in lighter colors. To acknowledge the semi-qualitative nature of this comparison we only use four colors, corresponding to four quartiles of similarity. Overall, this method, while not intended to provide a robust statistical measure, allows for a useful ordering based on the similarity of line pairs.

IM~Lup has the highest fraction of line pairs with similar morphologies, with most lines showing a central depression, followed by a wide ring and plateau-like distribution. GM~Aur, AS~209, and HD~163296 also show a relatively large fraction of similar lines. This is due to the consistent emission structures within each disk, namely a central cavity and narrow inner ring (GM~Aur); a single emission ring (AS~209); and multiple co-spatial rings of comparable widths (HD~163296). In contrast, MWC~480 shows the most dissimilar line pairs, which reflects broad differences in radial morphologies between different types of species. For instance, the hydrocarbons are in the form of a single ring with a central gap, while the CO lines and HCN 3--2 and HC$_3$N 29--28 have smoothly-decreasing, centrally-peaked profiles. In several disks, we also find that one or two lines are markedly different from the others: DCN 3--2 in IM~Lup, c-C$_3$H$_2$ 7--6 in GM~Aur, and H$_2$CO 3--2 in HD~163296. These differences reflect the mutually dissimilar emission structures of each line, i.e., double rings (DCN 3--2), a radially-offset emission ring (c-C$_3$H$_2$ 7--6), and a large central gap (H$_2$CO 3--2).

Within each disk, the CO isotopologues are the mutually most consistent, which is not surprising considering all lines originate from the same species, but yet informative since it suggests that excitation and optical depth effects do not dominate differences in radial profiles. Species belonging to the same molecular families, e.g., C$_2$H and c-C$_3$H$_2$; HCN and HC$_3$N, as well as different transitions from a single species, e.g., C$_2$H 3--2, 1--0; HCN 3--2, 1--0, are also typically similar to one another. This is an intuitive result since the radial profiles appear similar, with consistent multi-ring emission structures, e.g., C$_2$H and c-C$_3$H$_2$ in HD~163296 (Figure \ref{fig:C2H_Radial_Profiles}), or emission shoulders that occur at similar radial locations, e.g., double shoulders in HCN 3--2 and HC$_3$N 29--28 in MWC~480 (Figure \ref{fig:HCN_Radial_Profiles}). As for the CO isotopologues, this broad similarity in the radial profiles of similar molecules or different transitions of the same species indicates that excitation effects are not causing substantial differences in their radial morphologies. This is unsurprising as typical differences in upper state energies are only a few 10s of K, with the exception of HC$_3$N 29--28 (E$_{\rm{u}}~\approx$~190~K).

Overall, the correlation patterns between the five disks appear quite complex. A few lines are consistently well-correlated, but many aspects of disk chemistry are disk specific. For instance, the relationship between HCN and H$_2$CO spans from strongly dissimilar in HD~163296 to strongly similar in IM~Lup and GM~Aur with only a modest association in AS~209 and MWC~480, while C$_2$H and CO lines are well-correlated in IM~Lup and GM~Aur, but are dissimilar in AS~209, HD~163296, and MWC~480.

\subsection{Relationship between chemical and continuum substructures}
\label{sec:versus_continuum}

\begin{figure*}[!ht]
\centering
\includegraphics[width=\linewidth]{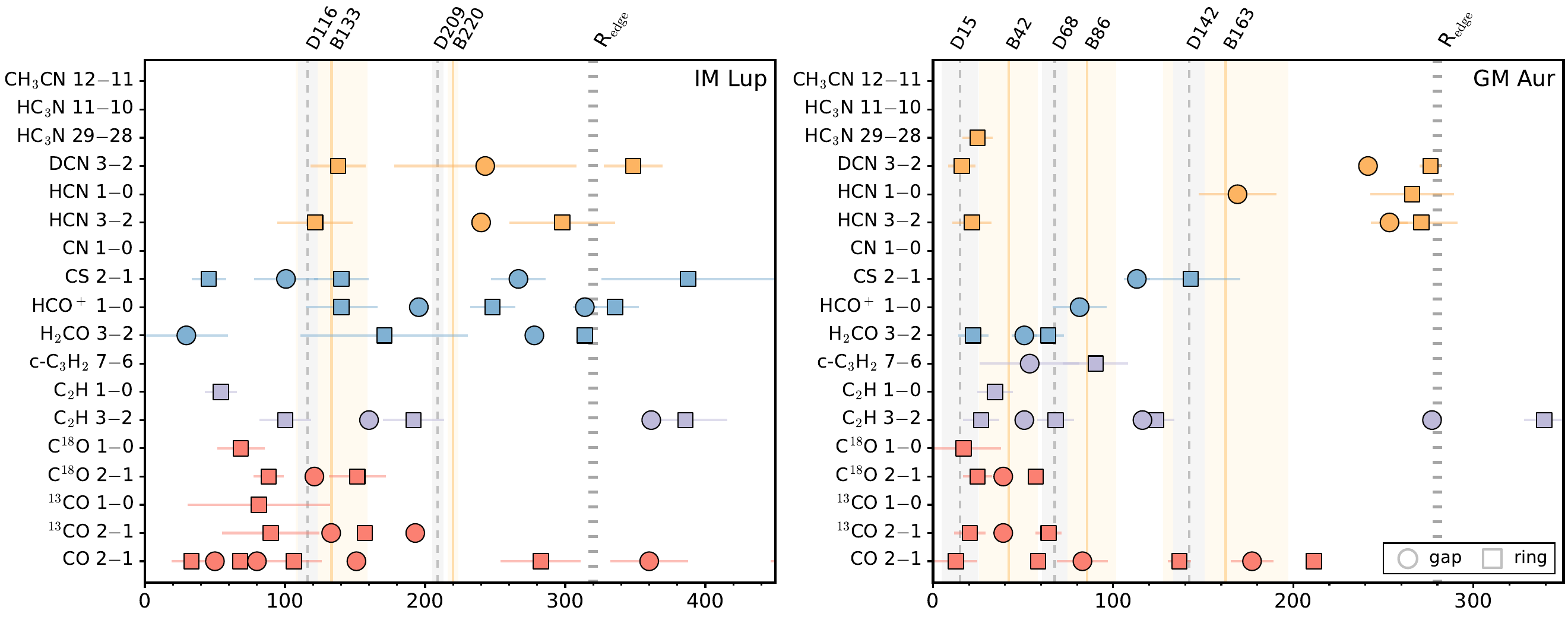}\\
\includegraphics[width=\linewidth]{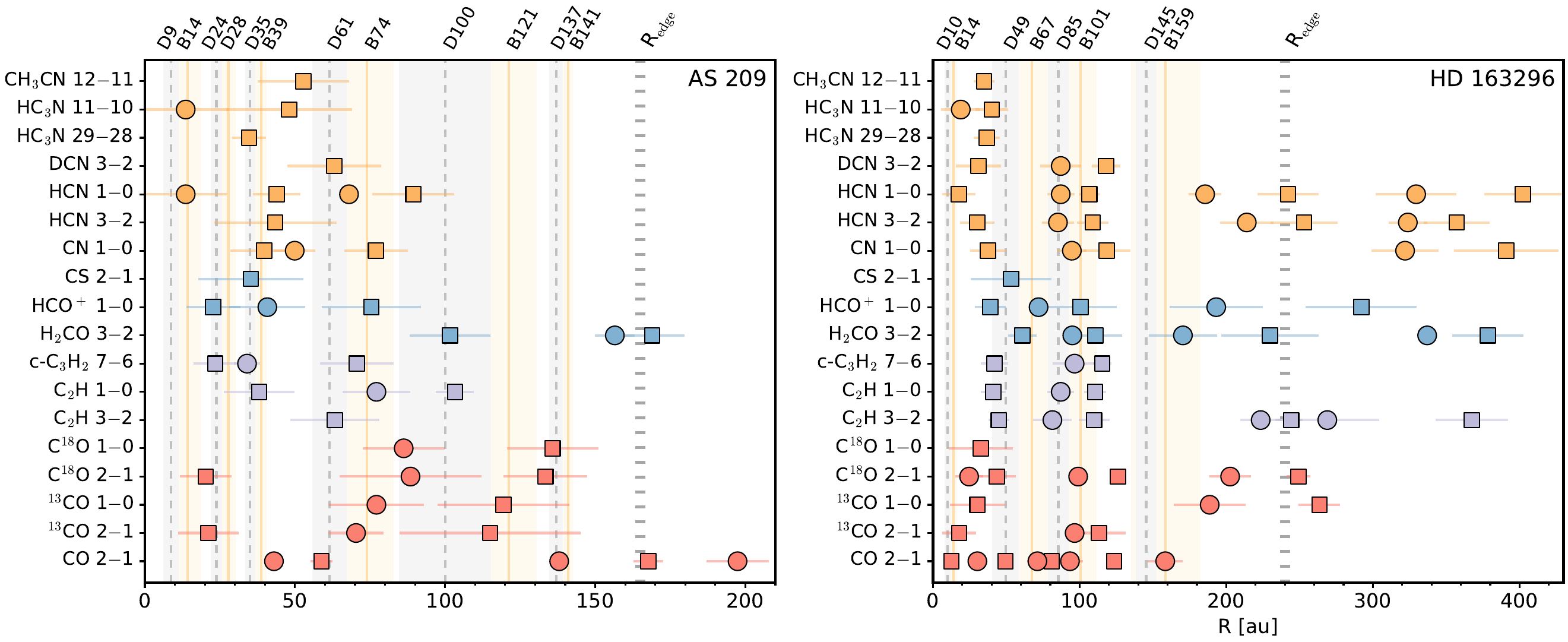}\\
\includegraphics[width=\linewidth]{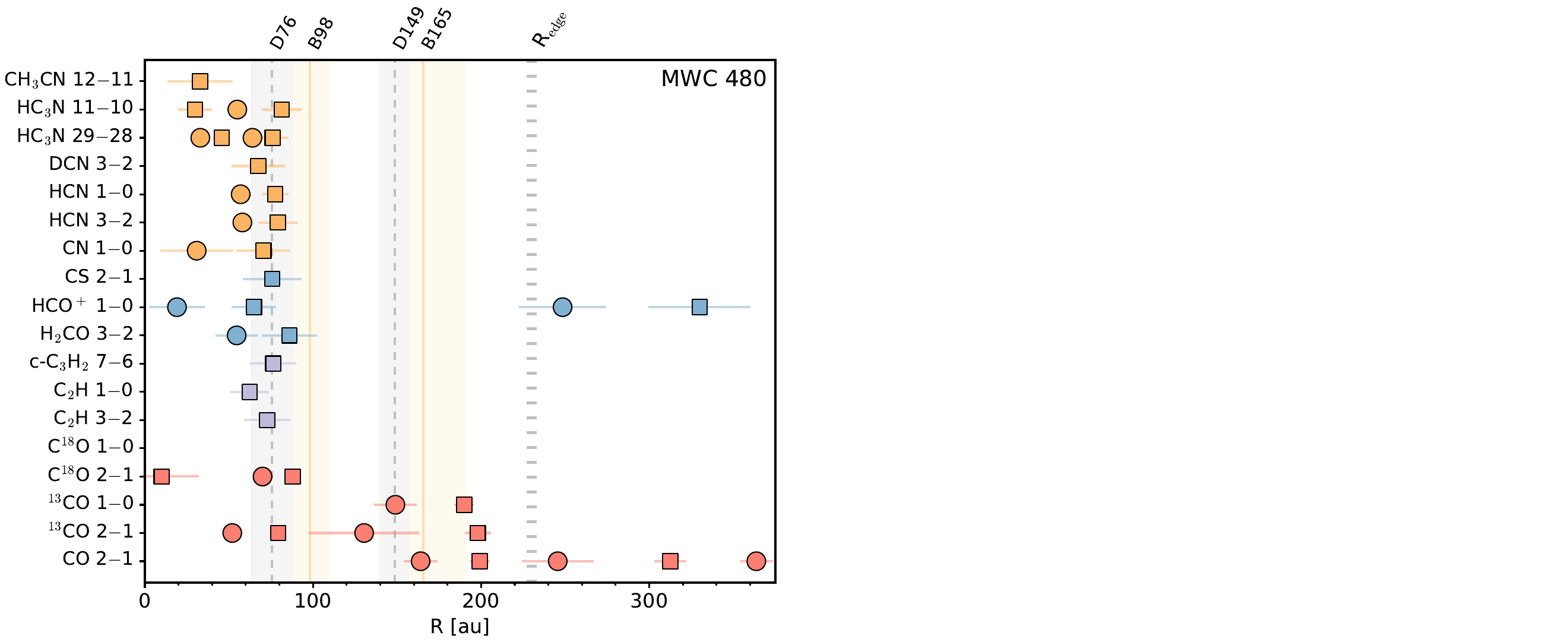}
\caption{Radial locations of chemical and millimeter continuum substructures in the MAPS sample. Line emission rings and gaps are shown as squares and circles, respectively. Species are color-coded, as in Figures \ref{fig:disk_sizes}, \ref{fig:Spearman_disk_size}, and \ref{fig:r0_histogram}. Gray dashed lines mark continuum gaps and orange solid lines denote continuum rings. Shading indicates the widths of continuum substructures. All millimeter continuum features are labeled according to Table \ref{tab:ContinuumSubstrProp}. Broad dotted lines mark the location of the edge of millimeter continuum disk. Chemical substructures at large radii beyond the millimeter continuum, which are only seen in CO 2--1, are omitted. The widths of error bars for chemical substructures represent $\sigma$ instead of the full FWHM, i.e., FWHM / 2.355, for visual clarity.}
\label{fig:Substructure_vs_Continuum_Features}
\end{figure*}

One primary goal of this work is to assess the relationship between continuum and chemical substructures at high spatial resolution in protoplanetary disks. Figure \ref{fig:Substructure_vs_Continuum_Features} shows the radial locations of chemical substructures versus those of annular continuum substructures in the MAPS disks. Although there is no one-to-one correlation between continuum and line emission substructures, several suggestive trends emerge. Below, we first provide a source-by-source description in Subsection \ref{sec:source_specific_trends}. Then, in Subsection \ref{sec:trends_across_disk_and_among_species}, we discuss the spatial links between dust and chemical substructures across the entire MAPS sample as well as comment on the relative frequency of such associations and likely physical origins. In Subsection \ref{sec:outer_edge_of_mm_continuum_disk}, we identify and discuss molecular emission features that are coincident with the outer edge of the millimeter continuum disks.

\subsubsection{Source-specific trends} \label{sec:source_specific_trends}

In IM~Lup, a few line emission rings, e.g., $^{13}$CO 2--1, HCN 3--2, DCN 3--2, HCO$^+$ 1--0, are coincident with the inner continuum ring (B133). However, the majority of chemical substructures are widely distributed in radial locations with no particular association with continuum substructures. A few chemical substructures (e.g., HCN 3--2, DCN 3--2, H$_2$CO 3--2) at larger radii are spatially associated with the outer edge of the continuum disk. 

In GM~Aur, chemical substructures are closely associated with the inner three continuum features (D15, B42, D68). Specifically, we see alternating pairs of ring-gap associations between the continuum and chemical substructures: line emission rings are associated with the D15 dust gap, line emission gaps with the dust gap at B42, and another set of line emission rings at the dust gap at D68. Few chemical substructures are present beyond ${\sim}$120~au with the notable exception of a set of outer emission rings in HCN 3--2, 1--0, and DCN 3--2, each of which are coincident with the outer continuum edge. 

In AS~209, the majority of chemical substructures are spatially coincident with continuum substructures. The inner line emission gaps in HCN 1--0 and HC$_3$N 11--10 are both radially coincident with the B14 dust ring. Line emission peaks in $^{13}$CO 2--1, C$^{18}$O 2--1, c-C$_3$H$_2$ 7--6, HCO$^+$ 1--0 are aligned with the D24 dust gap and CS 2--1, CN 1--0, and HC$_3$N 29--28 rings aligned with the D35 dust gap. Similarly, several emission rings (c-C$_3$H$_2$, HCO$^+$, CN) are located near the dust ring at B74. However, not all line substructures are aligned with continuum substructures. Several nitrile rings (HCN 3--2, 1--0, CH$_3$CN 12--11, HC$_3$N 11-10) fall between the dust ring at B39 and dust gap at D61. Interesting, there are relatively few chemical substructures within the broad D100 dust gap. No chemical substructures, with the exception of CO 2--1, are located beyond the continuum edge. 

In HD~163296, nearly all chemical substructures show some spatial association with those of the continuum within ${\sim}$120~au, while few features are associated with the outer dust ring-gap pair (D145-B159). Chemical rings and gaps within 120~au are also radially coincident with one another. This consistency in radial location is particularly striking for those substructures associated with the hydrocarbons and nitriles. Line emission rings are coincident with the dust gap at D49, while gaps in line emission are coincident with the dust gap at D85. Another set of chemical rings aligns with the B101 dust ring. Over 20\% --- the highest fraction in the MAPS disks --- of chemical substructures are located at or beyond the edge of the millimeter continuum disk.

In MWC~480, the majority of chemical substructures are radially coincident with the D76 dust gap with line emission rings showing the closest spatial associations. However, in some cases, e.g., HC$_3$N 29--28, line emission rings and gaps both overlap within the width of the D76 dust gap. Besides CO isotopologues and HCO$^+$, no other chemical substructures are seen outside of the inner D76-B98 continuum feature. The numerous CO substructures at large radii generally do not show any trends with continuum features with the exception of CO 2--1, $^{13}$CO 2--1, 1--0 rings around 200~au, which is at the outer edge of the B165 dust ring.

\subsubsection{Spatial links between chemical and dust substructures across MAPS disks} \label{sec:trends_across_disk_and_among_species}

While, in detail, each disk displays a different relationship between chemical and continuum substructures, several broader trends emerge. Most notably, the majority of chemical substructures across the MAPS disks show some degree of spatial association with continuum substructures for radii less than ${\sim}$100-150~au. Beyond these radii, the fraction of chemical substructures that can be linked to dust substructures is quite small. Chemical substructures in some lines are also present at or near the outer continuum edge in several disks, which is discussed in detail in the following subsection.

\begin{figure}[!ht]
\centering
\includegraphics[width=\linewidth]{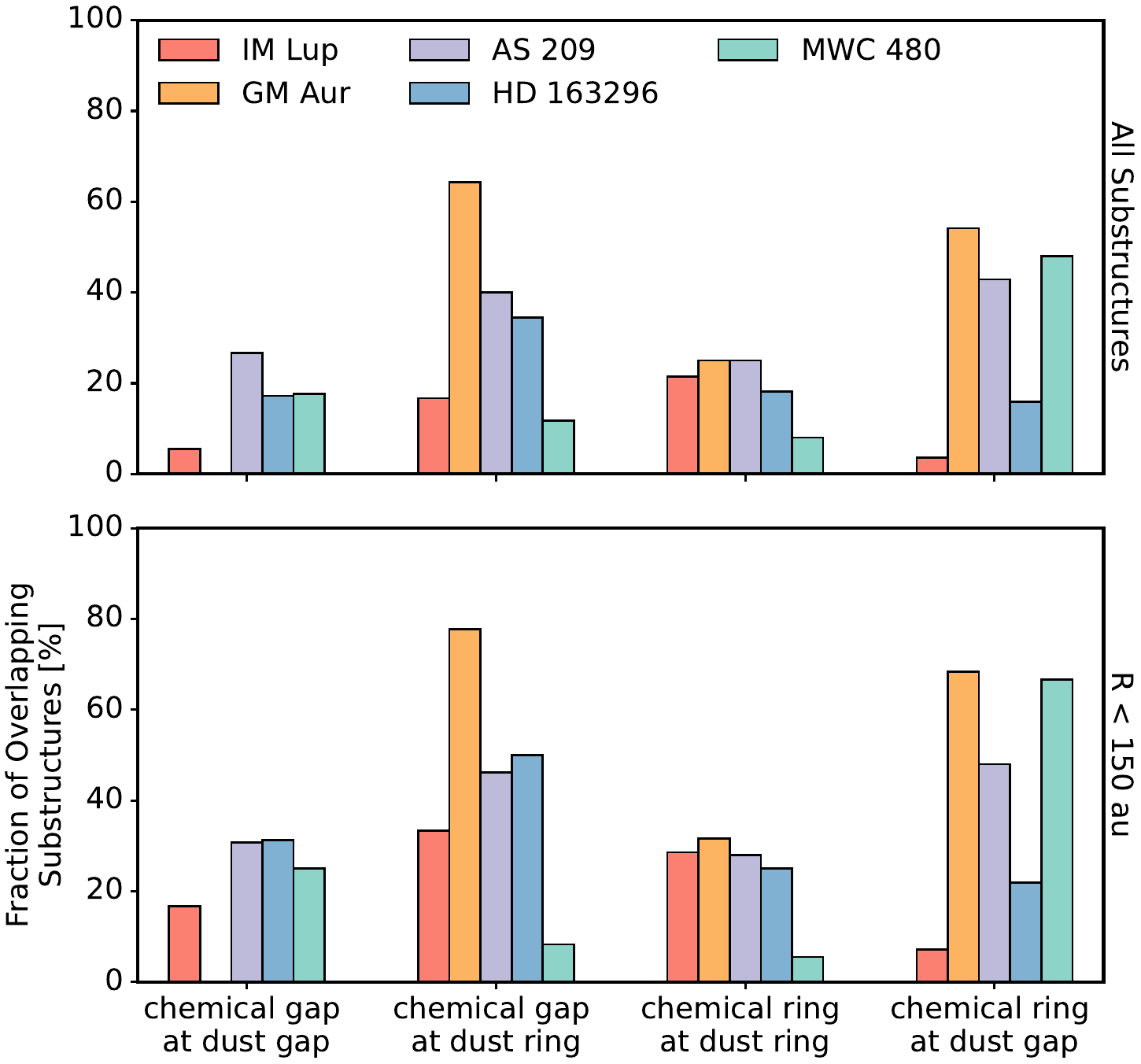}
\caption{Fraction of chemical and millimeter continuum substructures that spatially overlap for all chemical structures (top) and those with radial locations ${<}$150~au (bottom). Overlapping features are those where the radial location of a line emission substructure falls within the width of a continuum feature.}
\label{fig:Gaps_Rings_Stats}
\end{figure}

To quantify how likely continuum and chemical substructures are to spatially correlate with one another, we calculated the relative occurrence rate of overlapping features. We considered features to be overlapping if the radial position of the chemical substructure (listed in Table \ref{tab:SubstrProp}) falls within the width of the continuum substructure. As the outer continuum ring-gap pair (D209-B220) in IM~Lup was visually-identified, we adopt a conservative width of ${\sim}$10~au for both of these features. Figure \ref{fig:Gaps_Rings_Stats} shows the spatial overlap fractions for all four possible pairs of substructure alignments. The top panel shows the overlap fractions for all chemical substructures. The highest fractions (up to ${\sim}$65\%) are between line emission rings-dust gaps and chemical gaps-dust rings, followed by chemical rings-dust rings (${\sim}$10-25\%).  Chemical gap-dust gap alignments are consistently the least common ($\lesssim$25\%). The disk-to-disk variation is high for gap-ring alignments, while ring-ring and gap-gap alignment frequencies are almost constant among the disks.GM~Aur shows the highest fractions of overlapping features, which is, in part, due to most species showing emission peaks in its central dust cavity. High overlap fractions are also seen in the MWC~480 disk due to the spatial association of many of its chemical substructures with the continuum gap at D76. 

The bottom panel of Figure \ref{fig:Gaps_Rings_Stats} only considers those chemical substructures with radial locations less than 150~au, which yields 10--20\% higher overlap fractions for nearly all disks and substructure alignments. For instance, all disks have over one-third of either (or both) chemical rings and dust gaps or chemical gaps and dust rings aligned in the inner 150~au. GM~Aur and MWC~480 chemical ring-dust gap alignments are ${\sim}$70\%. Thus, in general, chemical and dust substructures are closely associated in the inner 150~au of disks.

Dust gaps may cause chemical gaps if they are associated with gas depletion. Similarly, dust rings may result in chemical rings if they are associated with gas enhancements. Dust gaps and rings may also give rise to either chemical gaps or rings due to changes in radiation, ionization, gas-phase elemental abundances, and temperature, since different species are expected to be more rapidly formed or destroyed as these properties increase or decrease \citep[e.g.,][]{Facchini2018, Alarcon2020, Rab20}. However, many associations between line emission and continuum features are independent of gas-dust substructure correlations, i.e., only some dust gaps are obviously also depleted in gas \citep{zhang20}. For these cases, some process is needed to link the midplane and elevated disk layers, since millimeter dust grains emit from near the midplane \citep[e.g.,][]{Villenave20} and line emission from vertically flared surfaces \citep[e.g,][]{Podio20,Teague20_CN,van_Hoff20_warm, law20a}.

Such links may be due to vertical mixing \citep{Semenov11, Flock17, vanderMarel2021} or flows of molecular material from the disk surface to the midplane at the radial locations of dust gaps \citep{Teague_19Natur}. Theoretically, these links should be easier to establish in the inner disk regions, since most line emission heights are expected to increase with radius due to disk flaring. Moreover, dust scale heights may also be locally-enhanced, i.e., comparable to that of the gas \citep{Doi21}, in the inner disk, which would place the line and dust emitting regions in closer contact. This provides a natural explanation for the frequent dust and line emission associations seen within 150~au, but the relatively few at larger radii, at which point the increasingly flared surfaces become disconnected from disk midplanes. We would also expect a closer association between dust and those molecules, that for chemical or excitation reasons, emit closer to the midplane. In general, disks are expected to be highly stratified with different lines and species originating in different vertical layers \citep{Dartois03}, and thus different lines may become disconnected from the midplane dust at different radii depending on their particular emission heights.

In the case of chemical gap-dust ring association there is also another possible explanation: the absorption of line emission by dust. Continuum subtraction of the dust emission may result in gaps in molecular line emission in regions where the line emission is optically thick and absorbs most of the dust emission coming from the midplane \citep[e.g.,][]{Boehler17, Weaver18}. While this may be responsible for some of the observed spatial links, especially in the inner 50~au, the lack of consistent associations between line emission and dust substructure suggests that this is not a dominant effect. For instance, in HD~163296, one set of line emission rings aligns with the D49 dust gap, while another group of chemical rings is co-located with the B101 dust ring.

While gas and millimeter dust emit from distinct disk layers, gas emitting surfaces and dust scattered light features are, in some cases, vertically co-located \citep[see][]{law20a}. As IM~Lup, AS~209, and HD~163296 also have well-defined rings in scattered light \citep{Monnier17, Muro_Arena18, Avenhaus18, Rich20}, we searched for spatial associations between chemical substructures and these NIR rings. We found no strong links with the following two exceptions. The outermost set of gaps in CN 1--0, HCN 3--2, 1--0 in HD~163296 are approximately aligned with the NIR ring at 330~au \citep{Rich20}. In IM~Lup, the NIR ring at 240~au \citep{Avenhaus18} is co-located with a gap in HCN 3--2 and the center of the large gap between the double DCN rings. These spatial associations are intriguing, as NIR wavelengths probe micron-sized grains in elevated disk layers that help regulate UV flux, which is an important parameter for the formation of CN and HCN. A more detailed discussion of NIR features and these two molecules in the MAPS disks is found in \citet{bergner20}. Figure \ref{fig:NIR_ring-vs-chemsubstr} in Appendix \ref{sec:app:NIR_rings_appendix} shows the full comparison  between chemical substructures and NIR rings in these three disks.

\subsubsection{Outer edge of millimeter continuum disk} \label{sec:outer_edge_of_mm_continuum_disk}

Line emission features are often spatially associated with the edge of the millimeter continuum in disks \citep[e.g.,][]{Oberg15, Bergin16}. In particular, associations between line emission rings and continuum edges have been previously observed in the MAPS disks, e.g., for DCO$^+$ in IM~Lup \citep{Oberg15, Huang17} and HD~163296 \citep{Salinas17, Flaherty17}; H$^{13}$CO$^+$ \citep{Huang17} and H$_2$CO \citep{Carney17} in HD~163296, as well as for $^{13}$CO \citep{Schwarz16} and C$_2$H \citep{Bergin16} in TW~Hya; HCN \citep{Guzman15} and C$_2$H \citep{Bergin16} in DM~Tau; and DCN in LkCa~15 \citep{Huang17}. The MAPS data also show that an N$_2$D$^+$ 3--2 emission ring in IM~Lup, AS~209, and HD~163296 is associated with the outer continuum edge \citep{cataldi20}. Models explain these spatial links in the context of dust evolution leading to non-thermal desorption \citep{Oberg15}, a thermal inversion in the outer disk \citep{Cleeves16, Facchini17}, or higher UV penetration at this dust edge \citep{Bergin16}. To explore links between chemical substructure and disk edges,  we first visually estimate the outer edge of the millimeter continuum R$_{\rm{edge}}$ for each disk from the radial profiles in Figure \ref{fig:CO_wcont_radial_profiles}. The determined R$_{\rm{edge}}$ values are listed in Table \ref{tab:ContinuumSubstrProp} and are shown as broad dotted lines in Figure \ref{fig:Substructure_vs_Continuum_Features}.

We observe coincidences between R$_{\rm{edge}}$ and line emission rings in all MAPS sources except MWC~480. Rings from HCN and DCN in GM~Aur, IM~Lup, and HD~163296 are spatially correlated with R$_{\rm{edge}}$, as are H$_2$CO rings in IM~Lup, AS~209, and HD~163296. In fact, when combined with the previous survey of \citet{Pegues20}, these results indicate that at least 50\% of disks may show spatial associations between H$_2$CO rings and continuum edges \citep[see][for further details]{guzman20}. Moreover, this also suggests that similar fractions of HCN and DCN rings may be spatially linked to R$_{\rm{edge}}$, but this requires a large disk survey at high spatial resolution and sensitivity to confirm. 

HCN, DCN, and H$_2$CO are not directly chemically linked, and their joint appearance at the edge of the pebble disk suggests that some or perhaps all of the proposed chemical effects listed above are active at different levels in the different disks. HCN and DCN are expected to form through gas-phase chemistry, and the origins of the HCN and DCN line emission rings are likely due to higher UV penetration at the dust edge increasing the atomic carbon abundance \citep{Alarcon2020}. One potential caveat of this explanation is that we never observe a corresponding association between a ring in the photochemically sensitive CN molecule and R$_{\rm{edge}}$, but this may also be explained by CN emission originating from elevated disk layers \citep{Cazzoletti18, Teague20_CN, bergner20}. H$_2$CO may form in the gas-phase or through grain-surface chemistry via CO ice hydrogenation \citep[e.g.,][]{Loomis15}. At the millimeter dust edge, H$_2$CO could arise from either non-thermal desorption of H$_2$CO ice, or by gas-phase formation following thermal or non-thermal CO desorption, or by gas-phase formation fuelled by photo-produced atomic carbon \citep{Qi13_HD16, Oberg17, Pegues20, vanScheltinga21}. In disks where the HCN and H$_2$CO rings coincide, the latter seems the most likely explanation. By contrast, in the IM~Lup disk, the proximity of HCO$^+$, DCO$^+$, and DCN rings to R$_{\rm{edge}}$, indicates that in this case the edge of the pebble disk results in cold CO-driven chemistry, as well as an increased ionization rate, and an increased gas-phase deuteration chemistry. 

\begin{figure*}[!ht]
\centering
\includegraphics[width=\linewidth]{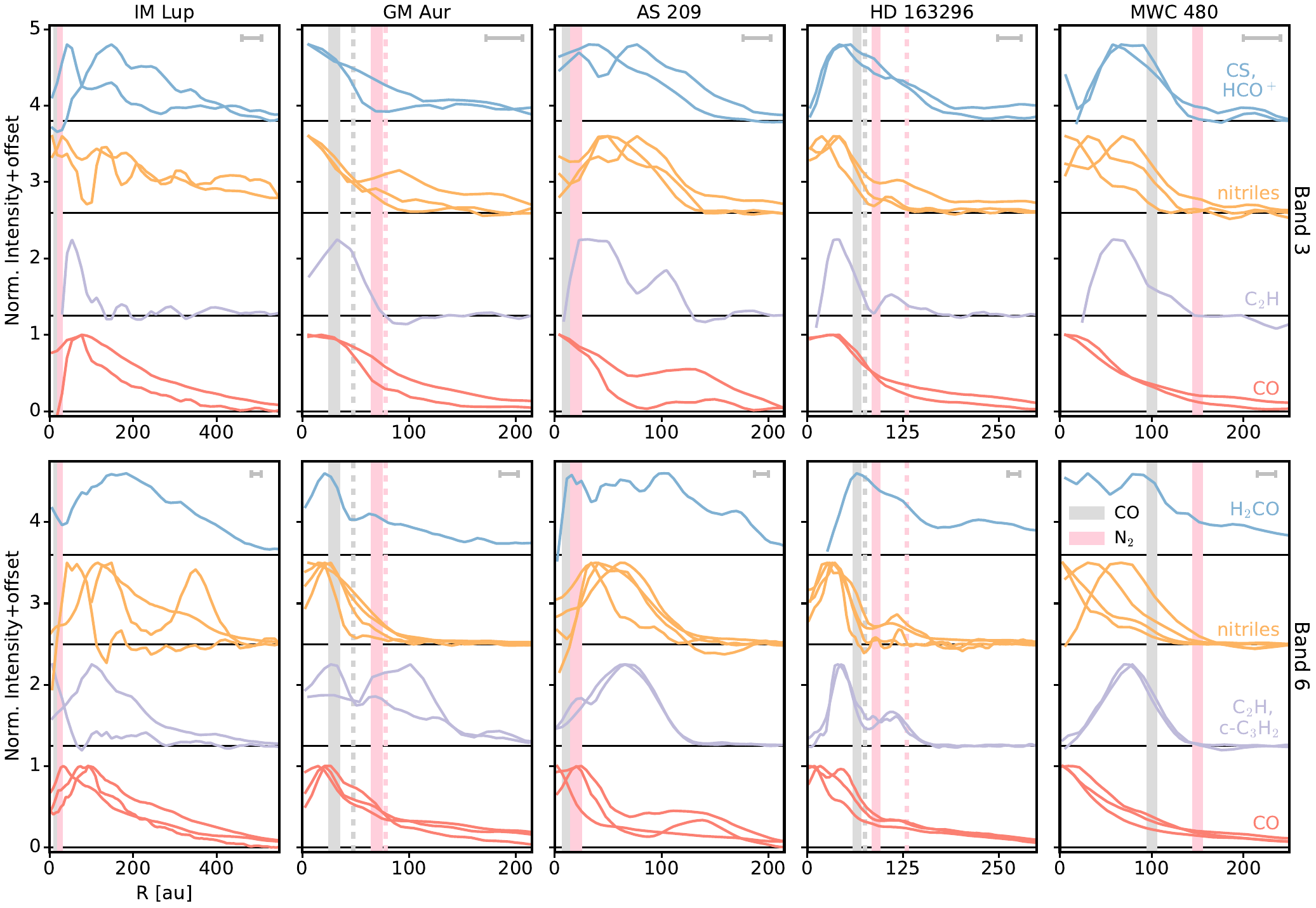}
\caption{Normalized radial intensity profiles for lines observed in Band 3 (top) and Band 6 (bottom). In the case of Band 6, most lines are at 0\farcs15, but some are tapered to 0\farcs30 (see Table \ref{tab:ProfSelection}). Each panel is subdivided by species and vertically offset for visual clarity. Species are color-coded, as in Figures \ref{fig:disk_sizes}, \ref{fig:Spearman_disk_size}, and \ref{fig:r0_histogram}. Shading shows the radial range of the CO (gray) and N$_2$ (pink) midplane snowlines derived with thermochemical models \citep{zhang20}, while the dashed lines indicate those determined from observations of N$_2$H$^+$ \citep{Qi15, Qi19}. Only the central radius of the N$_2$H$^+$-derived snowlines are shown and uncertainties are omitted for visual clarity but are typically on the order of 5--10~au. The FWHM of the synthesized beam is shown by a horizontal bar in the upper right corner of each panel.}
\label{fig:norm_all_rad_prof}
\end{figure*}

In summary, given the diversity in spatial associations with the outer disk edge across the MAPS disks, no single explanation can account for these trends. Instead, some combination of different processes must be at work and their relative importance appears to depend on the specific physical conditions of each disk.

\subsection{Relationship between chemical substructures and snowlines}
\label{sec:versus_temperature}

The condensation of key volatiles, such as CO$_2$, CO, and N$_2$, has been suggested as one possible origin for annular continuum substructures due to changes in the fragmentation and coagulation properties of dust grains at the location of molecular snowlines \citep[e.g.,][]{Zhang15, Okuzumi16}. The freezeout of different volatiles will also lead to changes in elemental and molecular composition in disks that may generate, or facilitate the growth of, chemical substructures at or near the location of particular snowlines. To explore whether the chemical substructures observed in MAPS are connected to molecular snowlines, we first show the normalized radial profiles of all lines considered in this study in Figure \ref{fig:norm_all_rad_prof}. We then shade the radial locations of the midplane snowlines of CO (gray) and N$_2$ (pink) derived from the thermochemical models of \citet{zhang20}. Snowlines determined using observations of N$_{2}$H$^{+}$ from \citet{Qi15, Qi19} are indicated as dashed lines but are only available for GM~Aur (CO and N$_2$) and HD~163296 (CO). The MAPS observations of N$_2$H$^+$ 3--2 and N$_2$D$^+$ 3--2 \citep[e.g.,][]{aikawa20, cataldi20} suggest that the N$_2$ snowline in HD~163296 is at approximately 130~au (Qi, private comm.). In all MAPS disks, the CO$_2$ snowline occurs at a radius of ${<}$10~au from the central star \citep{zhang20} and is thus always unresolved.

No strong spatial association between the locations of chemical substructures and snowlines is evident in Figure \ref{fig:norm_all_rad_prof}. Moreover, there also are no obvious trends in the relative locations of chemical substructures and snowlines. For instance, nearly all line emission substructures in IM~Lup and AS~209 are at radii exterior to the CO and N$_2$ snowlines, while the majority of features in MWC~480 are located interior to both snowlines. GM~Aur and HD~163296 show a more widely spread distribution of substructures that occur across snowlines. 

A few tentative associations are seen within individual disks, such as the alignment of the CO snowline with the edge of the central depression in IM~Lup or several line emission rings (C$_2$H 3--2, H$_2$CO 3--2) that are radially coincident with the N$_2$ snowline in GM~Aur. However, the relative uncertainty in snowline locations often makes discerning precise spatial links difficult. For example, in GM~Aur, the CO snowline from the models of \citet{zhang20} is at 30~au, which is radially coincident with line emission rings (CO, HCN, C$_2$H, H$_2$CO), but the CO snowline predicted from N$_2$H$^+$ \citep{Qi15} is at 48~au which instead is spatially co-located with several line emission gaps (H$_2$CO, C$_2$H, c-C$_3$H$_2$). A similar discrepancy is seen in HD~163296 with the N$_2$ snowline from \citet{zhang20} aligning with chemical gaps, while the snowline based on N$_2$H$^+$ aligns, or is slightly exterior to, several chemical rings. More empirical data on the snowline locations of CO and N$_2$ are needed to derive reliable statistics on links between snowlines and the locations of either chemical or dust substructures. However, even for the cases where snowline estimates based on N$_2$H$^+$ exist, at most a small fraction of chemical substructures are spatially coincident and therefore possibly caused by snowlines.

\begin{figure*}[!ht]
\centering
\includegraphics[width=0.95\linewidth]{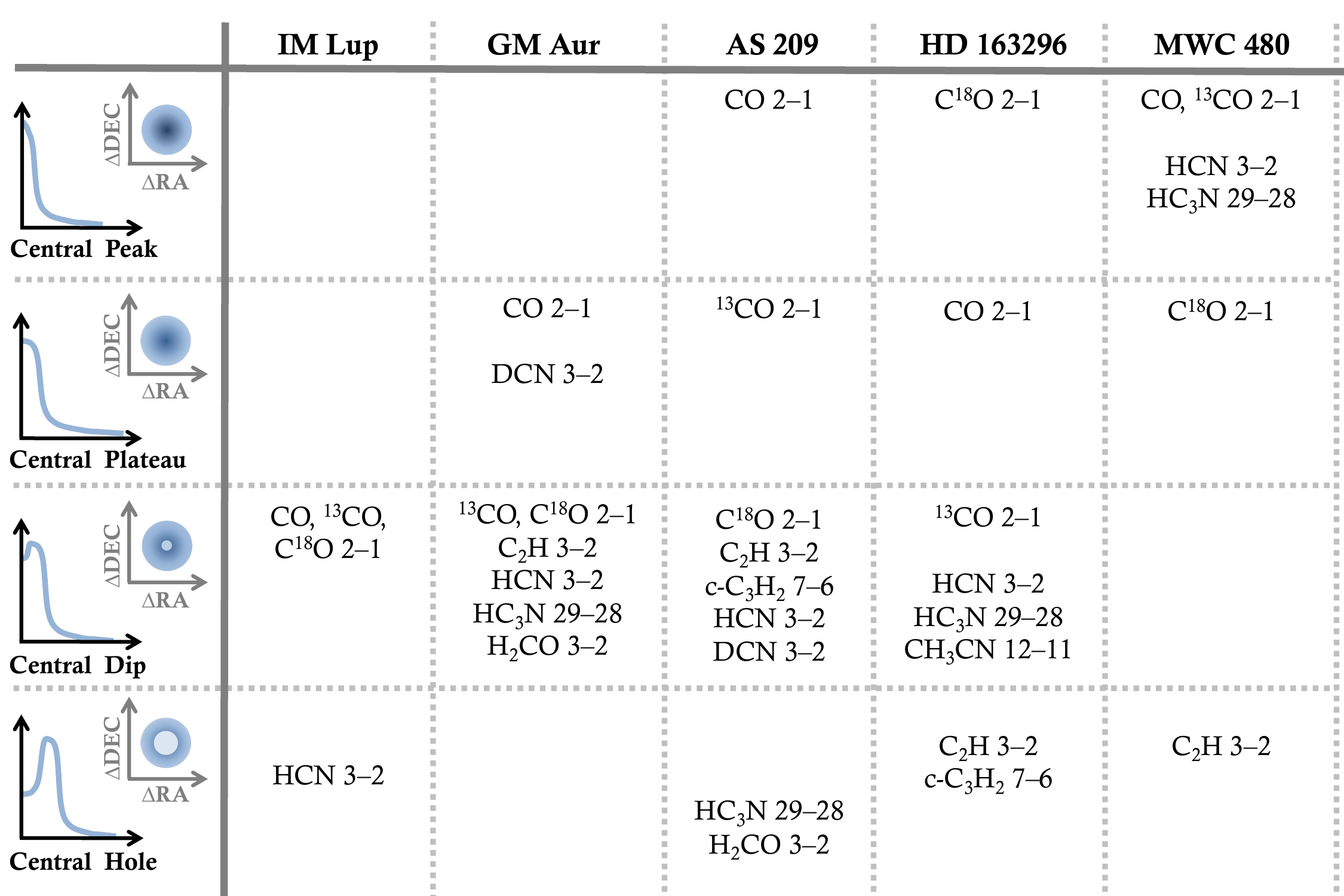}
\caption{Grid of radial morphologies in the inner 50~au of the MAPS disks. The rows classify lines by the shape of the central emission. Classifications are restricted to those lines observed in Band 6 and are based on the normalized intensity at the innermost radial bin in the radial profiles according to: I$_{\rm{norm}}{<}$0.2 (hole), 0.2$<$I$_{\rm{norm}}<$0.8 (dip), I$_{\rm{norm}}{>}$0.8 (peak or plateau). The difference between a peak and plateau was determined visually. Grid adopted and lightly modified from \citet{Pegues20}.}
\label{fig:inner_disk_behavior}
\end{figure*}

Thus, snowlines do not offer a universal explanation for the observed chemical substructures. However, this does not rule out that 2D snow-surfaces play a more important role in regulating disk chemical structures. Emission from molecular lines in disks often originates from an elevated emitting surface. A potentially informative, but intensive comparison would involve a 2D disk-specific model for each snow surface of interest and a diverse set of molecules with well-constrained emission surfaces.

\subsection{Line emission at ${<}$50~au}
\label{sec:inner_disk}

The high angular resolution of the MAPS observations provides access to the behavior of gas in the inner 50~au, which is directly relevant for the formation of planets, such as the giant planets in our own solar system. Here, we explore trends in line emission within this inner region, while \citet{bosman20_inner_au} provides a more detailed and quantitative examination of the CO lines within the innermost 20~au. We only consider those lines covered in Band 6, as they possess the highest angular resolutions (0\farcs15). We omitted those lines, although covered in Band 6, that required the use of the tapered images to achieve sufficient SNR (see Table \ref{tab:ProfSelection}). In Figure \ref{fig:inner_disk_behavior}, we adopted the qualitative grid employed by \citet{Pegues20} and classified radial morphologies by normalized intensities at the innermost radial bins of the radial profiles according to: I$_{\rm{norm}}{<}$0.2 (hole), 0.2$<$I$_{\rm{norm}}<$0.8 (dip), I$_{\rm{norm}}{>}$0.8 (peak or plateau). The difference between a peak and plateau was determined visually.

\begin{figure*}[!ht]
\centering
\includegraphics[width=\linewidth]{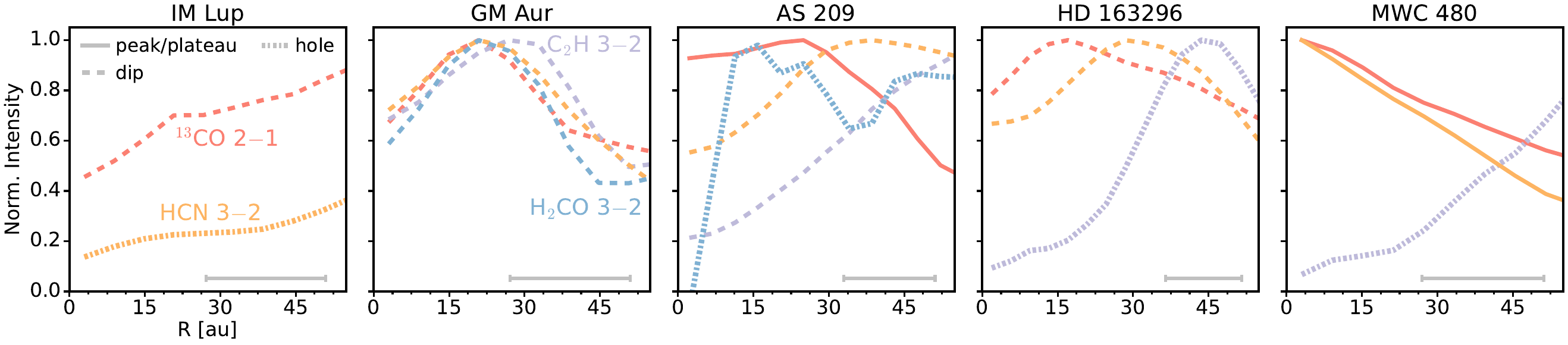}
\caption{Normalized radial intensity profiles of select lines covered in Band 6 at 0\farcs15 in the inner 50~au. Species are color-coded, as in Figures \ref{fig:disk_sizes}, \ref{fig:Spearman_disk_size}, \ref{fig:r0_histogram}, \ref{fig:norm_all_rad_prof}. Classifications, as in Figure \ref{fig:inner_disk_behavior}, of central peaks/plateaus, dips, and holes are shown as solid, dashed, and dotted lines, respectively. The FWHM of the synthesized beam is shown by a horizontal bar in the lower right corner of each panel.}
\label{fig:inner_plots}
\end{figure*}

Figure \ref{fig:inner_disk_behavior} shows that there is a wide range of radial morphologies within 50~au in different disks and lines. This is further illustrated in Figure \ref{fig:inner_plots}, which shows normalized radial profiles for representative lines in each disk. For instance, the majority of lines in GM~Aur show a central dip and are remarkably consistent in their central radial behavior, while AS~209 and HD~163296 have lines that exhibit all possible central behaviors from dips to peaks. MWC~480 has the highest fraction of lines showing a central peak. IM~Lup consistently shows a central dip or hole, but as only HCN 3--2 and the 2--1 transitions of the CO isotopologues had sufficient SNR at 0\farcs15, conclusions are necessarily limited. We note, however, that the majority of lines at 0\farcs30 in IM~Lup also show evidence of a central depression and as this central deficit has a large radial extent ($\gtrsim$100~au), it is resolved even at 0\farcs30 for many lines, e.g., C$_2$H 3--2, HCO$^+$ 1--0.

Figure \ref{fig:inner_plots} shows that if we survey a single disk in different lines, we often see a diversity of line emission profiles within the central 50~au. In MWC~480, for example, C$_2$H 3--2 steadily decreases in the form of a central hole, while both CO 2--1 and HCN 3--2 are centrally peaked. Similar species-specific differences are also present in AS~209 and HD~163296. If we survey all of the disks in a single line, we also see a diversity in radial morphologies. For instance, HCN 3--2 spans the range of possible central behaviors with a central peak in MWC~480, central dip in GM~Aur, AS~209, and HD~163296, and central hole in IM~Lup. Thus, the chemistry appears quite different between disks, which implies both that planets may assemble in chemically distinct environments when forming at the same radius around different stars, and that common molecular probes may trace different regions in different disks. 

The most common type of central behavior across lines and disks is that of a central dip. One reason for this may be the dust optical depth. As the dust becomes modestly optically thick interior to ${\sim}$30~au in IM~Lup, HD~163296, and MWC~480, and possibly in AS~209, \citep{sierra20}, we expect a reduction in intensity as the back side of the disk is increasingly hidden. However, this cannot fully explain the observed diversity in radial morphologies, as dust optical depth would block the back side of the disk leading to only a factor of two reduction (unless the scale height of the millimeter grains is higher as suggested in IM~Lup by \citet{bosman20_inner_au}). Instead, these features are likely due to a combination of differences in abundance, excitation, temperature, and chemistry in the inner regions of the MAPS disks.

The only non-CO molecules to show either central peaks or plateaus in their radial morphologies are nitriles. In particular, MWC~480 shows sharply-rising HCN and HC$_3$N profiles, while DCN 3--2 is in the form of a central plateau in GM~Aur. Nitriles may be important players in origins of life chemistry \citep{Powner09, Sutherland16} and it is interesting that, at least in some cases, nitriles peak toward the innermost regions of planet-forming disks. Indications that HCN emits from a surface relatively close to the disk midplane \citep{law20a} further suggest a potential connection to planet-forming material.

\section{Summary} \label{sec:conclusions}

We present a systematic analysis of radial substructures in a set of 18 molecular emission lines toward five protoplanetary disks observed at high angular resolution as part of the MAPS Large Program. We conclude the following:

\begin{enumerate}
  \item Over 200 unique radial chemical substructures, including rings, gaps, and emission plateaus, are identified. Substructures occur at nearly all radii where line emission is detected with widths ranging from ${<}$10~au to over 200~au. 
  \item All lines are azimuthally symmetric, at least to first order, with the exception of the spiral arms seen in CO 2--1 in GM~Aur. This is true even for those disks that have millimeter continua with known non-axisymmetric features (HD~163296) or spiral arms (IM~Lup).
  \item The large radial extent of CO and its inorganic and organic derivatives compared to O-poor molecules, such as the hydrocarbons C$_2$H, c-C$_3$H$_2$ and nitriles HC$_3$N, CH$_3$CN, indicates that the inner planet-forming 100~au of disks may be more C-rich than disk-averaged line emission suggests. Additionally, the complex nitriles and c-C$_3$H$_2$ are spatially compact and do not extend beyond the edge of the continuum disk. At radii beyond the pebble disk, these molecules may be more easily destroyed, e.g., due to radiation or increased gas-phase O-chemistry.
  \item Within 150~au, there is substantial spatial overlap between dust and chemical substructures, while in the outer disks, such overlaps are sparse. This suggests a scenario where the dust-emitting midplane and warm molecular layers are linked in the inner disk regions but become increasingly disconnected at larger disk radii.
  \item Some chemical substructures are spatially associated with the outer edge of the millimeter continuum. In particular, HCN, DCN, and H$_2$CO emission rings are commonly coincident with the outer continuum edge in the MAPS disks.
  \item The vast majority of chemical substructures in the MAPS disks are not spatially associated with midplane snowlines, which indicates that snowlines do not directly cause most chemical substructures.
  \item In the inner 50~au, the MAPS disks exhibit a wide range of radial morphologies, including central peaks, plateaus, dips, and holes. This diversity is present both across disks and between lines, where it is often but not exclusively associated with molecules of different chemical families. The only non-CO molecules to show central peaked profiles are those of the nitriles HCN, HC$_3$N, and DCN.
\end{enumerate}

The MAPS observations reveal a striking diversity in the radial morphologies of molecular line emission in protoplanetary disks. Chemical substructures are ubiquitous and extremely varied with a wide range of radial locations, widths, and relative contrasts. This suggests that planets often form in diverse chemical environments both across disks and at different radii within the same disk. While the MAPS Large Program provides a comprehensive view of the chemical environments in which planet formation occurs, the observations are of a sample of only five carefully-selected disks. Unbiased surveys toward larger numbers of disks at comparably high spatial resolution are required to determine the universality of the trends identified here and to assess if typical planet-forming disks host chemical environments similar to those of the MAPS disks.

\section{Value-Added Data Products} \label{sec:VADPs_listing}

The MAPS Value-Added Data Products described in this work can be accessed through the ALMA Archive via \url{https://almascience.nrao.edu/alma-data/lp/maps}. An interactive browser for this repository is also available on the MAPS project homepage at \url{http://www.alma-maps.info}.

For each combination of transition \citep[Tables 2 \& 3,][]{oberg20}, disk \citep[Table 1,][]{oberg20}, and spatial resolution \citep[Table 5,][]{oberg20}, the following data products are available:

\begin{itemize}
    \item Zeroth moment map
    \item Rotation map
    \item Peak intensity map
    \item Radial intensity profile (including a combination of wedge sizes and with surfaces, when available)
    \item Emission surface, when available \citep[see][for more details]{law20a}
    \item Python script to generate the data products
\end{itemize}

For VADPs generated using \texttt{bettermoments}, namely the zeroth moment, rotation, and peak intensity maps, we also provide the corresponding uncertainty maps. For more information on the data products associated with the imaging process, see Section 9 in \citet{czekala20}. For a detailed description of the naming conventions of all VADPs, see Section 3.5 in \citet{oberg20}.  \newpage

\acknowledgments

The authors thank the anonymous referee for valuable comments that improved both the content and presentation of this work. This paper makes use of the following ALMA data: ADS/JAO.ALMA\#2018.1.01055.L. ALMA is a partnership of ESO (representing its member states), NSF (USA) and NINS (Japan), together with NRC (Canada), MOST and ASIAA (Taiwan), and KASI (Republic of Korea), in cooperation with the Republic of Chile. The Joint ALMA Observatory is operated by ESO, AUI/NRAO and NAOJ. The National Radio Astronomy Observatory is a facility of the National Science Foundation operated under cooperative agreement by Associated Universities, Inc.

C.J.L. acknowledges funding from the National Science Foundation Graduate Research Fellowship under Grant DGE1745303. R.T. and F.L. acknowledge support from the Smithsonian Institution as a Submillimeter Array (SMA) Fellow. K.I.\"O. acknowledges support from the Simons Foundation (SCOL \#321183) and an NSF AAG Grant (\#1907653). I.C. was supported by NASA through the NASA Hubble Fellowship grant \#HST-HF2-51405.001-A awarded by the Space Telescope Science Institute, which is operated by the Association of Universities for Research in Astronomy, Inc., for NASA, under contract NAS5-26555. S.M.A. and J.H. acknowledge funding support from the National Aeronautics and Space Administration under Grant No. 17-XRP17 2-0012 issued through the Exoplanets Research Program. J.H. acknowledges support for this work provided by NASA through the NASA Hubble Fellowship grant \#HST-HF2-51460.001-A awarded by the Space Telescope Science Institute, which is operated by the Association of Universities for Research in Astronomy, Inc., for NASA, under contract NAS5-26555. Y.A. acknowledges support by NAOJ ALMA Scientific Research Grant Code 2019-13B, and Grant-in-Aid for Scientific Research 18H05222 and 20H05847. J.B. acknowledges support by NASA through the NASA Hubble Fellowship grant \#HST-HF2-51427.001-A awarded by the Space Telescope Science Institute, which is operated by the Association of Universities for Research in Astronomy, Incorporated, under NASA contract NAS5-26555. E.A.B., A.D.B, and F.A. acknowledge support from NSF AAG Grant (\#1907653). J.B.B. acknowledges support from NASA through the NASA Hubble Fellowship grant \#HST-HF2-51429.001-A, awarded by the Space Telescope Science Institute, which is operated by the Association of Universities for Research in Astronomy, Inc., for NASA, under contract NAS5-26555. Y.B. acknowledges funding from ANR (Agence Nationale de la Recherche) of France under contract number ANR-16-CE31-0013 (Planet-Forming-Disks). A.S.B. acknowledges the studentship funded by the Science and Technology Facilities Council of the United Kingdom (STFC). J.K.C. acknowledges support from the National Science Foundation Graduate Research Fellowship under Grant No. DGE 1256260 and the National Aeronautics and Space Administration FINESST grant, under Grant no. 80NSSC19K1534. G.C. is supported by NAOJ ALMA Scientific Research Grant Code 2019-13B. L.I.C. gratefully acknowledges support from the David and Lucille Packard Foundation and Johnson \& Johnson's WiSTEM2D Program. V.V.G. acknowledges support from FONDECYT Iniciaci\'on 11180904 and ANID project Basal AFB-170002. J.D.I. acknowledges support from the Science and Technology Facilities Council of the United Kingdom (STFC) under ST/T000287/1. R.L.G. acknowledges support from a CNES fellowship grant. Y.L. acknowledges the financial support by the Natural Science Foundation of China (Grant No. 11973090). F.M. acknowledges support from ANR of France under contract ANR-16-CE31-0013 (Planet-Forming-Disks) and ANR-15-IDEX-02 (through CDP ``Origins of Life"). H.N. acknowledges support by NAOJ ALMA Scientific Research Grant Code 2018-10B and Grant-in-Aid for Scientific Research 18H05441. K.R.S. acknowledges the support of NASA through Hubble Fellowship Program grant HST-HF2-51419.001, awarded by the Space Telescope Science Institute,which is operated by the Association of Universities for Research in Astronomy, Inc., for NASA, under contract NAS5-26555. A.S. acknowledges support from ANID/CONICYT Programa de Astronom\'ia Fondo ALMA-CONICYT 2018 31180052. T.T. is supported by JSPS KAKENHI Grant Numbers JP17K14244 and JP20K04017. Y.Y. is supported by IGPEES, WINGS Program, the University of Tokyo. M.L.R.H. acknowledges support from the Michigan Society of Fellows. C.W. acknowledges financial support from the University of Leeds, STFC and UKRI (grant numbers ST/R000549/1, ST/T000287/1, MR/T040726/1). K.Z. acknowledges the support of the Office of the Vice Chancellor for Research and Graduate Education at the University of Wisconsin – Madison with funding from the Wisconsin Alumni Research Foundation, and support of the support of NASA through Hubble Fellowship grant HST-HF2-51401.001. awarded by the Space Telescope Science Institute, which is operated by the Association of Universities for Research in Astronomy, Inc., for NASA, under contract NAS5-26555. A.S.B acknowledges the studentship funded by the Science and Technology Facilities Council of the United Kingdom (STFC).

%% To help institutions obtain information on the effectiveness of their telescopes the AAS Journals has created a group of keywords for telescope facilities. Following the acknowledgments section, use the following syntax and the \facility{} or \facilities{} macros to list the keywords of facilities used in the research for the paper.  Each keyword is check against the master list during copy editing. Individual instruments can be provided in parentheses, after the keyword, but they are not verified.

\vspace{5mm}
\facilities{ALMA}

%% Similar to \facility{}, there is the optional \software command to allow authors a place to specify which programs were used during the creation of the manuscript. Authors should list each code and include either a citation or url to the code inside ()s when available.

\software{Astropy \citep{astropy_2013,astropy_2018}, \texttt{bettermoments} \citep{Teague18_bettermoments}, CASA \citep{McMullin_etal_2007}, \texttt{GoFish} \citep{Teague19JOSS}, \texttt{LMFIT} \citep{LMFIT}, Matplotlib \citep{Hunter07}, NumPy \citep{vanderWalt_etal_2011}, SciPy \citep{Virtanen_etal_2020}}

\clearpage

\appendix
\vspace{-1.45cm}
\section{Generation of Hybrid Zeroth Moment Maps}
\label{sec:app:moment_map_generation}

Collapsing a three dimensional data cube into a two dimensional moment map requires the loss of information. In this process, the level of masking applied to the data prior to the collapse to the desired summary statistic will largely depend on its desired use.

The most typical mask used is a $\sigma$-clip, which removes all pixels from each channel of the data cube which are below a certain threshold value, e.g., a $2\sigma$ clip where $\sigma$ is the RMS of the cube. While this yields a relatively noise-free moment map, it can significantly reduce the measured integrated intensities. This is especially problematic if a large fraction of emission is expected at a per channel level of $\lesssim 2 \sigma$, as is the case for many lines covered by MAPS. Thus, while $\sigma$-clipped moment maps are adequate for qualitative descriptions of emission morphology, they cannot be used for quantitative comparisons.

An alternative approach is to adopt a Keplerian mask \citep[e.g.,][]{Rosenfeld13, Loomis15, Oberg15Natur}. For a source in Keplerian rotation, such as the gas in protoplanetary disks, the expected emission morphology is well characterized in position-position-velocity (PPV) space. Using a simple analytical model describing this rotation, it is possible to consider only regions of the cube where line emission should originate and thus avoid unknowingly removing low-level emission. These masks are also routinely used for CLEANing data \citep[see, e.g.,][]{czekala20}. The use of a sufficiently conservative mask, i.e., one that encloses all disk emission, provides the most accurate description of the integrated intensity. For this reason, we use the Keplerian CLEAN masks from \citet{czekala20} when collapsing the data cubes into zeroth moment maps. We use these maps for all quantitative work within MAPS, such as the radial profiles described in this work.

While this approach produced maps with the most accurate flux values, they do not always produce maps that are trivial to interpret visually. For weak lines, this Keplerian masking led to noise-dominated moment maps, primarily because the masks were purposely made large in position-position space, so as not to miss any emission. In addition, artifacts in the form of arc-shaped ridges were present in the central few arcsecs of the maps due to the extended velocity wings required for the CLEAN masks to fully capture $^{13}$CO emission in the inner disk \citep[][]{czekala20, bosman20_inner_au}. These features were due to these disk inner regions including a considerably larger number of channels (e.g., see Figure \ref{fig:MomentMapsMask}) in the integration and thus producing stepped emission distributions.

\begin{figure}
    \centering
    \includegraphics[width=0.875\textwidth]{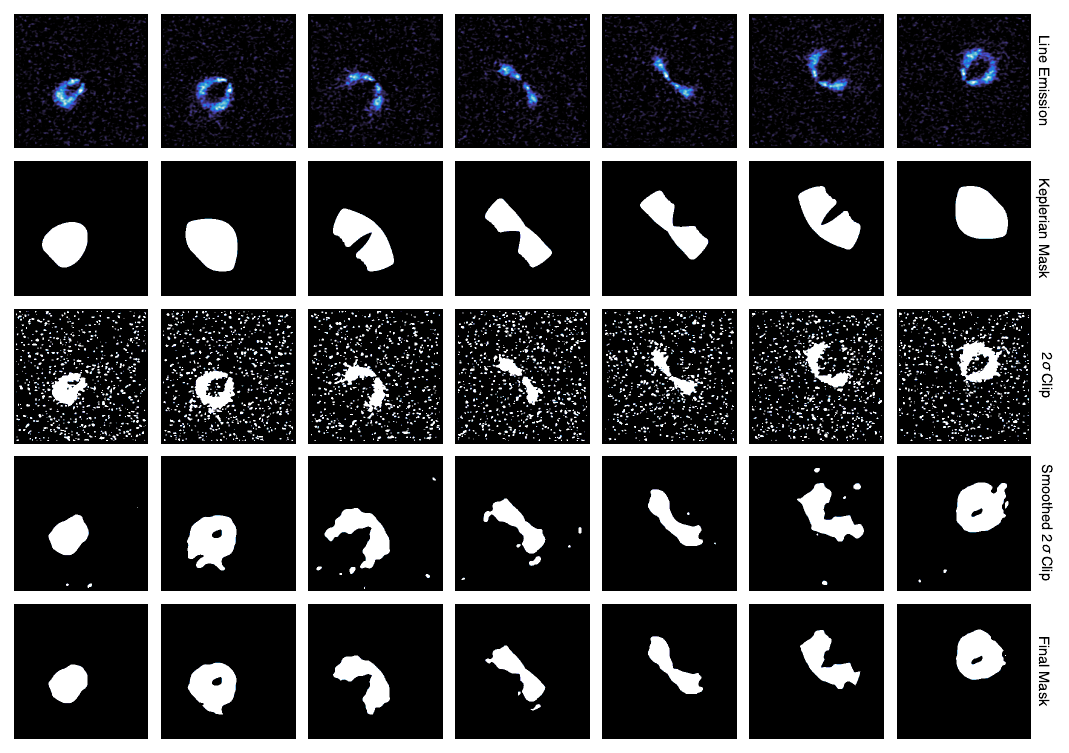}
    \vspace{-0.2cm}
    \caption{Comparisons of the observed HCO$^+$ 1--0 emission from the CLEANed images (top row) in HD~163296 with four types of masks: the Keplerian mask (second row); a threshold mask clipping all values below $2\sigma$ (third row); the threshold mask enlarged by first convolving the emission with a 2D Gaussian kernel based on the synthesized beam before applying the $2\sigma$ clip (fourth row); and the combination of the Keplerian and smoothed $2\sigma$ clip masks (bottom row).}
    \label{fig:app:mask_comparison}
\end{figure}

To account for this, we generated a set of ``hybrid" zeroth moment maps using a combination of these techniques: we combined a Keplerian mask with a smoothed, intensity-threshold map, as demonstrated in Figure~\ref{fig:app:mask_comparison}. The top row shows HCO$^+$ 1--0 emission in the HD~163296 disk. The second row shows the Keplerian mask used for cleaning. Clearly, this follows the same morphology as the emission in the PPV space of the cube, but overestimates the radial and azimuthal extent of the emission in any given channel. The third row shows a mask based on a $2\sigma$ threshold. While this does a better job of tracing the emission in each channel, it simultaneously misses the low level emission at the edges of the line, while also adding in background noise at large radii. The fourth row shows an extended clip mask, which was generated by convolving the data with a circular Gaussian kernel with a FWHM equal to that of the synthesized beam before the 2$\sigma$ clip was applied to the smoothed data. Applying the $\sigma$-clip to smoothed data has the following benefits: (1) the convolution removes noise peaks at large offsets from the disk center; and (2) where the emission is strong (i.e., values much larger than the adopted clip-threshold), the resultant mask is broader than one generated from the un-smoothed data. This leverages the proximity to real (strong) emission to allow for weaker emission to be included in the moment map which would typically be lost with a $\sigma$-clip. The bottom row shows the final mask, which is a combination of the Keplerian and smoothed $2\sigma$ masks, which together, removes persistent patches of background noise on the scale of the beam.

\begin{figure}[b]
    \centering
    \includegraphics[width=0.6\textwidth]{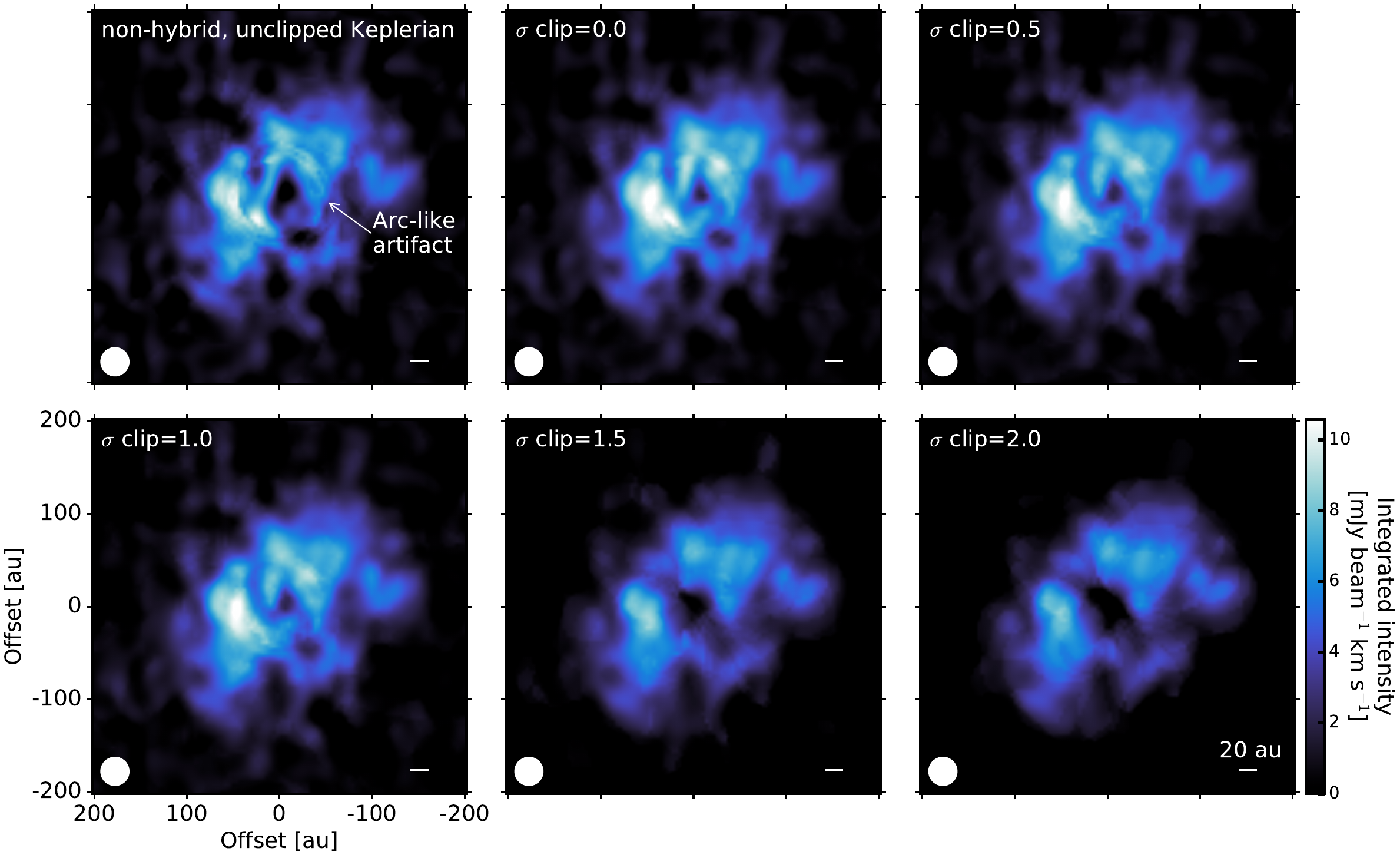}
    \caption{Comparison of unclipped, Keplerian masked zeroth moment map with hybrid zeroth moment maps using different sigma clips for CS 2--1 in HD~163296. The color bar is the same for all panels. The synthesized beam and a scale bar indicating 20~au is shown in the lower left and right corner, respectively, of each panel.}
    \label{fig:app:CS_HD16_example}
\end{figure}

For all lines, disks, and spatial resolutions covered in MAPS \citep{oberg20}, we generated hybrid zeroth moment maps with a set of $\sigma$ clips: 0, 0.5, 1.0, 1.5, and 2.0. For each map, we visually selected the lowest clip value that best mitigated the arc-like artifacts, while not substantially lowering the flux. Figure \ref{fig:app:CS_HD16_example} illustrates this process for CS 2--1 in HD~163296. Circular artifacts are evident in the unclipped, Keplerian masked zeroth moment map, as well as in the hybrid maps with $\sigma$ clips of 0 and 0.5. However, the higher $\sigma$ clips of 1.5 and 2 substantially reduce the faint emission in the inner disk, which causes the inner dip in line emission to appear artificially radially extended. In this case, a choice of $\sigma$=1.0 shows an optimal balance between mitigating the central artifacts and retaining sufficient flux to accurately illustrate the spatial distribution of CS 2--1 emission. 

To generate these maps, we used the following commands with \texttt{bettermoments} \citep{Teague18_bettermoments}: ``bettermoments cube.fits -method zeroth -mask cube.mask.fits -clip -100 X -smooth 3 -smooththreshold 1", where X is the value of the $\sigma$ clip.

\vspace{-5cm}

\newpage
\clearpage

\section{Disk-specific zeroth moment map} \label{sec:app:disk_specific_zeroth_moment_map}

\begin{figure*}[!ht]
\centering
\textbf{Fig. Set 27. MAPS Zeroth Moment Maps}\par\medskip
\figurenum{27.1}
\includegraphics[width=\linewidth]{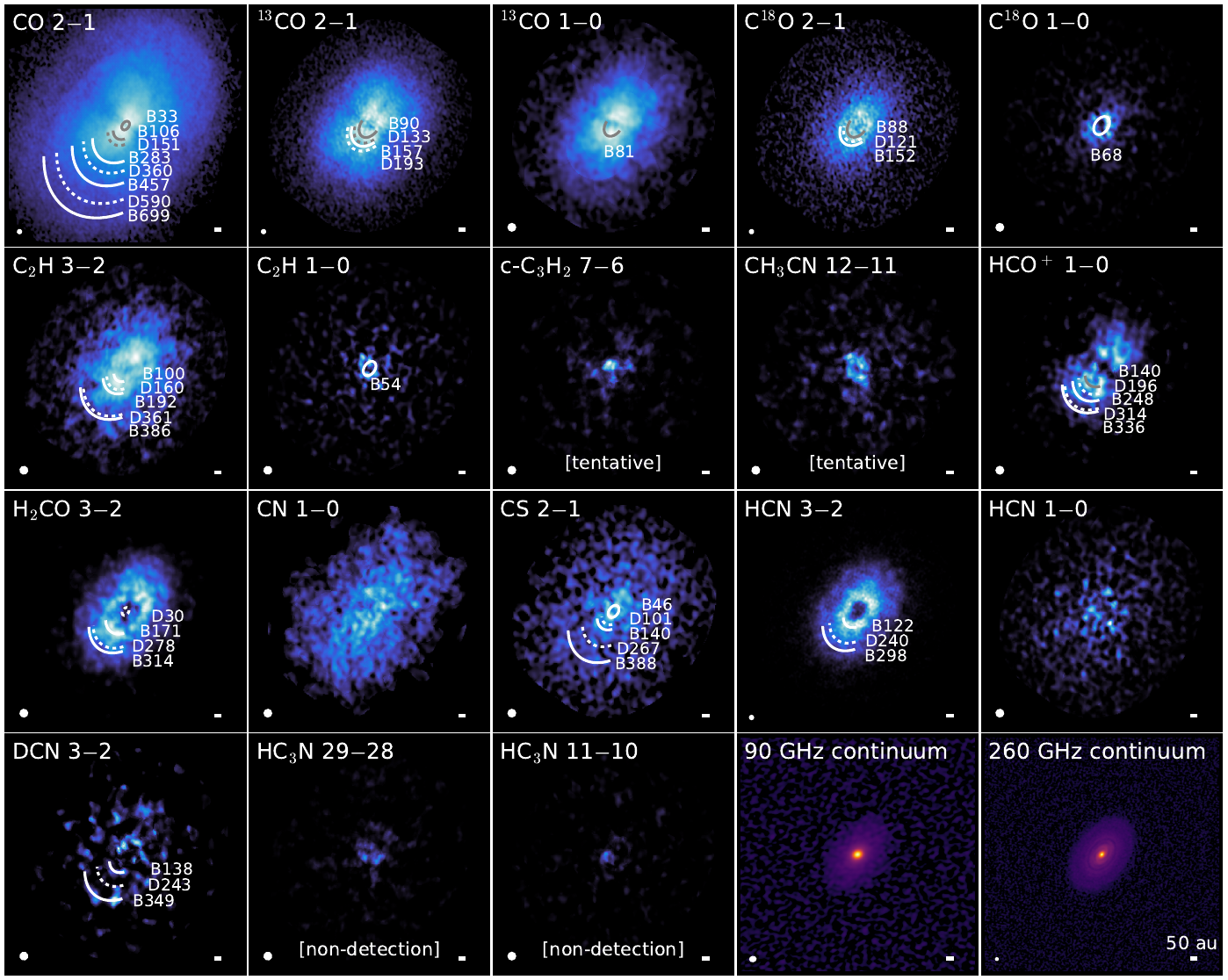}
\caption{Zeroth moment maps for all lines targeted in this study and the 90~GHz and 260~GHz continuum in IM~Lup. All panels show the same spatial scale. Otherwise, as in Figure \ref{fig:CO_Moment0}.}
\label{fig:IM_Lup_mom0}
\end{figure*}

To illustrate the diversity of radial emission morphologies within each MAPS disk, Figure \ref{fig:IM_Lup_mom0} shows a gallery of zeroth moment maps for all 18 lines considered in this study, along with the 90~GHz and 260~GHz continuum images, for the IM~Lup disk. A complete gallery of zeroth moment maps for all MAPS disks is shown in Figure Set 1, which is available in the electronic edition of the journal.

\begin{figure*}[!ht]
\figurenum{27.2}
\centering
\includegraphics[width=\linewidth]{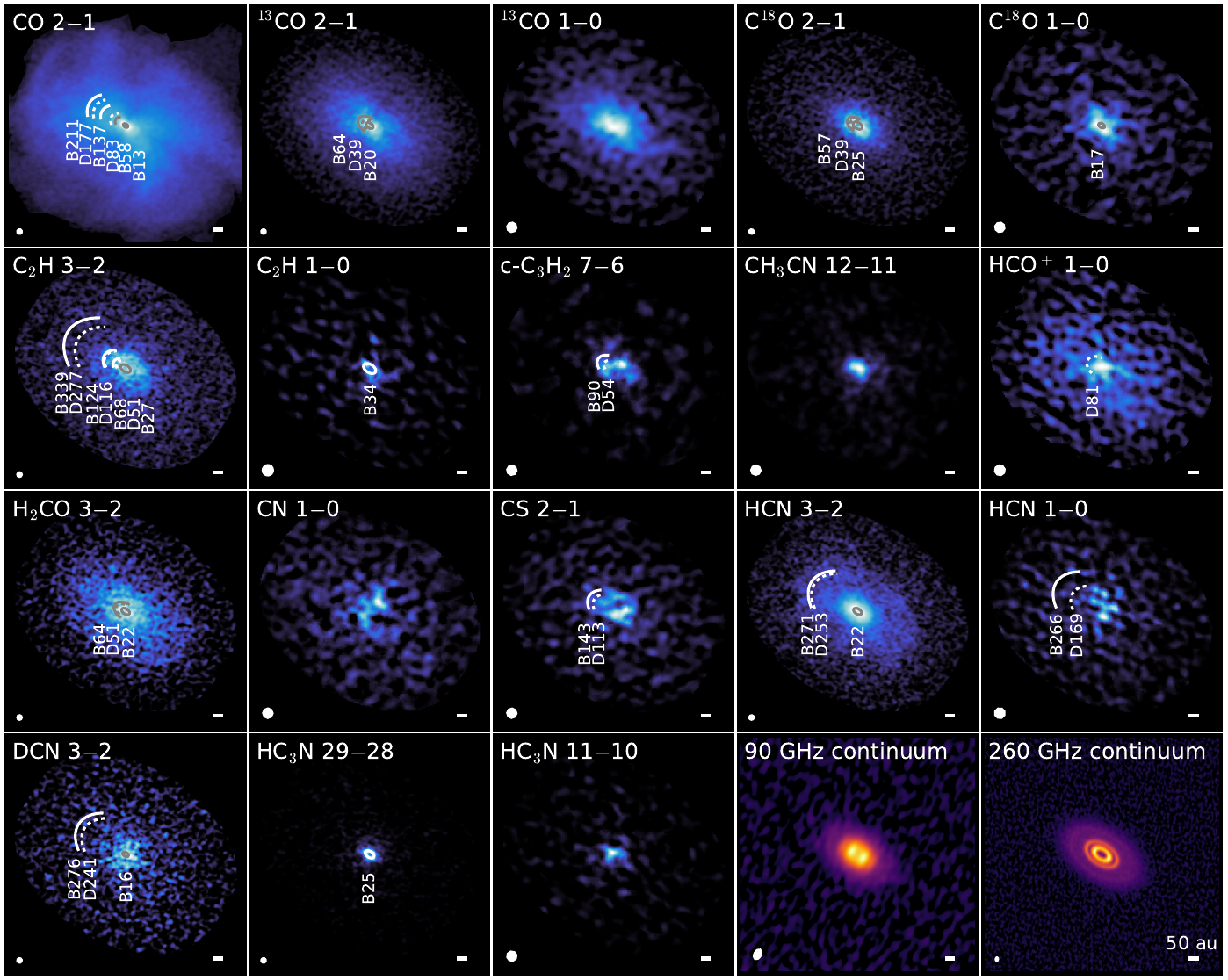}
\caption{Zeroth moment maps for all lines targeted in this study and the 90~GHz and 260~GHz continuum in GM~Aur. Otherwise, as in Figure \ref{fig:CO_Moment0}.}
\label{fig:GM_Aur_mom0}
\end{figure*}

\begin{figure*}[!ht]
\figurenum{27.3}
\centering
\includegraphics[width=\linewidth]{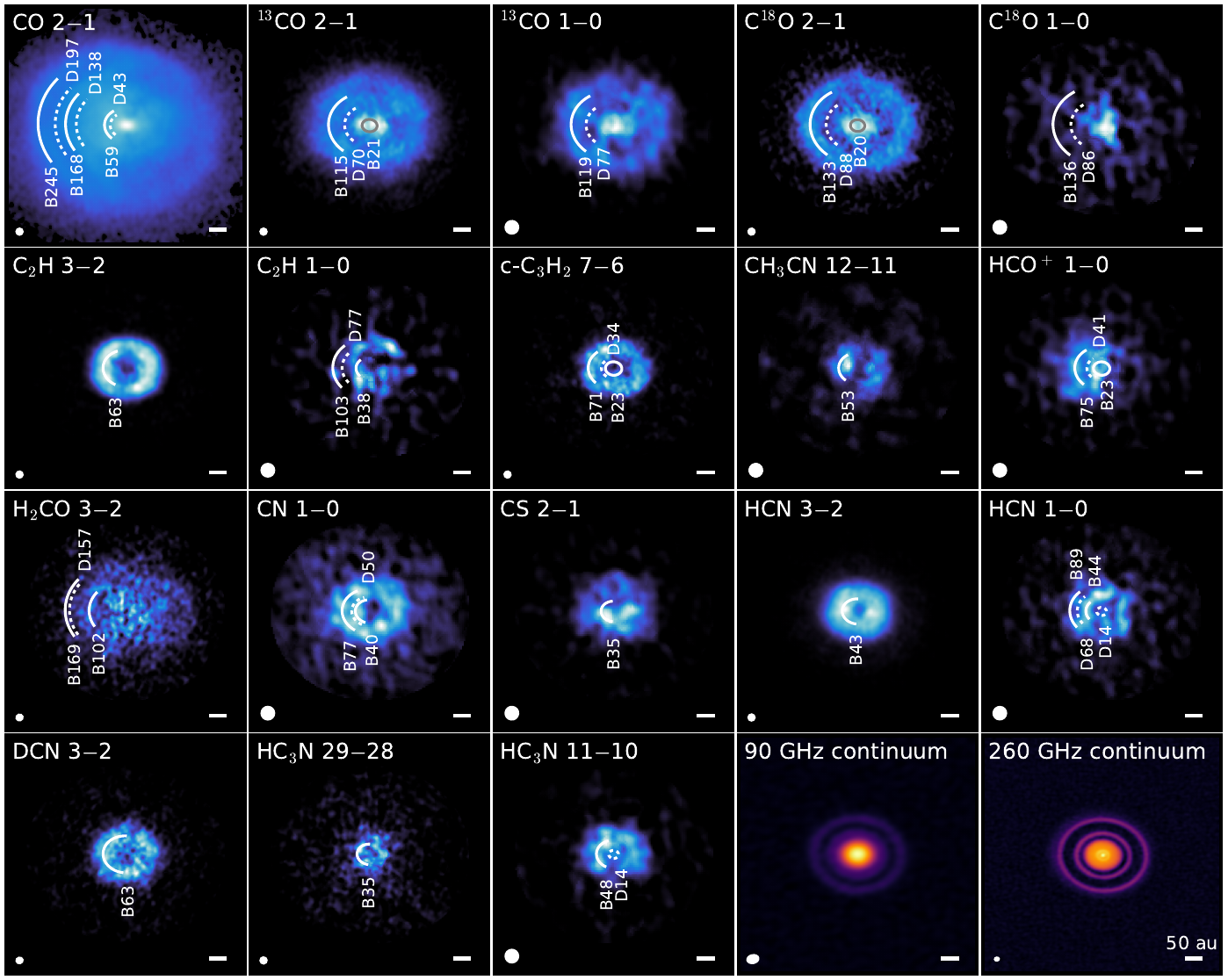}
\caption{Zeroth moment maps for all lines targeted in this study and the 90~GHz and 260~GHz continuum in AS~209. Otherwise, as in Figure \ref{fig:CO_Moment0}.}
\label{fig:AS_209_mom0}
\end{figure*}

\begin{figure*}[!ht]
\figurenum{27.4}
\centering
\includegraphics[width=\linewidth]{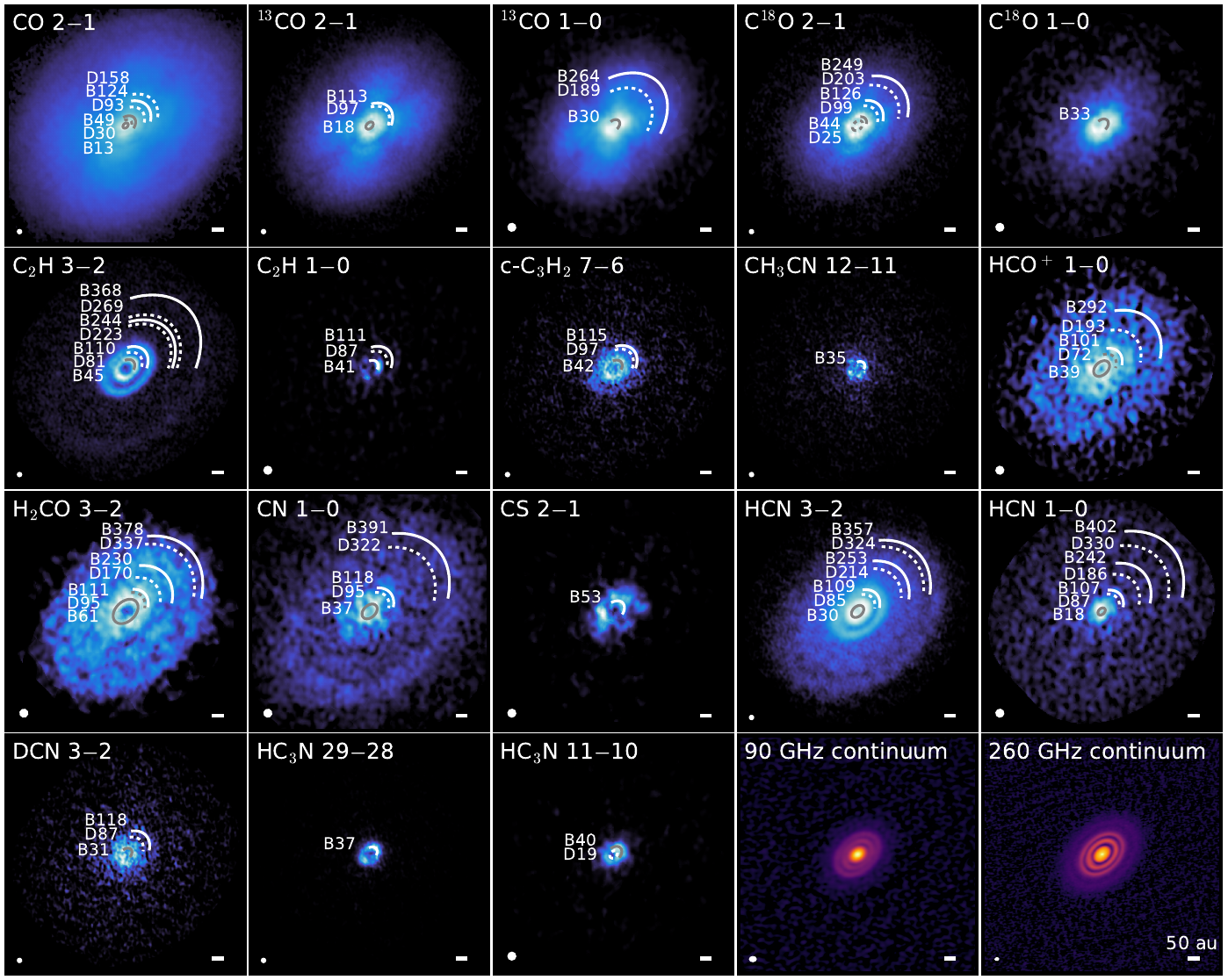}
\caption{Zeroth moment maps for all lines targeted in this study and the 90~GHz and 260~GHz continuum in HD~163296. Otherwise, as in Figure \ref{fig:CO_Moment0}.}
\label{fig:HD_163296_mom0}
\end{figure*}

\begin{figure*}[!ht]
\figurenum{27.5}
\centering
\includegraphics[width=\linewidth]{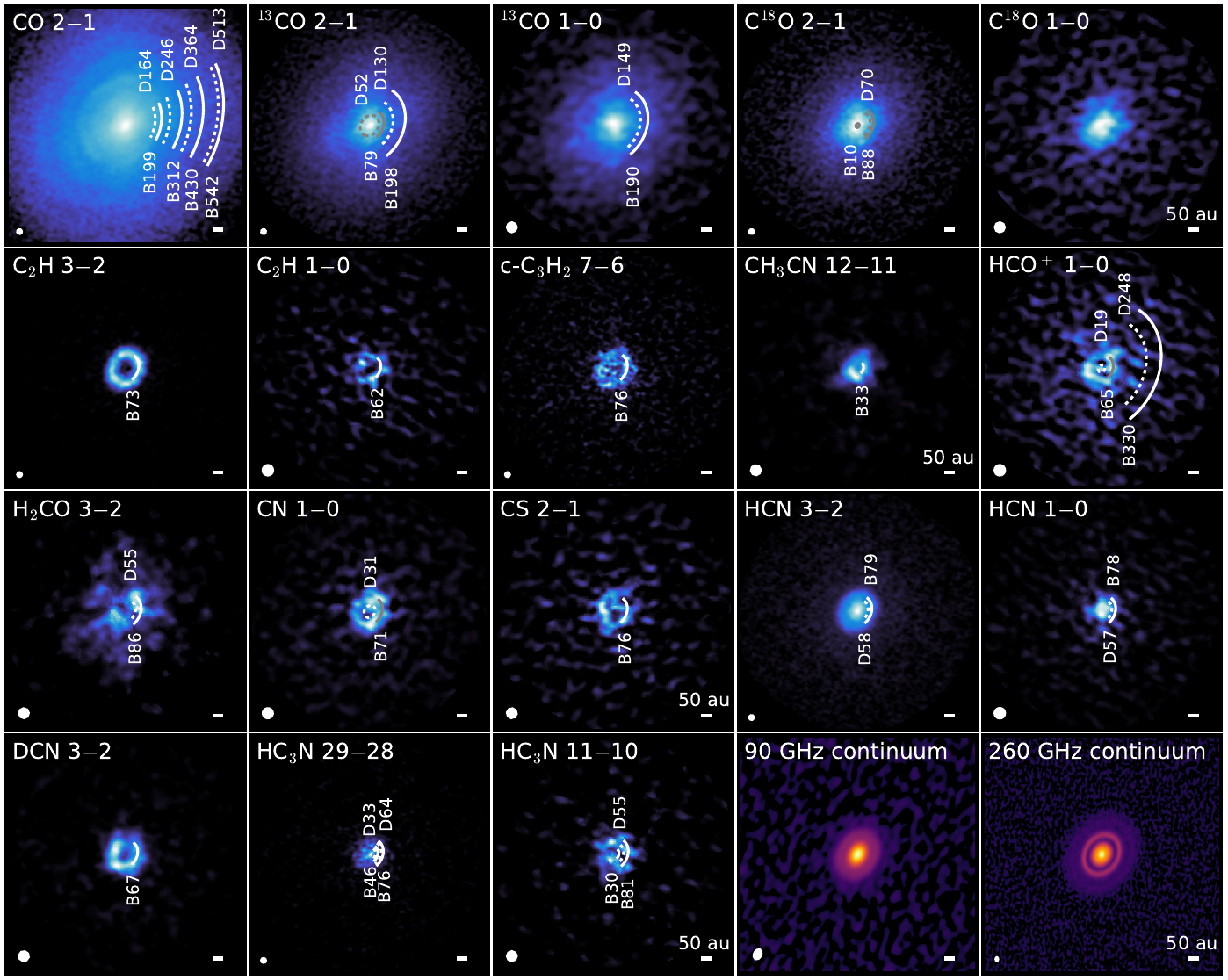}
\caption{Zeroth moment maps for all lines targeted in this study and the 90~GHz and 260~GHz continuum in MWC~480. Otherwise, as in Figure \ref{fig:CO_Moment0}.}
\label{fig:MWC_480_mom0}
\end{figure*}

%%\begin{figure*}[!ht]
%%\centering
%%\includegraphics[width=\linewidth]{Figures/Figure_GM_Aur_hor.pdf}
%%\caption{Zeroth moment maps for all lines targeted in this study and the 90~GHz and 260~GHz continuum in GM~Aur. All panels show the same spatial scale. Otherwise, as in Figure \ref{fig:CO_Moment0}.}
%%\label{fig:GM_Aur_mom0}
%%\end{figure*}

%%\begin{figure*}[!ht]
%%\centering
%%\includegraphics[width=\linewidth]{Figures/Figure_AS_209_hor.pdf}
%%\caption{Zeroth moment maps for all lines targeted in this study and the 90~GHz and 260~GHz continuum in AS~209. All panels show the same spatial scale. Otherwise, as in Figure \ref{fig:CO_Moment0}.}
%%\label{fig:AS_209_mom0}
%%\end{figure*}

%%\begin{figure*}[!ht]
%%\centering
%%\includegraphics[width=\linewidth]{Figures/Figure_HD_163296_hor.pdf}
%%\caption{Zeroth moment maps for all lines targeted in this study and the 90~GHz and 260~GHz continuum in HD~163296. All panels show the same spatial scale. Otherwise, as in Figure \ref{fig:CO_Moment0}.}
%%\label{fig:HD_163296_mom0}
%%\end{figure*}

%%\begin{figure*}[!ht]
%%\centering
%%\includegraphics[width=\linewidth]{Figures/Figure_MWC_480_hor.pdf}
%%\caption{Zeroth moment maps for all lines targeted in this study and the 90~GHz and 260~GHz continuum in MWC~480. All panels show the same spatial scale. Otherwise, as in Figure \ref{fig:CO_Moment0}.}
%%\label{fig:MWC_480_mom0}
%%\end{figure*}

\newpage
\clearpage
\section{Peak intensity maps} \label{sec:app:peak_intensity_maps}

\begin{figure*}[!ht]
\centering
\textbf{Fig. Set 28. MAPS Peak Intensity Maps}\par\medskip
\figurenum{28.1}
\includegraphics[width=\linewidth]{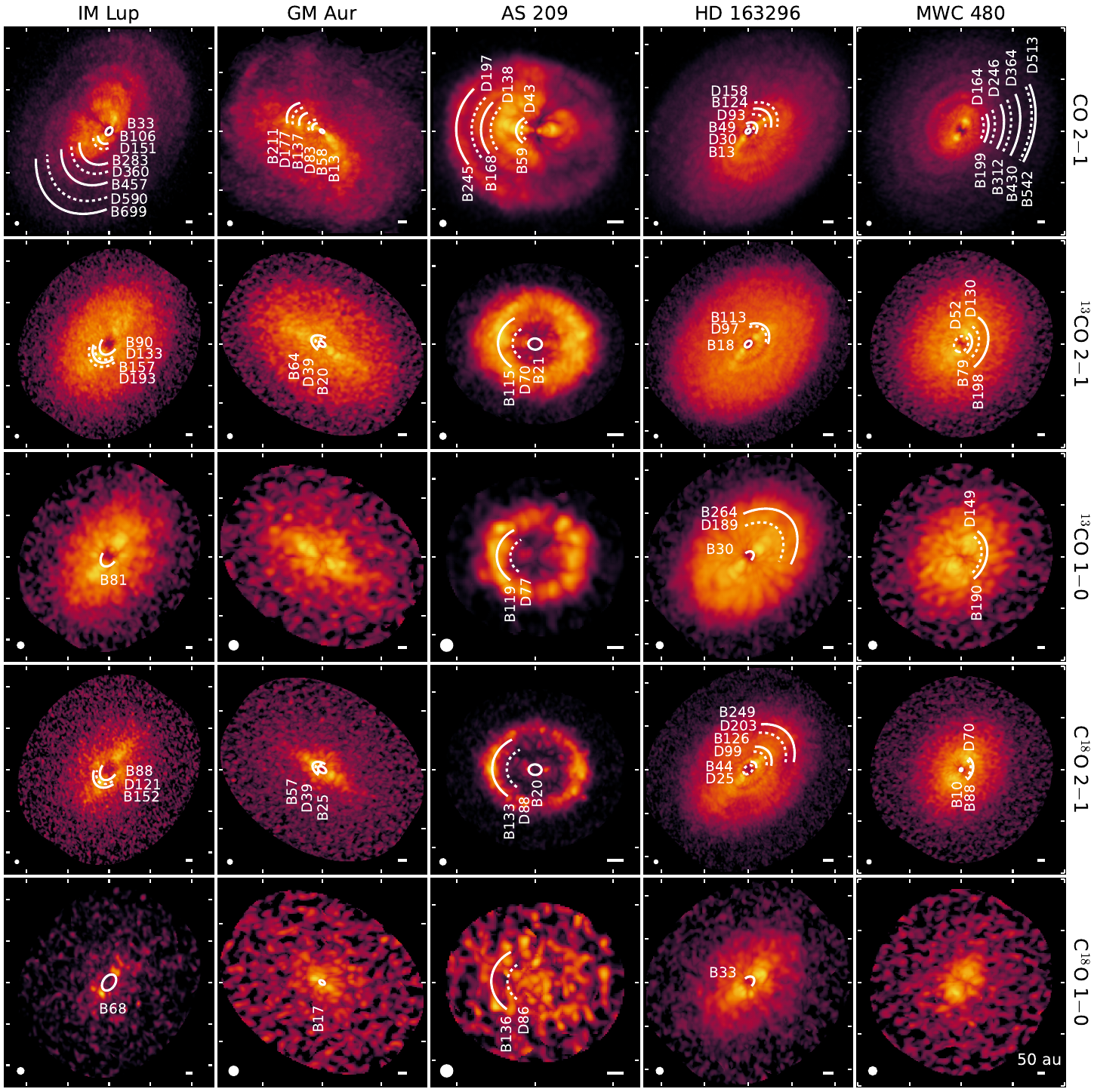}
\caption{Peak intensity maps of CO, $^{13}$CO, and C$^{18}$O lines. Otherwise, as in Figure \ref{fig:CO_Moment0}.}
\label{fig:CO_Fnu}
\end{figure*}

Figure \ref{fig:CO_Fnu} shows peak intensity maps for the CO lines in all MAPS disks. A complete gallery of peak intensity maps for all 18 lines considered here is shown in Figure Set 2, which is available in the electronic edition of the journal.

\begin{figure*}[!ht]
\centering
\figurenum{28.2}
\includegraphics[width=\linewidth]{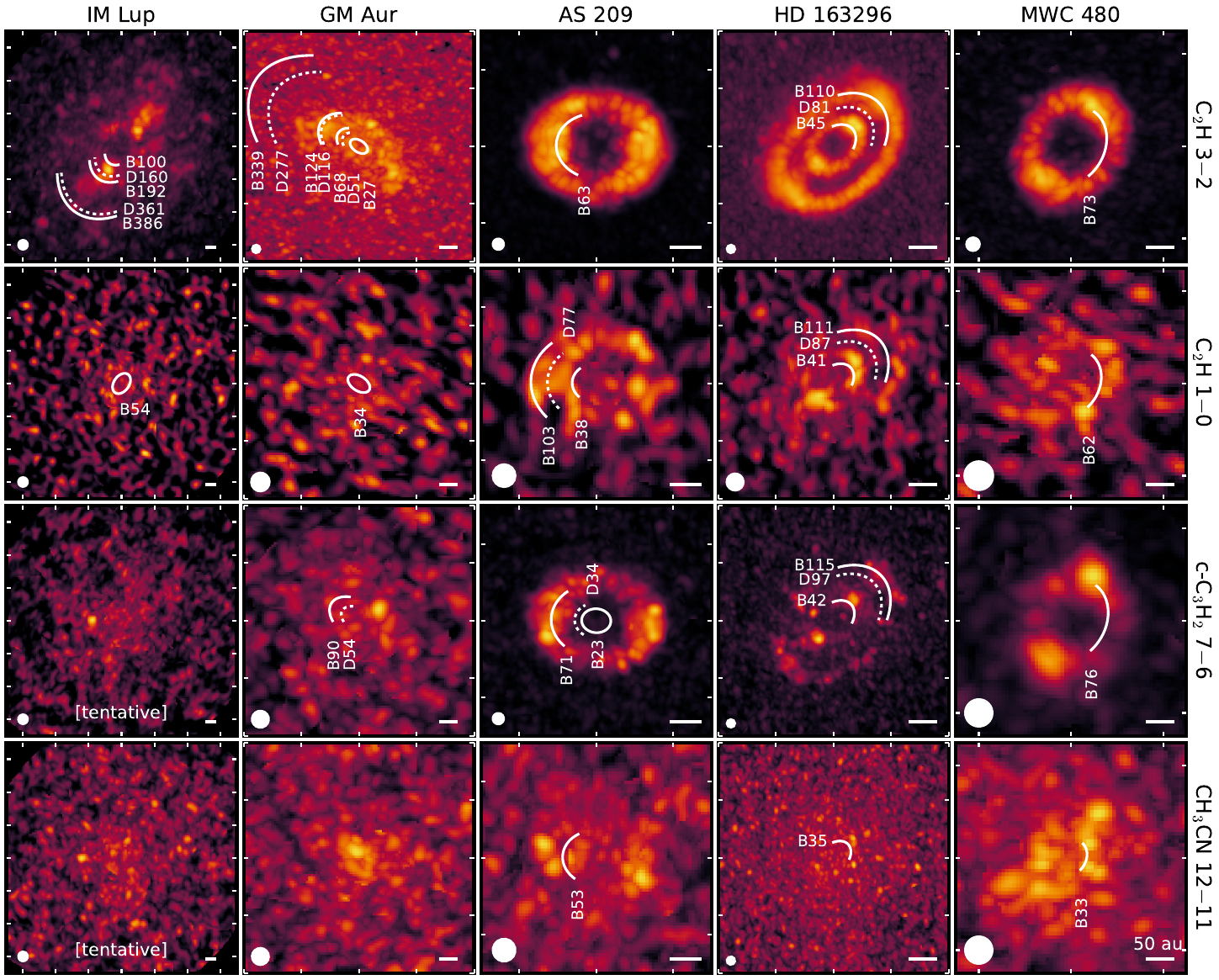}
\caption{Peak intensity maps of C$_2$H, c-C$_3$H$_2$, and CH$_3$CN lines. Otherwise, as in Figure \ref{fig:CO_Moment0}.}
\label{fig:CO_2_Fnu}
\end{figure*}

\begin{figure*}[!ht]
\centering
\figurenum{28.3}
\includegraphics[width=\linewidth]{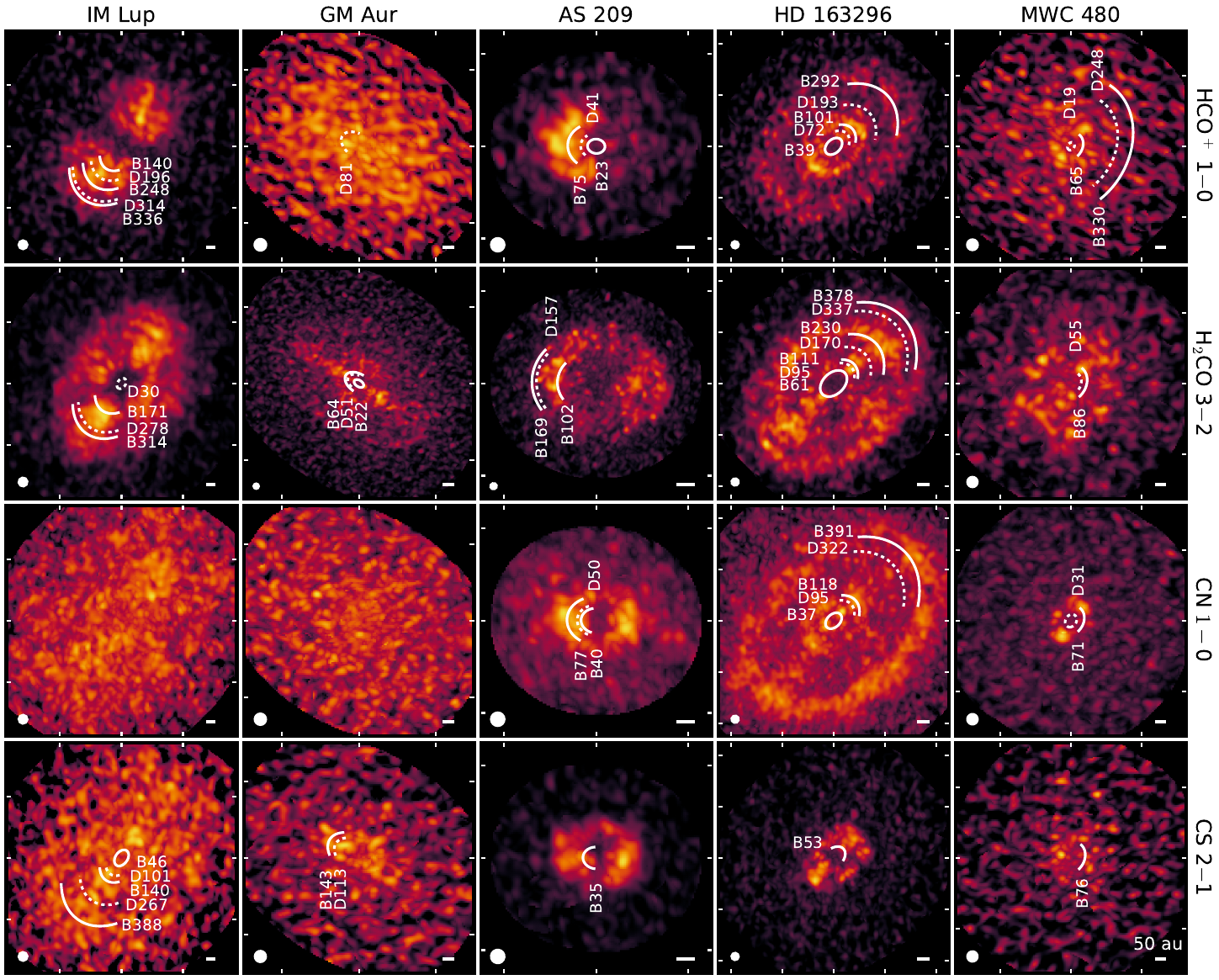}
\caption{Peak intensity maps of HCO$^+$, H$_2$CO, CN, and CS lines. Otherwise, as in Figure \ref{fig:CO_Moment0}.}
\label{fig:CO_3_Fnu}
\end{figure*}

\begin{figure*}[!ht]
\centering
\figurenum{28.4}
\includegraphics[width=\linewidth]{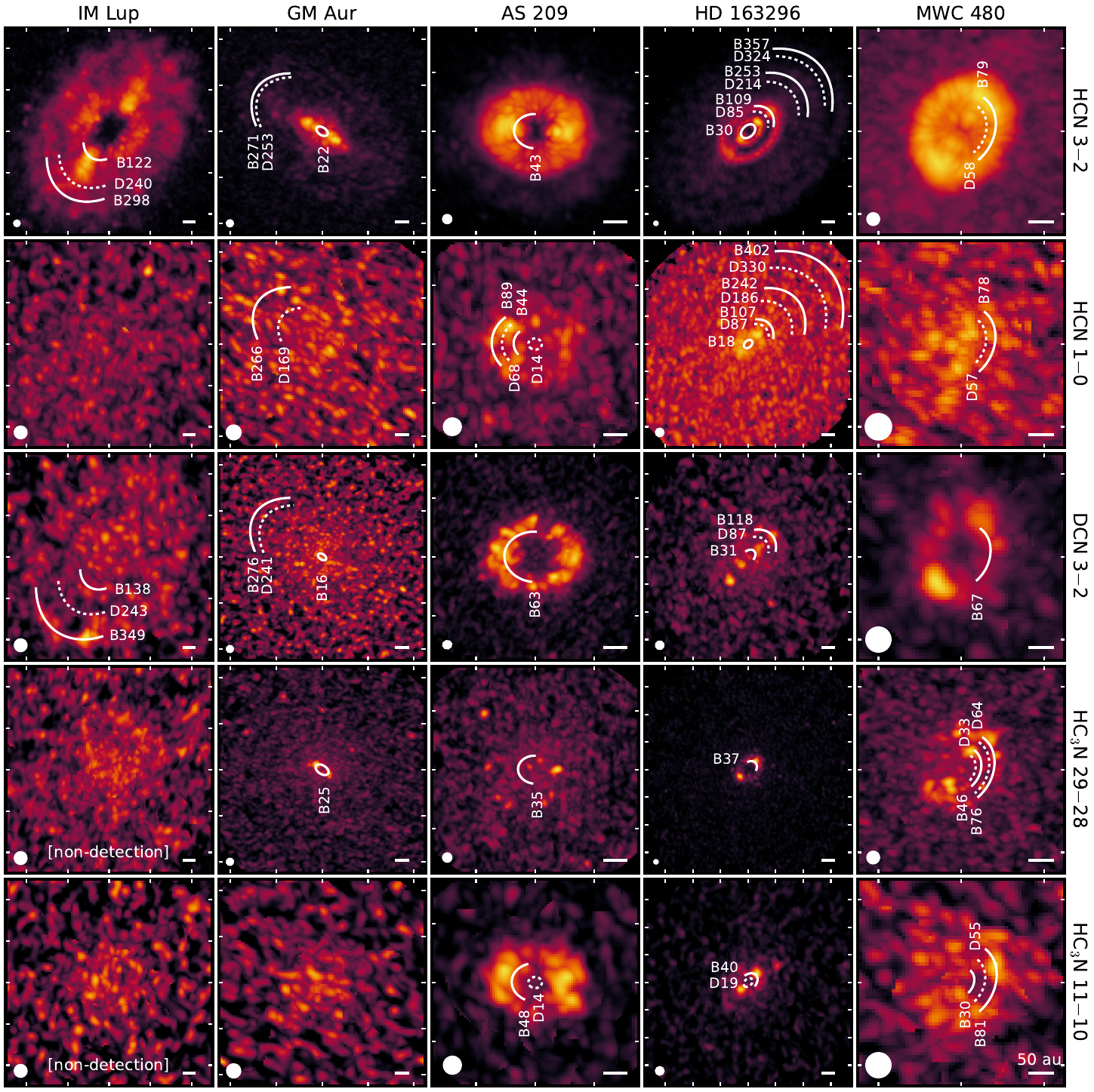}
\caption{Peak intensity maps of HCN, DCN, and HC$_3$N lines. Otherwise, as in Figure \ref{fig:CO_Moment0}.}
\label{fig:CO_4_Fnu}
\end{figure*}

%%\begin{figure*}[!ht]
%%\centering
%%\includegraphics[width=\linewidth]{Moment_Maps/Fnu/C2H_Fnu_Transpose.png}
%%\caption{Peak intensity maps of C$_2$H, c-C$_3$H$_2$, CH$_3$CN lines. Otherwise, as in Figure \ref{fig:C2H_Moment0}. [\edit1{missing panels are due to missing VADPs}]}
%%\label{fig:C2H_Fnu}
%%\end{figure*}

%%\begin{figure*}[!ht]
%%\centering
%%\includegraphics[width=\linewidth]{Moment_Maps/Fnu/H2CO_Fnu_Transpose.png}
%%\caption{Peak intensity maps of HCO$^+$, H$_2$CO, CN, and CS lines. Otherwise, as in Figure \ref{fig:H2CO_Moment0}.}
%%\label{fig:H2CO_Fnu}
%%\end{figure*}

%%\begin{figure*}[!ht]
%%\centering
%%\includegraphics[width=\linewidth]{Moment_Maps/Fnu/HCN_Fnu_Transpose.png}
%%\caption{Peak intensity maps of HCN, DCN, and HC$_3$N lines. Otherwise, as in Figure \ref{fig:HCN_Moment0}.}
%%\label{fig:HCN_Fnu}
%%\end{figure*}

\newpage
\clearpage

\section{Surface Assumptions on Radial Profiles} \label{sec:app:surface_and_wedges_assumptions}

\begin{figure*}[!ht]
\centering
\figurenum{29}
\includegraphics[width=\linewidth]{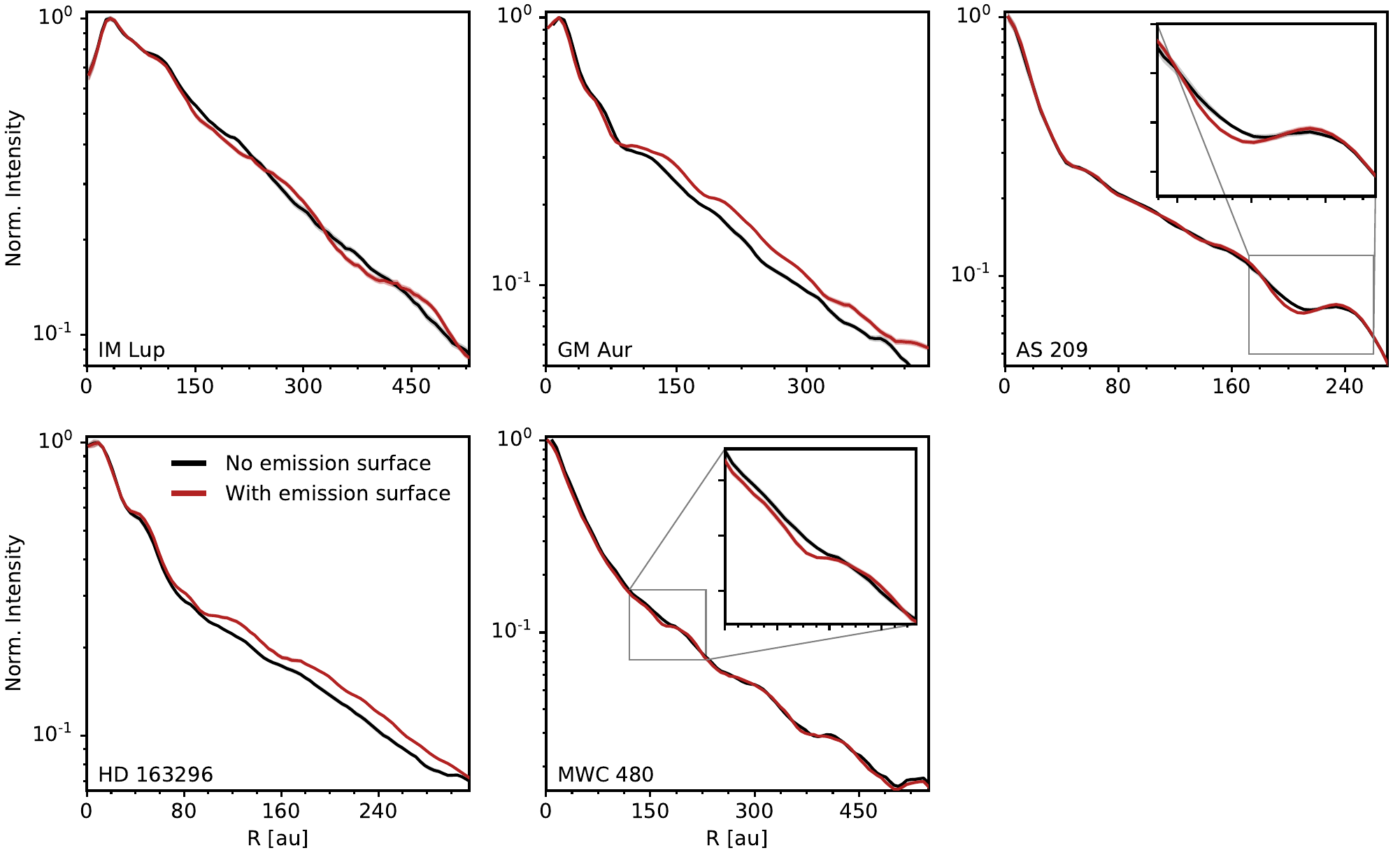}
\caption{Normalized radial intensity profiles for CO 2--1 that are deprojected with (red) and without (black) the emission surfaces derived in \citet{law20a}.}
\label{fig:CO_surfaces}
\end{figure*}

Figure \ref{fig:CO_surfaces} shows the effects of considering emission surfaces, when they could be constrained \citep{law20a}, during the radial profile deprojection process. The CO 2--1 lines are shown for all MAPS disks, as they possess the most elevated emitting surfaces. In all cases, the shape of radial intensity profile is not altered but the inclusion of an emitting surface often leads to the emergence of additional low-contrast features or sharpens the contrasts of existing substructures.

\newpage
\clearpage
\section{Logarithmically-scaled CO radial intensity profiles} \label{sec:app:log_CO_profiles}

\begin{figure*}[!ht]
\centering
\figurenum{30}
\includegraphics[width=\linewidth]{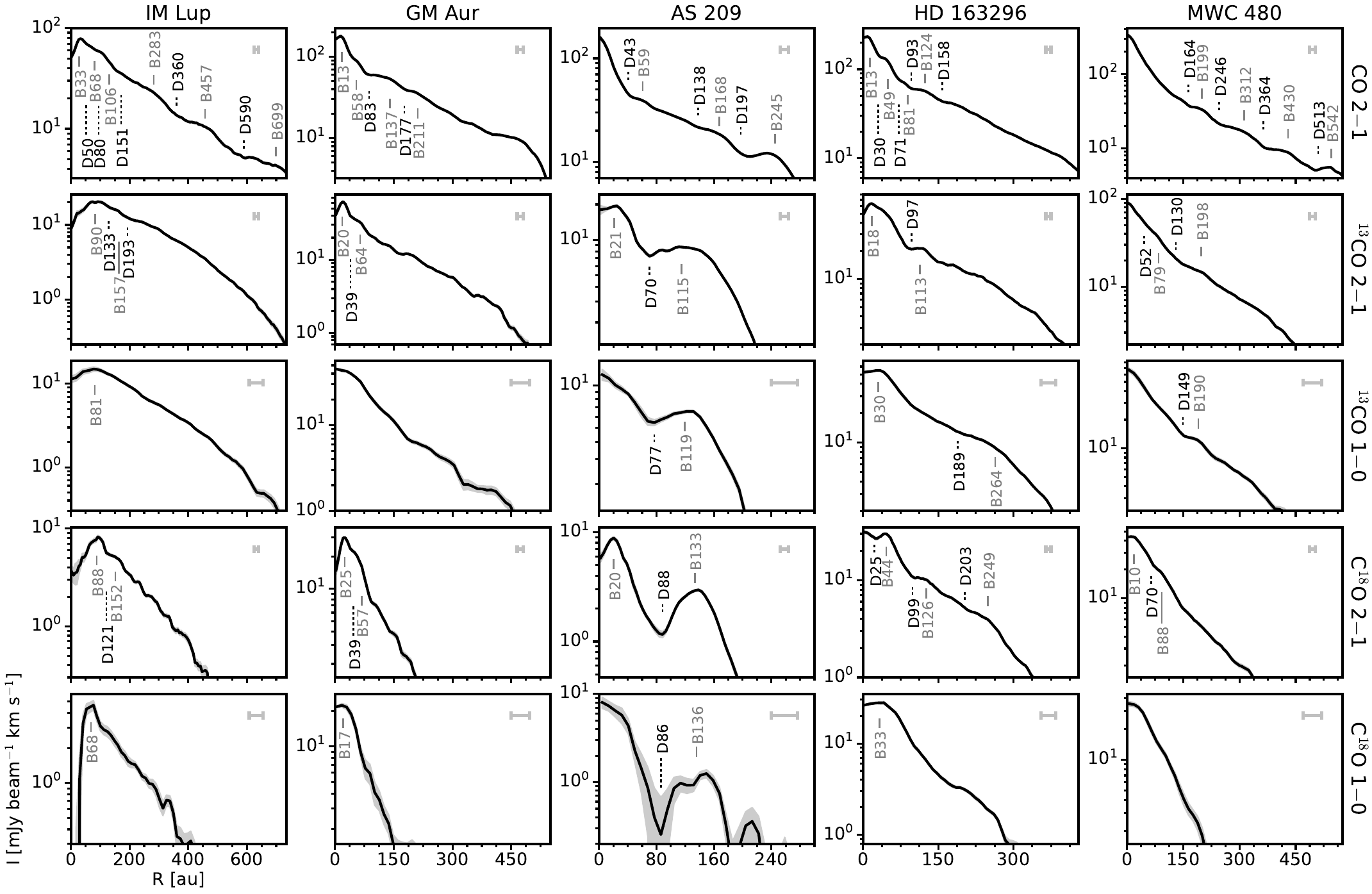}
\caption{Deprojected radial intensity profiles of CO lines, as in Figure \ref{fig:CO_Radial_Profiles}, but logarithmically-scaled to highlight low-contrast substructures.}
\label{fig:CO_log_scale}
\end{figure*}

To better illustrate low-contrast substructures and those at large radii present in the CO isotopologues, Figure \ref{fig:CO_log_scale} shows logarithmically-scaled radial intensity profiles.

\section{Full Tables of Fitted Substructure Characteristics}
\label{sec:app:Table_of_Features}

For readability, Tables \ref{tab:SubstrProp}--\ref{tab:ContinuumSubstrProp} are listed in this Appendix. 

\subsection{Detected features}

A full list of radial chemical substructures is found in Table \ref{tab:SubstrProp}.

\startlongtable
\centerwidetable
\begin{deluxetable*}{lllccccc}
\tablecaption{Properties of Radial Chemical Substructures\label{tab:SubstrProp}}
\tablewidth{0pt}
\tablehead{
\colhead{Source} & \colhead{Line} & \colhead{Feature} & \colhead{r$_0$} & \colhead{r$_0$} & \colhead{Method} & \colhead{Width} & \colhead{Depth}
\vspace{0.2cm}\\ \colhead{}     & \colhead{}     & \colhead{} & \colhead{[mas]} & \colhead{[au]} & & \colhead{[au]} & \colhead{}
\vspace{-0.1cm}\\ \colhead{(1)}     & \colhead{(2)}     & \colhead{(3)} & \colhead{(4)} & \colhead{(5)}  & \colhead{(6)} & \colhead{(7)} & \colhead{(8)} }
\startdata
IM Lup		&CO 2$-$1	&B33$^{\ddag}$	&	210.7~$\pm$~1.0	&	33.3~$\pm$~0.2	&	G	&	67~$\pm$~2	&	$-$\\
			&		&D50	&	${\sim}$316	&	${\sim}$50	&	V	&	$-$	&	$-$\\
			&		&B68	&	${\sim}$430	&	${\sim}$68	&	V	&	$-$	&	$-$\\
			&		&D80	&	${\sim}$506	&	${\sim}$80	&	V	&	$-$	&	$-$\\
			&		&B106	&	673.8~$\pm$~1.0	&	106.5~$\pm$~0.2	&	G	&	93~$\pm$~1	&	$-$\\
			&		&D151	&	${\sim}$956	&	${\sim}$151	&	V	&	$-$	&	$-$\\
			&		&B283	&	1788.6~$\pm$~6.8	&	282.6~$\pm$~1.1	&	G	&	134~$\pm$~3	&	$-$\\
			&		&D360	&	2278.2~$\pm$~11.5	&	360.0~$\pm$~1.8	&	G	&	55~$\pm$~18	&	$-$\\
			&		&B457	&	2893.5~$\pm$~5.8	&	457.2~$\pm$~0.9	&	G	&	49~$\pm$~7	&	$-$\\
			&		&D590	&	${\sim}$3734	&	${\sim}$590	&	V	&	$-$	&	$-$\\
			&		&B699	&	4424.7~$\pm$~0.8	&	699.1~$\pm$~0.1	&	G	&	31~$\pm$~2	&	$-$\\
			&$^{13}$CO 2$-$1	&B90	&	568.7~$\pm$~5.4	&	89.8~$\pm$~0.9	&	G	&	164~$\pm$~6	&	$-$\\
			&		&D133	&	${\sim}$842	&	${\sim}$133	&	V	&	$-$	&	$-$\\
			&		&B157	&	${\sim}$994	&	${\sim}$157	&	V	&	$-$	&	$-$\\
			&		&D193	&	${\sim}$1222	&	${\sim}$193	&	V	&	$-$	&	$-$\\
			&$^{13}$CO 1$-$0	&B81$^{\ddag}$	&	514.8~$\pm$~34.1	&	81.3~$\pm$~5.4	&	G	&	240~$\pm$~5	&	$-$\\
			&C$^{18}$O 2$-$1	&B88	&	559.1~$\pm$~2.6	&	88.3~$\pm$~0.4	&	G	&	51~$\pm$~3	&	$-$\\
			&		&D121	&	${\sim}$766	&	${\sim}$121	&	V	&	$-$	&	$-$\\
			&		&B152	&	959.5~$\pm$~10.6	&	151.6~$\pm$~1.7	&	G	&	96~$\pm$~8	&	$-$\\
			&C$^{18}$O 1$-$0	&B68	&	433.4~$\pm$~8.1	&	68.5~$\pm$~1.3	&	G	&	80~$\pm$~5	&	$-$\\
			&C$_2$H 3$-$2	&B100	&	634.1~$\pm$~11.4	&	100.2~$\pm$~1.8	&	G	&	87~$\pm$~4	&	$-$\\
			&		&D160	&	${\sim}$1013	&	${\sim}$160	&	V	&	$-$	&	$-$\\
			&		&B192	&	1214.1~$\pm$~17.5	&	191.8~$\pm$~2.8	&	G	&	103~$\pm$~8	&	$-$\\
			&		&D361$^{\dag}$	&	2287.5~$\pm$~75.0	&	361.4~$\pm$~11.8	&	R	&	12~$\pm$~2	&	0.96~$\pm$~0.07\\
			&		&B386	&	2441.6~$\pm$~4.7	&	385.8~$\pm$~0.7	&	G	&	141~$\pm$~8	&	$-$\\
			&C$_2$H 1$-$0	&B54	&	343.3~$\pm$~26.8	&	54.2~$\pm$~4.2	&	G	&	55~$\pm$~16	&	$-$\\
			&H$_2$CO 3$-$2	&D30$^{\ddag}$	&	187.5~$\pm$~75.0	&	29.6~$\pm$~11.8	&	R	&	$>$59	&	0.37~$\pm$~0.12\\
			&		&B171	&	1082.0~$\pm$~40.7	&	171.0~$\pm$~6.4	&	G	&	281~$\pm$~13	&	$-$\\
			&		&D278	&	${\sim}$1759	&	${\sim}$278	&	V	&	$-$	&	$-$\\
			&		&B314	&	${\sim}$1987	&	${\sim}$314	&	V	&	$-$	&	$-$\\
			&HCO$^+$ 1$-$0	&B140	&	888.2~$\pm$~6.9	&	140.3~$\pm$~1.1	&	G	&	120~$\pm$~6	&	$-$\\
			&		&D196$^{\dag}$	&	1237.5~$\pm$~75.0	&	195.5~$\pm$~11.8	&	R	&	8~$\pm$~8	&	0.96~$\pm$~0.06\\
			&		&B248	&	1570.7~$\pm$~7.1	&	248.2~$\pm$~1.1	&	G	&	76~$\pm$~1	&	$-$\\
			&		&D314$^{\dag}$	&	1987.5~$\pm$~75.0	&	314.0~$\pm$~11.8	&	R	&	17~$\pm$~2	&	0.92~$\pm$~0.07\\
			&		&B336	&	2124.5~$\pm$~4.1	&	335.7~$\pm$~0.7	&	G	&	80~$\pm$~1	&	$-$\\
			&CS 2$-$1	&B46	&	288.6~$\pm$~2.7	&	45.6~$\pm$~0.4	&	G	&	58~$\pm$~0.2	&	$-$\\
			&		&D101$^{\dag}$	&	637.5~$\pm$~75.0	&	100.7~$\pm$~11.8	&	R	&	46~$\pm$~3	&	0.86~$\pm$~0.11\\
			&		&B140	&	887.8~$\pm$~5.8	&	140.3~$\pm$~0.9	&	G	&	93~$\pm$~4	&	$-$\\
			&		&D267$^{\dag}$	&	1687.5~$\pm$~75.0	&	266.6~$\pm$~11.8	&	R	&	39~$\pm$~2	&	0.54~$\pm$~0.12\\
			&		&B388	&	2454.2~$\pm$~24.1	&	387.8~$\pm$~3.8	&	G	&	292~$\pm$~15	&	$-$\\
			&HCN 3$-$2	&B122	&	769.0~$\pm$~3.4	&	121.5~$\pm$~0.5	&	G	&	126~$\pm$~2	&	$-$\\
			&		&D240	&	${\sim}$1519	&	${\sim}$240	&	V	&	$-$	&	$-$\\
			&		&B298	&	1885.2~$\pm$~5.7	&	297.9~$\pm$~0.9	&	G	&	177~$\pm$~2	&	$-$\\
			&DCN 3$-$2	&B138	&	873.5~$\pm$~84.8	&	138.0~$\pm$~13.4	&	G	&	92~$\pm$~40	&	$-$\\
			&		&D243	&	1537.5~$\pm$~75.0	&	242.9~$\pm$~11.8	&	R	&	130~$\pm$~2	&	0.12~$\pm$~0.15\\
			&		&B349	&	2205.9~$\pm$~9.2	&	348.5~$\pm$~1.4	&	G	&	98~$\pm$~15	&	$-$\\
\hline
GM Aur			&CO 2$-$1	&B13$^{\ddag}$	&	80.1~$\pm$~2.3	&	12.7~$\pm$~0.4	&	G	&	56~$\pm$~1	&	$-$\\
			&		&B58$^{\dag}$	&	366.9~$\pm$~1.5	&	58.3~$\pm$~0.2	&	G	&	14~$\pm$~2	&	$-$\\
			&		&D83	&	521.7~$\pm$~3.8	&	83.0~$\pm$~0.6	&	G	&	28~$\pm$~6	&	$-$\\
			&		&B137	&	860.5~$\pm$~10.4	&	136.8~$\pm$~1.7	&	G	&	31~$\pm$~3	&	$-$\\
			&		&D177$^b$	&	1113.8~$\pm$~2.2	&	177.1~$\pm$~0.4	&	G	&	24~$\pm$~3	&	$-$\\
			&		&B211$^{\dag}{}^b$	&	1329.7~$\pm$~3.0	&	211.4~$\pm$~0.5	&	G	&	20~$\pm$~7	&	$-$\\
			&$^{13}$CO 2$-$1	&B20$^{\ddag}$	&	128.5~$\pm$~2.6	&	20.4~$\pm$~0.4	&	G	&	41~$\pm$~3	&	$-$\\
			&		&D39	&	${\sim}$245	&	${\sim}$39	&	V	&	$-$	&	$-$\\
			&		&B64	&	403.8~$\pm$~1.5	&	64.2~$\pm$~0.2	&	G	&	34~$\pm$~3	&	$-$\\
			&C$^{18}$O 2$-$1	&B25	&	155.9~$\pm$~0.6	&	24.8~$\pm$~0.1	&	G	&	38~$\pm$~3	&	$-$\\
			&		&D39	&	${\sim}$245	&	${\sim}$39	&	V	&	$-$	&	$-$\\
			&		&B57	&	${\sim}$358	&	${\sim}$57	&	V	&	$-$	&	$-$\\
			&C$^{18}$O 1$-$0	&B17$^{\ddag}$	&	106.7~$\pm$~11.9	&	17.0~$\pm$~1.9	&	G	&	98~$\pm$~1	&	$-$\\
			&C$_2$H 3$-$2	&B27	&	168.3~$\pm$~1.9	&	26.8~$\pm$~0.3	&	G	&	47~$\pm$~1	&	$-$\\
			&		&D51$^{\dag}$	&	318.7~$\pm$~37.5	&	50.7~$\pm$~6.0	&	R	&	12~$\pm$~0.2	&	0.82~$\pm$~0.07\\
			&		&B68	&	428.5~$\pm$~0.4	&	68.1~$\pm$~0.1	&	G	&	48~$\pm$~2	&	$-$\\
			&		&D116$^{\dag}$	&	731.2~$\pm$~37.5	&	116.3~$\pm$~6.0	&	R	&	7~$\pm$~4	&	0.93~$\pm$~0.09\\
			&		&B124	&	779.7~$\pm$~1.8	&	124.0~$\pm$~0.3	&	G	&	47~$\pm$~4	&	$-$\\
			&		&D277	&	${\sim}$1742	&	${\sim}$277	&	V	&	67~$\pm$~1	&	$-$\\
			&		&B339	&	2133.6~$\pm$~3.2	&	339.2~$\pm$~0.5	&	G	&	53~$\pm$~8	&	$-$\\
			&C$_2$H 1$-$0	&B34$^{\dag}$	&	216.6~$\pm$~10.9	&	34.4~$\pm$~1.7	&	G	&	47~$\pm$~12	&	$-$\\
			&c-C$_3$H$_2$ 7$-$6	&D54	&	337.5~$\pm$~75.0	&	53.7~$\pm$~11.9	&	R	&	$>$55	&	0.58~$\pm$~0.21\\
			&		&B90	&	567.6~$\pm$~7.3	&	90.3~$\pm$~1.2	&	G	&	85~$\pm$~5	&	$-$\\
			&H$_2$CO 3$-$2	&B22	&	140.3~$\pm$~2.1	&	22.3~$\pm$~0.3	&	G	&	40~$\pm$~7	&	$-$\\
			&		&D51$^{\dag}$	&	318.7~$\pm$~37.5	&	50.7~$\pm$~6.0	&	R	&	14~$\pm$~0.5	&	0.86~$\pm$~0.08\\
			&		&B64	&	402.4~$\pm$~3.0	&	64.0~$\pm$~0.5	&	G	&	40~$\pm$~5	&	$-$\\
			&HCO$^+$ 1$-$0	&D81$^{\dag}$	&	511.9~$\pm$~78.8	&	81.4~$\pm$~12.5	&	R	&	${\sim}$30	&	$-$\\
			&CS 2$-$1	&D113$^{\dag}$	&	711.7~$\pm$~74.9	&	113.2~$\pm$~11.9	&	R	&	15~$\pm$~14	&	$-$\\
			&		&B143	&	899.6~$\pm$~102.3	&	143.0~$\pm$~16.3	&	G	&	129~$\pm$~28	&	$-$\\
			&HCN 3$-$2	&B22$^{\ddag}$	&	136.2~$\pm$~1.7	&	21.7~$\pm$~0.3	&	G	&	52~$\pm$~1	&	$-$\\
			&		&D253$^{\dag}$	&	1593.7~$\pm$~37.5	&	253.4~$\pm$~6.0	&	R	&	20~$\pm$~1	&	$-$\\
			&		&B271	&	1705.2~$\pm$~3.6	&	271.1~$\pm$~0.6	&	G	&	95~$\pm$~2	&	$-$\\
			&HCN 1$-$0	&D169$^{\dag}$	&	1063.1~$\pm$~78.7	&	169.0~$\pm$~12.5	&	R	&	43~$\pm$~5	&	$-$\\
			&		&B266	&	1672.5~$\pm$~46.3	&	265.9~$\pm$~7.4	&	G	&	109~$\pm$~54	&	$-$\\
			&DCN 3$-$2	&B16$^{\ddag}$	&	100.6~$\pm$~2.1	&	16.0~$\pm$~0.3	&	G	&	35~$\pm$~3	&	$-$\\
			&		&D241	&	1518.7~$\pm$~37.5	&	241.5~$\pm$~6.0	&	R	&	$-$	&	$-$\\
			&		&B276	&	1736.7~$\pm$~13.6	&	276.1~$\pm$~2.2	&	G	&	28~$\pm$~16	&	$-$\\
			&HC$_3$N 29$-$28	&B25	&	155.8~$\pm$~0.6	&	24.8~$\pm$~0.1	&	G	&	39~$\pm$~1	&	$-$\\
\hline
AS 209			&CO 2$-$1	&D43	&	${\sim}$355	&	${\sim}$43	&	V	&	$-$	&	$-$\\
			&		&B59$^{\dag}$	&	486.2~$\pm$~1.3	&	58.8~$\pm$~0.2	&	G	&	17~$\pm$~3	&	$-$\\
			&		&D138	&	${\sim}$1140	&	${\sim}$138	&	V	&	$-$	&	$-$\\
			&		&B168	&	1385.9~$\pm$~7.8	&	167.7~$\pm$~0.9	&	G	&	23~$\pm$~4	&	$-$\\
			&		&D197	&	1631.4~$\pm$~0.4	&	197.4~$\pm$~0.04	&	G	&	21~$\pm$~2	&	$-$\\
			&		&B245	&	2026.3~$\pm$~1.4	&	245.2~$\pm$~0.2	&	G	&	44~$\pm$~1	&	$-$\\
			&$^{13}$CO 2$-$1	&B21$^{\ddag}$	&	174.7~$\pm$~2.7	&	21.1~$\pm$~0.3	&	G	&	48~$\pm$~0.3	&	$-$\\
			&		&D70	&	581.2~$\pm$~37.5	&	70.3~$\pm$~4.5	&	R	&	19~$\pm$~1	&	0.84~$\pm$~0.02\\
			&		&B115	&	950.3~$\pm$~1.0	&	115.0~$\pm$~0.1	&	G	&	142~$\pm$~1	&	$-$\\
			&$^{13}$CO 1$-$0	&D77$^{\dag}$	&	637.5~$\pm$~75.0	&	77.1~$\pm$~9.1	&	R	&	32~$\pm$~2	&	0.83~$\pm$~0.04\\
			&		&B119	&	987.2~$\pm$~3.4	&	119.5~$\pm$~0.4	&	G	&	103~$\pm$~2	&	$-$\\
			&C$^{18}$O 2$-$1	&B20	&	167.4~$\pm$~2.2	&	20.2~$\pm$~0.3	&	G	&	40~$\pm$~3	&	$-$\\
			&		&D88	&	731.2~$\pm$~37.5	&	88.5~$\pm$~4.5	&	R	&	47~$\pm$~0.1	&	0.40~$\pm$~0.03\\
			&		&B133	&	1102.8~$\pm$~1.0	&	133.4~$\pm$~0.1	&	G	&	66~$\pm$~1	&	$-$\\
			&C$^{18}$O 1$-$0	&D86$^{\dag}$	&	712.5~$\pm$~75.0	&	86.2~$\pm$~9.1	&	R	&	27~$\pm$~8	&	0.28~$\pm$~0.45\\
			&		&B136	&	1122.4~$\pm$~30.8	&	135.8~$\pm$~3.7	&	G	&	72~$\pm$~12	&	$-$\\
			&C$_2$H 3$-$2	&B63	&	522.7~$\pm$~1.7	&	63.2~$\pm$~0.2	&	G	&	70~$\pm$~1	&	$-$\\
			&C$_2$H 1$-$0	&B38	&	314.8~$\pm$~10.4	&	38.1~$\pm$~1.3	&	G	&	55~$\pm$~9	&	$-$\\
			&		&D77$^{\dag}$	&	637.5~$\pm$~75.0	&	77.1~$\pm$~9.1	&	R	&	22~$\pm$~1	&	0.48~$\pm$~0.30\\
			&		&B103$^{\dag}$	&	853.9~$\pm$~4.4	&	103.3~$\pm$~0.5	&	G	&	30~$\pm$~2	&	$-$\\
			&c-C$_3$H$_2$ 7$-$6	&B23	&	193.9~$\pm$~7.7	&	23.5~$\pm$~0.9	&	G	&	34~$\pm$~6	&	$-$\\
			&		&D34$^{\dag}$	&	281.2~$\pm$~37.5	&	34.0~$\pm$~4.5	&	R	&	9~$\pm$~0.3	&	0.87~$\pm$~0.15\\
			&		&B71	&	583.4~$\pm$~2.1	&	70.6~$\pm$~0.3	&	G	&	57~$\pm$~2	&	$-$\\
			&H$_2$CO 3$-$2	&B102	&	839.8~$\pm$~1.0	&	101.6~$\pm$~0.1	&	G	&	63~$\pm$~2	&	$-$\\
			&		&D157$^{\dag}$	&	1293.6~$\pm$~37.5	&	156.5~$\pm$~4.5	&	R	&	13~$\pm$~1	&	0.92~$\pm$~0.06\\
			&		&B169	&	1396.4~$\pm$~8.0	&	169.0~$\pm$~1.0	&	G	&	51~$\pm$~1	&	$-$\\
			&HCO$^+$ 1$-$0	&B23	&	188.3~$\pm$~16.8	&	22.8~$\pm$~2.0	&	G	&	42~$\pm$~1	&	$-$\\
			&		&D41$^{\dag}$	&	337.2~$\pm$~74.9	&	40.8~$\pm$~9.1	&	R	&	25~$\pm$~2	&	0.58~$\pm$~0.07\\
			&		&B75	&	623.8~$\pm$~6.3	&	75.5~$\pm$~0.8	&	G	&	78~$\pm$~1	&	$-$\\
			&CS 2$-$1	&B35$^{\ddag}$	&	291.5~$\pm$~5.7	&	35.3~$\pm$~0.7	&	G	&	82~$\pm$~0.3	&	$-$\\
			&CN 1$-$0	&B40	&	328.6~$\pm$~19.7	&	39.8~$\pm$~2.4	&	G	&	53~$\pm$~3	&	$-$\\
			&		&D50$^{\dag}$	&	412.5~$\pm$~75.0	&	49.9~$\pm$~9.1	&	R	&	14~$\pm$~8	&	0.91~$\pm$~0.15\\
			&		&B77	&	636.1~$\pm$~2.2	&	77.0~$\pm$~0.3	&	G	&	50~$\pm$~1	&	$-$\\
			&HCN 3$-$2	&B43$^{\ddag}$	&	359.1~$\pm$~9.0	&	43.4~$\pm$~1.1	&	G	&	96~$\pm$~2	&	$-$\\
			&HCN 1$-$0	&D14	&	112.5~$\pm$~75.0	&	13.6~$\pm$~9.1	&	R	&	$>$27	&	0.39~$\pm$~0.08\\
			&		&B44	&	363.1~$\pm$~4.1	&	43.9~$\pm$~0.5	&	G	&	37~$\pm$~1	&	$-$\\
			&		&D68	&	${\sim}$562	&	${\sim}$68	&	V	&	$-$	&	$-$\\
			&		&B89	&	739.2~$\pm$~1.6	&	89.4~$\pm$~0.2	&	G	&	64~$\pm$~1	&	$-$\\
			&DCN 3$-$2	&B63	&	521.7~$\pm$~6.1	&	63.1~$\pm$~0.7	&	G	&	74~$\pm$~3	&	$-$\\
			&HC$_3$N 29$-$28	&B35	&	287.1~$\pm$~0.4	&	34.7~$\pm$~0.05	&	G	&	27~$\pm$~3	&	$-$\\
			&HC$_3$N 11$-$10	&D14$^{\ddag}$	&	112.5~$\pm$~75.0	&	13.6~$\pm$~9.1	&	R	&	$>$29	&	0.67~$\pm$~0.11\\
			&		&B48$^{\ddag}$	&	397.5~$\pm$~21.6	&	48.1~$\pm$~2.6	&	G	&	99~$\pm$~5	&	$-$\\
			&CH$_3$CN 12$-$11	&B53	&	436.2~$\pm$~102.1	&	52.8~$\pm$~12.4	&	G	&	72~$\pm$~25	&	$-$\\
\hline
HD 163296			&CO 2$-$1	&B13$^{\dag}$	&	124.7~$\pm$~3.7	&	12.6~$\pm$~0.4	&	G	&	9~$\pm$~3	&	$-$\\
			&		&D30$^{\dag}$	&	299.4~$\pm$~4.6	&	30.2~$\pm$~0.5	&	G	&	13~$\pm$~2	&	$-$\\
			&		&B49	&	489.8~$\pm$~1.9	&	49.5~$\pm$~0.2	&	G	&	16~$\pm$~1	&	$-$\\
			&		&D71$^{\dag}$	&	704.8~$\pm$~1.9	&	71.2~$\pm$~0.2	&	G	&	14~$\pm$~2	&	$-$\\
			&		&B81	&	${\sim}$802	&	${\sim}$81	&	V	&	$-$	&	$-$\\
			&		&D93	&	924.7~$\pm$~6.2	&	93.4~$\pm$~0.6	&	G	&	17~$\pm$~3	&	$-$\\
			&		&B124	&	1222.9~$\pm$~5.8	&	123.5~$\pm$~0.6	&	G	&	25~$\pm$~3	&	$-$\\
			&		&D158	&	1568.1~$\pm$~2.7	&	158.4~$\pm$~0.3	&	G	&	24~$\pm$~3	&	$-$\\
			&$^{13}$CO 2$-$1	&B18$^{\ddag}$	&	176.1~$\pm$~5.6	&	17.8~$\pm$~0.6	&	G	&	54~$\pm$~2	&	$-$\\
			&		&D97$^{\dag}$	&	956.2~$\pm$~37.5	&	96.6~$\pm$~3.8	&	R	&	9~$\pm$~1	&	0.98~$\pm$~0.02\\
			&		&B113	&	1120.3~$\pm$~2.2	&	113.2~$\pm$~0.2	&	G	&	85~$\pm$~1	&	$-$\\
			&$^{13}$CO 1$-$0	&B30$^{\ddag}$	&	299.9~$\pm$~6.6	&	30.3~$\pm$~0.7	&	G	&	89~$\pm$~1	&	$-$\\
			&		&D189	&	1868.3~$\pm$~6.4	&	188.7~$\pm$~0.6	&	G	&	49~$\pm$~14	&	$-$\\
			&		&B264	&	2609.2~$\pm$~31.0	&	263.5~$\pm$~3.1	&	G	&	67~$\pm$~10	&	$-$\\
			&C$^{18}$O 2$-$1	&D25	&	243.7~$\pm$~37.5	&	24.6~$\pm$~3.8	&	R	&	19~$\pm$~1	&	0.89~$\pm$~0.03\\
			&		&B44	&	431.1~$\pm$~2.1	&	43.5~$\pm$~0.2	&	G	&	61~$\pm$~3	&	$-$\\
			&		&D99$^{\dag}$	&	980.1~$\pm$~3.8	&	99.0~$\pm$~0.4	&	G	&	13~$\pm$~3	&	$-$\\
			&		&B126	&	1248.5~$\pm$~4.7	&	126.1~$\pm$~0.5	&	G	&	20~$\pm$~6	&	$-$\\
			&		&D203	&	2006.3~$\pm$~22.1	&	202.6~$\pm$~2.2	&	G	&	28~$\pm$~7	&	$-$\\
			&		&B249	&	2468.4~$\pm$~2.4	&	249.3~$\pm$~0.2	&	G	&	39~$\pm$~2	&	$-$\\
			&C$^{18}$O 1$-$0	&B33$^{\ddag}$	&	322.9~$\pm$~8.7	&	32.6~$\pm$~0.9	&	G	&	102~$\pm$~6	&	$-$\\
			&C$_2$H 3$-$2	&B45	&	444.3~$\pm$~1.0	&	44.9~$\pm$~0.1	&	G	&	34~$\pm$~1	&	$-$\\
			&		&D81	&	806.2~$\pm$~37.5	&	81.4~$\pm$~3.8	&	R	&	27~$\pm$~0.5	&	0.49~$\pm$~0.01\\
			&		&B110	&	1087.6~$\pm$~2.4	&	109.8~$\pm$~0.2	&	G	&	49~$\pm$~0.1	&	$-$\\
			&		&D223$^a$	&	2212.5~$\pm$~75.0	&	223.5~$\pm$~7.6	&	R	&	28~$\pm$~8	&	0.55~$\pm$~0.15\\
			&		&B244$^a$	&	2418.6~$\pm$~0.6	&	244.3~$\pm$~0.1	&	G	&	37~$\pm$~2	&	$-$\\
			&		&D269$^a$	&	2662.5~$\pm$~75.0	&	268.9~$\pm$~7.6	&	R	&	71~$\pm$~56	&	0.23~$\pm$~0.06\\
			&		&B368$^a$	&	3638.7~$\pm$~4.1	&	367.5~$\pm$~0.4	&	G	&	116~$\pm$~3	&	$-$\\
			&C$_2$H 1$-$0	&B41	&	407.4~$\pm$~2.0	&	41.1~$\pm$~0.2	&	G	&	40~$\pm$~1	&	$-$\\
			&		&D87$^{\dag}$	&	862.5~$\pm$~75.0	&	87.1~$\pm$~7.6	&	R	&	18~$\pm$~1	&	0.08~$\pm$~0.15\\
			&		&B111	&	1094.7~$\pm$~8.3	&	110.6~$\pm$~0.8	&	G	&	35~$\pm$~2	&	$-$\\
			&c-C$_3$H$_2$ 7$-$6	&B42	&	414.4~$\pm$~21.5	&	41.9~$\pm$~2.2	&	G	&	44~$\pm$~6	&	$-$\\
			&		&D97	&	956.2~$\pm$~37.5	&	96.6~$\pm$~3.8	&	R	&	${\sim}$30	&	$-$\\
			&		&B115	&	1142.0~$\pm$~0.9	&	115.3~$\pm$~0.1	&	G	&	29~$\pm$~1	&	$-$\\
			&H$_2$CO 3$-$2	&B61	&	603.3~$\pm$~15.7	&	60.9~$\pm$~1.6	&	G	&	46~$\pm$~5	&	$-$\\
			&		&D95	&	${\sim}$941	&	${\sim}$95	&	V	&	$-$	&	$-$\\
			&		&B111	&	1095.8~$\pm$~14.2	&	110.7~$\pm$~1.4	&	G	&	85~$\pm$~3	&	$-$\\
			&		&D170	&	1687.5~$\pm$~75.0	&	170.4~$\pm$~7.6	&	R	&	47~$\pm$~0.2	&	0.78~$\pm$~0.07\\
			&		&B230	&	2275.3~$\pm$~11.4	&	229.8~$\pm$~1.2	&	G	&	157~$\pm$~1	&	$-$\\
			&		&D337	&	${\sim}$3337	&	${\sim}$337	&	V	&	$-$	&	$-$\\
			&		&B378	&	3746.9~$\pm$~4.0	&	378.4~$\pm$~0.4	&	G	&	114~$\pm$~3	&	$-$\\
			&HCO$^+$ 1$-$0	&B39	&	386.6~$\pm$~0.9	&	39.1~$\pm$~0.1	&	G	&	50~$\pm$~4	&	$-$\\
			&		&D72	&	${\sim}$713	&	${\sim}$72	&	V	&	$-$	&	$-$\\
			&		&B101	&	996.7~$\pm$~6.3	&	100.7~$\pm$~0.6	&	G	&	115~$\pm$~1	&	$-$\\
			&		&D193	&	1912.5~$\pm$~75.0	&	193.2~$\pm$~7.6	&	R	&	63~$\pm$~1	&	0.77~$\pm$~0.10\\
			&		&B292	&	2892.3~$\pm$~14.8	&	292.1~$\pm$~1.5	&	G	&	178~$\pm$~9	&	$-$\\
			&CS 2$-$1	&B53$^{\ddag}$	&	528.2~$\pm$~25.5	&	53.3~$\pm$~2.6	&	G	&	129~$\pm$~4	&	$-$\\
			&CN 1$-$0	&B37	&	371.0~$\pm$~0.7	&	37.5~$\pm$~0.1	&	G	&	57~$\pm$~2	&	$-$\\
			&		&D95$^{\dag}$	&	937.5~$\pm$~75.0	&	94.7~$\pm$~7.6	&	R	&	20~$\pm$~3	&	0.90~$\pm$~0.06\\
			&		&B118	&	1173.0~$\pm$~6.1	&	118.5~$\pm$~0.6	&	G	&	75~$\pm$~0.3	&	$-$\\
			&		&D322	&	3187.5~$\pm$~75.0	&	321.9~$\pm$~7.6	&	R	&	46~$\pm$~0.4	&	0.62~$\pm$~0.04\\
			&		&B391	&	3871.6~$\pm$~6.0	&	391.0~$\pm$~0.6	&	G	&	168~$\pm$~3	&	$-$\\
			&HCN 3$-$2	&B30	&	299.3~$\pm$~3.8	&	30.2~$\pm$~0.4	&	G	&	55~$\pm$~1	&	$-$\\
			&		&D85	&	843.8~$\pm$~37.5	&	85.2~$\pm$~3.8	&	R	&	22~$\pm$~0.3	&	0.75~$\pm$~0.02\\
			&		&B109	&	1079.2~$\pm$~0.0	&	109.0~$\pm$~0.00	&	G	&	51~$\pm$~1	&	$-$\\
			&		&D214	&	2118.8~$\pm$~37.5	&	214.0~$\pm$~3.8	&	R	&	36~$\pm$~0.2	&	0.95~$\pm$~0.02\\
			&		&B253	&	2506.2~$\pm$~2.8	&	253.1~$\pm$~0.3	&	G	&	108~$\pm$~2	&	$-$\\
			&		&D324	&	3206.2~$\pm$~37.5	&	323.8~$\pm$~3.8	&	R	&	26~$\pm$~0.3	&	0.83~$\pm$~0.03\\
			&		&B357	&	3536.7~$\pm$~3.4	&	357.2~$\pm$~0.3	&	G	&	106~$\pm$~1	&	$-$\\
			&HCN 1$-$0	&B18$^{\ddag}$	&	173.9~$\pm$~3.0	&	17.6~$\pm$~0.3	&	G	&	54~$\pm$~1	&	$-$\\
			&		&D87$^{\dag}$	&	862.5~$\pm$~75.0	&	87.1~$\pm$~7.6	&	R	&	19~$\pm$~0.3	&	0.42~$\pm$~0.15\\
			&		&B107	&	1055.5~$\pm$~1.7	&	106.6~$\pm$~0.2	&	G	&	33~$\pm$~4	&	$-$\\
			&		&D186$^{\dag}$	&	1837.5~$\pm$~75.0	&	185.6~$\pm$~7.6	&	R	&	22~$\pm$~5	&	0.44~$\pm$~0.15\\
			&		&B242	&	2398.0~$\pm$~41.4	&	242.2~$\pm$~4.2	&	G	&	99~$\pm$~21	&	$-$\\
			&		&D330	&	3262.5~$\pm$~75.0	&	329.5~$\pm$~7.6	&	R	&	55~$\pm$~2	&	0.30~$\pm$~0.19\\
			&		&B402	&	3983.2~$\pm$~29.2	&	402.3~$\pm$~2.9	&	G	&	124~$\pm$~10	&	$-$\\
			&DCN 3$-$2	&B31$^{\ddag}$	&	307.0~$\pm$~90.4	&	31.0~$\pm$~9.1	&	G	&	72~$\pm$~14	&	$-$\\
			&		&D87$^{\dag}$	&	862.5~$\pm$~75.0	&	87.1~$\pm$~7.6	&	R	&	28~$\pm$~0.2	&	0.59~$\pm$~0.12\\
			&		&B118	&	1168.3~$\pm$~8.5	&	118.0~$\pm$~0.9	&	G	&	45~$\pm$~1	&	$-$\\
			&HC$_3$N 29$-$28	&B37	&	362.5~$\pm$~1.5	&	36.6~$\pm$~0.1	&	G	&	43~$\pm$~1	&	$-$\\
			&HC$_3$N 11$-$10	&D19	&	187.5~$\pm$~75.0	&	18.9~$\pm$~7.6	&	R	&	$>$27	&	0.79~$\pm$~0.26\\
			&		&B40	&	397.1~$\pm$~22.1	&	40.1~$\pm$~2.2	&	G	&	52~$\pm$~3	&	$-$\\
			&CH$_3$CN 12$-$11	&B35	&	344.5~$\pm$~1.2	&	34.8~$\pm$~0.1	&	G	&	33~$\pm$~0.3	&	$-$\\
\hline
MWC 480			&CO 2$-$1	&D164$^{\dag}$	&	1013.8~$\pm$~5.1	&	164.0~$\pm$~0.8	&	G	&	20~$\pm$~3	&	$-$\\
			&		&B199	&	1231.3~$\pm$~0.7	&	199.2~$\pm$~0.1	&	G	&	28~$\pm$~3	&	$-$\\
			&		&D246	&	1517.7~$\pm$~7.1	&	245.6~$\pm$~1.1	&	G	&	43~$\pm$~4	&	$-$\\
			&		&B312	&	1931.2~$\pm$~3.4	&	312.5~$\pm$~0.6	&	G	&	45~$\pm$~4	&	$-$\\
			&		&D364$^{\dag}$	&	2247.7~$\pm$~2.6	&	363.7~$\pm$~0.4	&	G	&	20~$\pm$~2	&	$-$\\
			&		&B430	&	2655.7~$\pm$~1.6	&	429.7~$\pm$~0.3	&	G	&	62~$\pm$~3	&	$-$\\
			&		&D513$^{\dag}$	&	${\sim}$3171	&	${\sim}$513	&	V	&	20~$\pm$~2	&	0.97~$\pm$~0.01\\
			&		&B542	&	3350.7~$\pm$~1.8	&	542.1~$\pm$~0.3	&	G	&	55~$\pm$~3	&	$-$\\
			&$^{13}$CO 2$-$1	&D52	&	${\sim}$321	&	${\sim}$52	&	V	&	$-$	&	$-$\\
			&		&B79$^{\dag}$	&	490.4~$\pm$~5.5	&	79.3~$\pm$~0.9	&	G	&	19~$\pm$~8	&	$-$\\
			&		&D130	&	806.0~$\pm$~37.0	&	130.4~$\pm$~6.0	&	G	&	66~$\pm$~33	&	$-$\\
			&		&B198	&	1223.9~$\pm$~2.2	&	198.0~$\pm$~0.4	&	G	&	37~$\pm$~10	&	$-$\\
			&$^{13}$CO 1$-$0	&D149$^{\dag}$	&	921.0~$\pm$~6.2	&	149.0~$\pm$~1.0	&	G	&	25~$\pm$~7	&	$-$\\
			&		&B190$^{\dag}$	&	1174.1~$\pm$~6.3	&	190.0~$\pm$~1.0	&	G	&	28~$\pm$~13	&	$-$\\
			&C$^{18}$O 2$-$1	&B10$^{\ddag}$	&	61.8~$\pm$~14.1	&	10.0~$\pm$~2.3	&	G	&	104~$\pm$~7	&	$-$\\
			&		&D70	&	${\sim}$433	&	${\sim}$70	&	V	&	$-$	&	$-$\\
			&		&B88	&	${\sim}$544	&	${\sim}$88	&	V	&	$-$	&	$-$\\
			&C$_2$H 3$-$2	&B73	&	449.9~$\pm$~3.4	&	72.8~$\pm$~0.5	&	G	&	64~$\pm$~2	&	$-$\\
			&C$_2$H 1$-$0	&B62	&	385.7~$\pm$~16.1	&	62.4~$\pm$~2.6	&	G	&	54~$\pm$~8	&	$-$\\
			&c-C$_3$H$_2$ 7$-$6	&B76	&	472.0~$\pm$~5.4	&	76.4~$\pm$~0.9	&	G	&	65~$\pm$~3	&	$-$\\
			&H$_2$CO 3$-$2	&D55$^{\dag}$	&	337.5~$\pm$~75.0	&	54.6~$\pm$~12.1	&	R	&	25~$\pm$~3	&	0.75~$\pm$~0.17\\
			&		&B86	&	531.5~$\pm$~10.0	&	86.0~$\pm$~1.6	&	G	&	77~$\pm$~16	&	$-$\\
			&HCO$^+$ 1$-$0	&D19$^{\dag}$	&	118.1~$\pm$~78.7	&	19.1~$\pm$~12.7	&	R	&	33~$\pm$~0.2	&	0.17~$\pm$~0.24\\
			&		&B65	&	401.0~$\pm$~4.8	&	64.9~$\pm$~0.8	&	G	&	62~$\pm$~4	&	$-$\\
			&		&D248	&	1535.6~$\pm$~78.7	&	248.5~$\pm$~12.7	&	R	&	52~$\pm$~2	&	0.13~$\pm$~0.16\\
			&		&B330	&	2039.4~$\pm$~8.8	&	330.0~$\pm$~1.4	&	G	&	143~$\pm$~34	&	$-$\\
			&CS 2$-$1	&B76	&	468.1~$\pm$~34.7	&	75.7~$\pm$~5.6	&	G	&	82~$\pm$~12	&	$-$\\
			&CN 1$-$0	&D31	&	190.6~$\pm$~76.2	&	30.8~$\pm$~12.3	&	R	&	$>$44	&	0.57~$\pm$~0.17\\
			&		&B71	&	436.1~$\pm$~8.1	&	70.6~$\pm$~1.3	&	G	&	76~$\pm$~4	&	$-$\\
			&HCN 3$-$2	&D58	&	${\sim}$358	&	${\sim}$58	&	V	&	$-$	&	$-$\\
			&		&B79	&	489.0~$\pm$~5.2	&	79.1~$\pm$~0.8	&	G	&	54~$\pm$~1	&	$-$\\
			&HCN 1$-$0	&D57	&	${\sim}$352	&	${\sim}$57	&	V	&	$-$	&	$-$\\
			&		&B78$^{\dag}$	&	479.5~$\pm$~20.8	&	77.6~$\pm$~3.4	&	G	&	36~$\pm$~2	&	$-$\\
			&DCN 3$-$2	&B67	&	417.1~$\pm$~2.1	&	67.5~$\pm$~0.3	&	G	&	75~$\pm$~3	&	$-$\\
			&HC$_3$N 29$-$28	&D33	&	${\sim}$204	&	${\sim}$33	&	V	&	$-$	&	$-$\\
			&		&B46$^{\dag}$	&	282.6~$\pm$~10.0	&	45.7~$\pm$~1.6	&	G	&	24~$\pm$~4	&	$-$\\
			&		&D64	&	${\sim}$396	&	${\sim}$64	&	V	&	$-$	&	$-$\\
			&		&B76	&	470.0~$\pm$~8.6	&	76.0~$\pm$~1.4	&	G	&	44~$\pm$~2	&	$-$\\
			&HC$_3$N 11$-$10	&B30$^{\dag}$	&	184.1~$\pm$~31.2	&	29.8~$\pm$~5.1	&	G	&	47~$\pm$~4	&	$-$\\
			&		&D55	&	${\sim}$340	&	${\sim}$55	&	V	&	$-$	&	$-$\\
			&		&B81	&	503.4~$\pm$~22.5	&	81.5~$\pm$~3.6	&	G	&	56~$\pm$~10	&	$-$\\
			&CH$_3$CN 12$-$11	&B33$^{\ddag}$	&	202.6~$\pm$~16.2	&	32.8~$\pm$~2.6	&	G	&	91~$\pm$~2	&	$-$\\
\enddata
\tablecomments{Column descriptions: (1) Name of host star. (2) Name of line. (3) Substructure label: ``B" (``bright") prefix refers to rings and ``D" (``dark") refers to gaps. (4) Radial location of substructure in mas (the uncertainties in mas are simply scaled from the fitting procedure and do not account for the uncertainty in the distance to the source). (5) Radial location of substructure in au. (6) Method used to derive radial location of substructure: ``G" indicates Gaussian-fitting, ``R" indicates identification of local extrema in the radial profiles, and ``V" indicates identification through visual inspection. (7) Width of substructure. (8) Depth of gap, defined as the intensity ratio of adjacent ring-gap pairs (see Section \ref{sec:widths_and_contrast}). All uncertainties are 1$\sigma$.}
\tablenotetext{$\dag$}{Width of feature is narrower than the FWHM of the synthesized beam (Table \ref{tab:ProfSelection}) and should be considered an upper limit.}
\tablenotetext{$\ddag$}{Width of feature results in an unphysical, negative inner radius, i.e., r$_0 - 0.5 \times$ FWHM $<$ 0.}
\tablenotetext{a}{Fit using the 0\farcs3 tapered radial profile with a $\pm$30$^{\circ}$ wedge due to the low SNR of these features (see Section \ref{sec:tentative_substructures}).}
\tablenotetext{b}{Potentially nonaxisymmetric substructures from spiral arms \citep[see ][]{huang20_spiral_arms}.}
\end{deluxetable*}

\subsection{Tentative features}

A list of tentative radial chemical substructures is found in Table \ref{tab:tentative_radial_substructures}.

\startlongtable
\centerwidetable
\begin{deluxetable*}{lllccccc}
\tablecaption{List of Tentative and Asymmetric Radial Substructures\label{tab:tentative_radial_substructures}}
\tablewidth{0pt}
\tablehead{
\colhead{Source} & \colhead{Line} & \colhead{Feature} & \colhead{Nearest ID} & \colhead{r$_0$} & \colhead{r$_0$} & \colhead{Comments}  
\vspace{0.2cm}\\ \colhead{}         & \colhead{} & \colhead{} & \colhead{} & \colhead{[mas]} & \colhead{[au]} & 
\vspace{-0.1cm}\\ \colhead{(1)}     & \colhead{(2)}     & \colhead{(3)} & \colhead{(4)}   & \colhead{(5)} & \colhead{(6)} & \colhead{(7)}   }
\startdata
IM~Lup & HCN 1$-$0 & Plateau & $\ldots$ & 1.27--3.32 & 200--525 & &  \\
GM~Aur & HCO$^+$ 1--0 & Plateau & $\ldots$ & 0.63--2.70 & 100--430 & \\
       & H$_2$CO 3--2 & Plateau & $\ldots$ & 0.94--2.70 & 150--430 & \\
       & CN 1$-$0 & Shoulder &$\ldots$ & 70 & 0.44 & coincident with B64, H$_2$CO 3--2 \\
       & & Shoulder & $\ldots$ & 170 & 1.68 & coincident with B143, CS 2--1  \\
       &  & Plateau & $\ldots$ & 1.38--2.83 & 220--450 & \\
AS~209 & HCO$^+$ 1--0 & Shoulder & $\ldots$ & 0.99 & 120 & \\
       & HCN 3$-$2 & Radial asymmetry & B43 & 0.36 & 43 & outward-sloping tail \\
       & HC$_3$N 29--28 & Shoulder & $\ldots$ & 0.74 & 90 & coincident with B89, HCN 1--0 \\
& HC$_3$N 11--10 & Radial asymmetry & B48 & 0.40 & 48 & outward-sloping tail \\
HD~163296 & C$_2$H 3--2 & Shoulder & $\ldots$ & 0.10 & 10 \\
          & CS 2$-1$ & Radial asymmetry & B53 & 0.52 & 53 & outward-sloping tail \\
          & HC$_3$N 11--10 & Shoulder & $\ldots$ & 0.99 & 100 & coincident with B109, HCN 3--2; B107, HCN 1--0 \\
MWC~480 & C$_2$H 3--2 & Shoulder & $\ldots$ & 0.06 & 10 & \\
        & CN 1--0 & Plateau & $\ldots$ & 1.55--3.40 & 250--550 \\
        & DCN 3--2 & Radial asymmetry & B67 & 0.41 & 67 & outward-sloping tail \\ 
\enddata
\tablecomments{Column descriptions: (1) Name of host star. (2) Name of line. (3) Tentative substructure. ``Shoulder" refers to localized emission plateau, while ``plateau" refer to emission over a large radial extent. ``Radial asymmetry" indicates an asymmetrical feature, as identified in the radial profiles. (4) Name of nearest identified substructure. (5) Radial location of substructure in mas. (6) Radial location of substructure in au. Radial locations are visually determined and approximate in nature. Formal uncertainties are not listed. (7) Comments about the tentative substructure.}
\end{deluxetable*}

\subsection{Continuum substructures}

A list of annular continuum substructures is found in Table \ref{tab:ContinuumSubstrProp}.

\centerwidetable
\begin{deluxetable*}{lllccccc}
\tablecaption{Properties of Annular Continuum Substructures\label{tab:ContinuumSubstrProp}}
\tablewidth{0pt}
\tablehead{
\colhead{Source} & \colhead{Feature} & \colhead{r$_0$} & \colhead{r$_0$} & \colhead{Method} & \colhead{Width} & \colhead{Depth} & \colhead{Ref} 
\vspace{0.2cm}\\ \colhead{}     & \colhead{} & \colhead{[mas]} & \colhead{[au]} & & \colhead{[au]} & \colhead{} & 
\vspace{-0.1cm}\\ \colhead{(1)}     & \colhead{(2)}     & \colhead{(3)} & \colhead{(4)} & \colhead{(5)}  & \colhead{(6)} & \colhead{(7)} & \colhead{(8)}  }
\startdata
IM Lup		&D116	&	735~$\pm$~23	&	116.18~$\pm$~3.57	&	R	&	13.1~$\pm$~0.2	&	0.83~$\pm$~0.01	&	D117 (1)\\
			&B133	&	844~$\pm$~0.2	&	133.28~$\pm$~0.03	&	G	&	50.2~$\pm$~0.3	&	$-$	&	B134 (1)\\
			&D209	&	${\sim}$1323	&	${\sim}$209	&	V	&	$-$	&	$-$	&	\\
			&B220	&	${\sim}$1392	&	${\sim}$220	&	V	&	$-$	&	$-$	&	\\
		&R$_{\rm{edge}}$	&	${\sim}$2025	&	${\sim}$320	&	V	&	$-$	&	$-$	&	\\
\hline
GM Aur		&\textit{D15}	&	\textit{94}	&	\textit{15.00}	&	\textit{$-$}	&	\textit{20.0}	&	\textit{$-$}	&	D15 (2)\\
			&B42	&	264~$\pm$~1	&	41.93~$\pm$~0.18	&	G	&	31.8~$\pm$~0.3	&	$-$	&	B40 (2)\\
			&D68	&	425~$\pm$~29	&	67.56~$\pm$~4.66	&	R	&	13.5~$\pm$~0.2	&	0.73~$\pm$~0.05	&	D67 (2)\\
			&B86	&	539~$\pm$~1	&	85.66~$\pm$~0.22	&	G	&	31.1~$\pm$~1.0	&	$-$	&	B84 (2)\\
			&D142	&	894~$\pm$~29	&	142.10~$\pm$~4.66	&	R	&	16.9~$\pm$~0.5	&	0.96~$\pm$~0.01	&	D145 (2)\\
			&B163	&	1022~$\pm$~2	&	162.51~$\pm$~0.27	&	G	&	69.0~$\pm$~0.4	&	$-$	&	B168 (2)\\
		&R$_{\rm{edge}}$	&	${\sim}$1761	&	${\sim}$280	&	V	&	$-$	&	$-$	&	\\
\hline
AS 209		&\textit{D9}	&	\textit{72}	&	\textit{8.69}	&	\textit{$-$}	&	\textit{4.7}	&	\textit{$-$}	&	D9 (1)\\
			&\textit{B14}	&	\textit{117}	&	\textit{14.20}	&	\textit{$-$}	&	\textit{8.9}	&	\textit{$-$}	&	B14 (1)\\
			&\textit{D24}	&	\textit{197}	&	\textit{23.84}	&	\textit{$-$}	&	\textit{3.4}	&	\textit{$-$}	&	D24 (1)\\
			&\textit{B28}	&	\textit{230}	&	\textit{27.80}	&	\textit{$-$}	&	\textit{4.7}	&	\textit{$-$}	&	B28 (1)\\
			&\textit{D35}	&	\textit{290}	&	\textit{35.04}	&	\textit{$-$}	&	\textit{3.0}	&	\textit{$-$}	&	D35 (1)\\
			&\textit{B39}	&	\textit{320}	&	\textit{38.70}	&	\textit{$-$}	&	\textit{3.4}	&	\textit{$-$}	&	B39 (1)\\
			&D61	&	508~$\pm$~25	&	61.43~$\pm$~3.00	&	R	&	11.1~$\pm$~0.8	&	0.38~$\pm$~0.03	&	D61 (1)\\
			&B74	&	611~$\pm$~2	&	73.98~$\pm$~0.20	&	G	&	17.2~$\pm$~1.1	&	$-$	&	B74 (1)\\
			&D100$^a$	&	826~$\pm$~0.0	&	100.00~$\pm$~0.00	&	R	&	30.2~$\pm$~0.7	&	0.06~$\pm$~0.01	&	D90, B97, D105 (1)\\
			&B121	&	1001~$\pm$~0.2	&	121.18~$\pm$~0.02	&	G	&	18.5~$\pm$~0.8	&	$-$	&	B120 (1)\\
			&\textit{D137}	&	\textit{1132}	&	\textit{137.00}	&	\textit{$-$}	&	\textit{4.2}	&	\textit{$-$}	&	\\
			&\textit{B141}	&	\textit{1165}	&	\textit{141.00}	&	\textit{$-$}	&	\textit{2.8}	&	\textit{$-$}	&	\\
		&R$_{\rm{edge}}$	&	${\sim}$1364	&	${\sim}$165	&	V	&	$-$	&	$-$	&	\\
\hline
HD 163296		&\textit{D10}	&	\textit{99}	&	\textit{10.00}	&	\textit{$-$}	&	\textit{3.2}	&	\textit{$-$}	&	D10 (1)\\
			&\textit{B14}	&	\textit{139}	&	\textit{14.00}	&	\textit{$-$}	&	\textit{3.6}	&	\textit{$-$}	&	B14 (1)\\
			&D49	&	490~$\pm$~30	&	49.46~$\pm$~3.00	&	R	&	17.3~$\pm$~0.5	&	0.23~$\pm$~0.03	&	D48 (1)\\
			&B67	&	668~$\pm$~0.4	&	67.44~$\pm$~0.04	&	G	&	21.4~$\pm$~0.7	&	$-$	&	B67 (1)\\
			&D85	&	846~$\pm$~30	&	85.44~$\pm$~3.00	&	R	&	13.3~$\pm$~0.5	&	0.40~$\pm$~0.02	&	D86 (1)\\
			&B101	&	998~$\pm$~1	&	100.78~$\pm$~0.09	&	G	&	20.5~$\pm$~0.4	&	$-$	&	B100 (1)\\
			&D145	&	1440~$\pm$~30	&	145.39~$\pm$~3.00	&	R	&	13.0~$\pm$~0.1	&	0.85~$\pm$~0.02	&	D145 (1)\\
			&B159	&	1571~$\pm$~0.3	&	158.70~$\pm$~0.03	&	G	&	46.5~$\pm$~0.4	&	$-$	&	B155 (1)\\
		&R$_{\rm{edge}}$	&	${\sim}$2376	&	${\sim}$240	&	V	&	$-$	&	$-$	&	\\
\hline
MWC 480		&D76	&	467~$\pm$~30	&	75.57~$\pm$~4.88	&	R	&	24.8~$\pm$~0.4	&	0.20~$\pm$~0.01	&	D73 (3)\\
			&B98	&	606~$\pm$~0.1	&	98.10~$\pm$~0.01	&	G	&	23.7~$\pm$~0.3	&	$-$	&	B98 (3)\\
			&D149	&	919~$\pm$~30	&	148.71~$\pm$~4.88	&	R	&	17.1~$\pm$~0.1	&	0.82~$\pm$~0.04	&	\\
			&B165	&	1023~$\pm$~1	&	165.48~$\pm$~0.08	&	G	&	51.6~$\pm$~0.9	&	$-$	&	\\
		&R$_{\rm{edge}}$	&	${\sim}$1422	&	${\sim}$230	&	V	&	$-$	&	$-$	&	\\
\hline
\enddata
\tablecomments{Column descriptions: (1) Name of host star. (2) Substructure label. (3) Radial location of substructure in mas (the uncertainties in mas are simply scaled from the fitting procedure and do not account for the uncertainty in the distance to the source). (4) Radial location of substructure in au. (5) Method used to derive radial location of substructure: ``G" indicates Gaussian-fitting, ``R" indicates identification of local extrema in radial profiles, and ``V" indicates identification through visual inspection. (6) Width of substructure. (7) Depth of gap, defined as the intensity ratio of adjacent ring-gap pairs (see Section \ref{sec:widths_and_contrast}). All uncertainties are 1$\sigma$. (8) Reference for previously known annular continuum substructures. Italics indicate continuum substructures unresolved in MAPS that were from adopted from previous observations at higher spatial resolution. References: (1) \citet{Huang18}; (2) \citet{Huang20}; (3) \citet{Long18}.}
\tablenotetext{a}{Observed as a single gap in MAPS, while in DSHARP it is resolved into three substructures: D90, B97, D105.}
\end{deluxetable*}

\section{Relationship between chemical and NIR substructures} \label{sec:app:NIR_rings_appendix}

\begin{figure*}[!ht]
\centering
\figurenum{31}
\includegraphics[width=\linewidth]{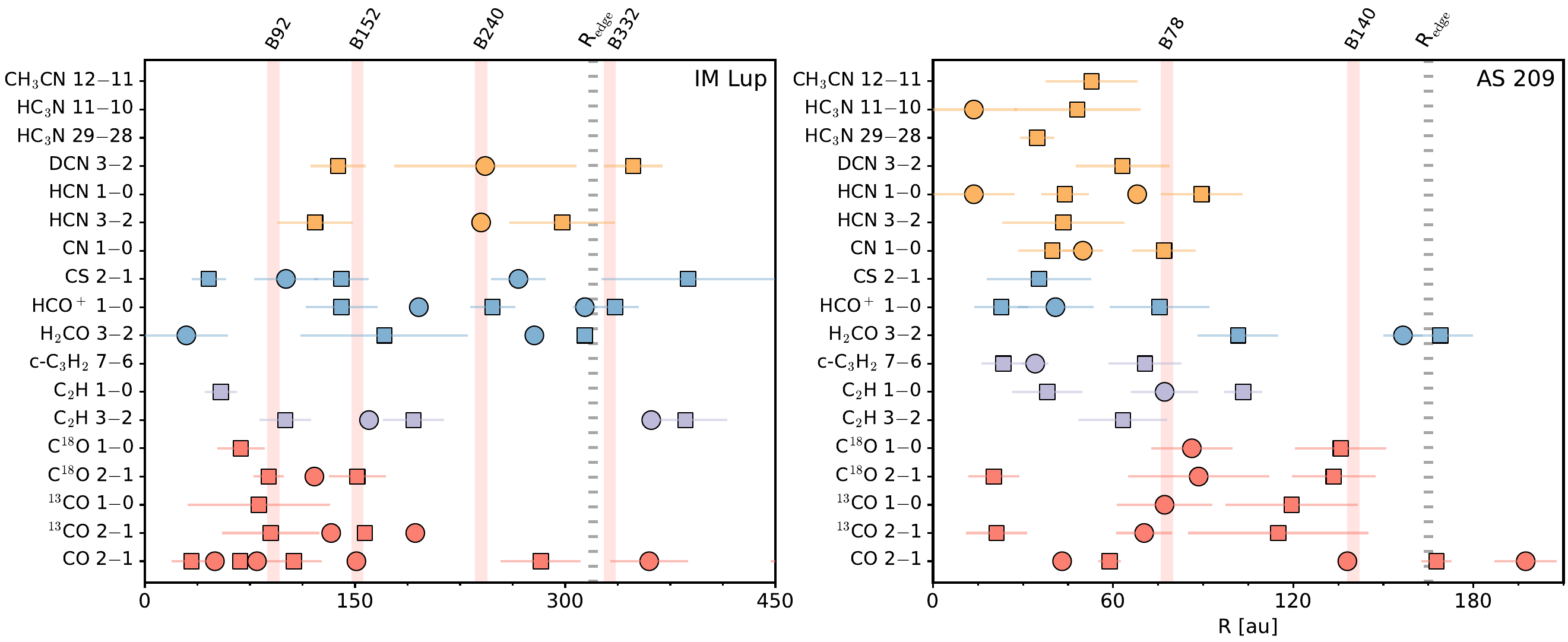}\\
\includegraphics[width=\linewidth]{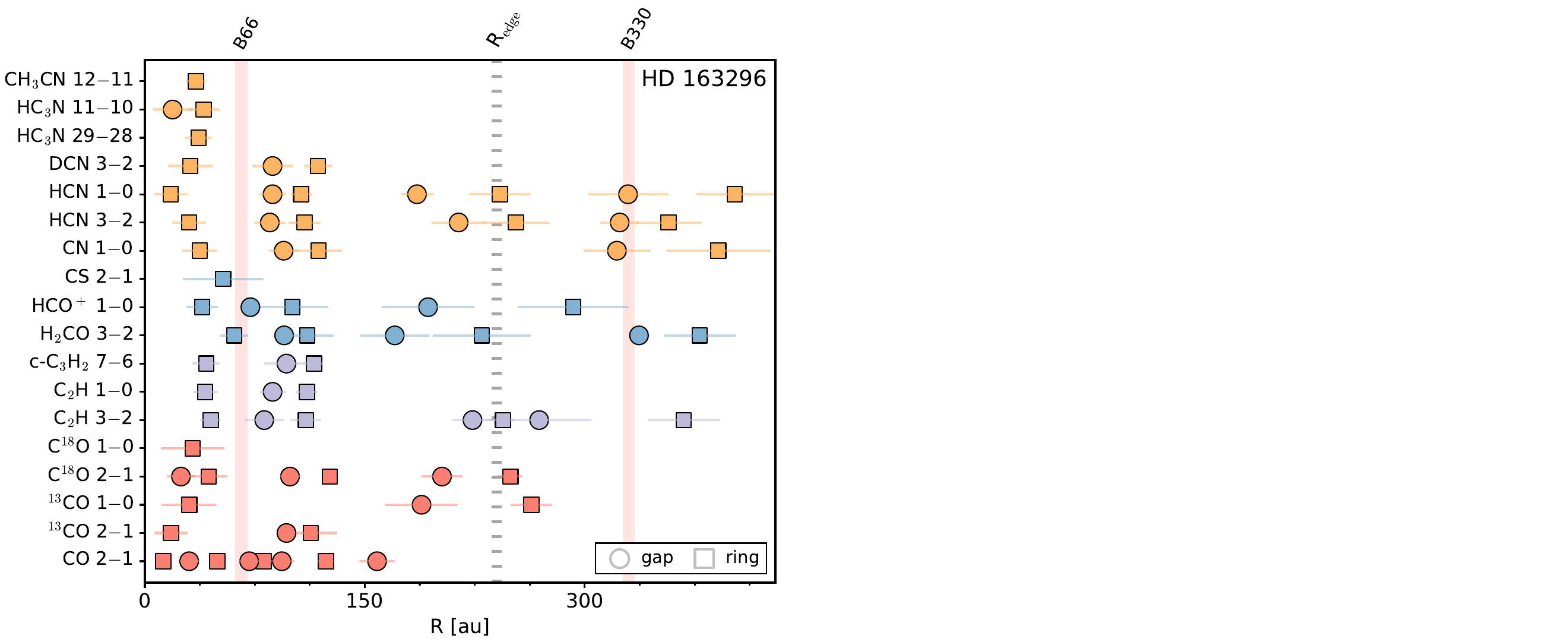}
\caption{Radial locations of chemical substructures and NIR rings in IM~Lup, AS~209, and HD~163296. Line emission rings and gaps are shown as squares and circles, respectively. Species are color-coded, as in Figures \ref{fig:disk_sizes}, \ref{fig:Spearman_disk_size}, and \ref{fig:r0_histogram}. Red lines mark the radial locations of NIR rings \citep{Monnier17, Avenhaus18}. Broad dotted lines mark the location of the edge of millimeter continuum disk. Chemical substructures at large radii beyond the outermost NIR rings, which are only seen in CO 2--1, are omitted. The widths of error bars for chemical substructures represent $\sigma$ instead of the full FWHM, i.e., FWHM / 2.355, for visual clarity.}
\label{fig:NIR_ring-vs-chemsubstr}
\end{figure*}

Figure \ref{fig:NIR_ring-vs-chemsubstr} shows the radial locations of chemical substructures versus NIR rings in the three MAPS disks with known NIR substructure.

\newpage
\clearpage

\bibliography{bibliography}{}
\bibliographystyle{aasjournal}

%% This command is needed to show the entire author+affiliation list when the collaboration and author truncation commands are used.  It has to go at the end of the manuscript.

%% Include this line if you are using the \added, \replaced, \deleted commands to see a summary list of all changes at the end of the article.

\end{document}